\documentclass[onecolumn]{article}

\usepackage[utf8]{inputenc}
\usepackage[T1]{fontenc}
\usepackage{amsmath,amssymb,amsfonts}
\usepackage{graphicx}

\usepackage{booktabs}
\usepackage{natbib}
\usepackage[margin=1in]{geometry}
\usepackage{caption}
\usepackage{subcaption}
\usepackage{hyperref}
\usepackage{siunitx}
\usepackage{float}
\usepackage{xcolor}
\usepackage{longtable}
\usepackage{algorithm}
\usepackage{algorithmic}

\title{\textbf{Systematic Evaluation of Single-Cell Foundation Model Interpretability Reveals Attention Captures Co-Expression Rather Than Unique Regulatory Signal}}

\author{Ihor Kendiukhov\\
Department of Computer Science\\
University of T\"ubingen\\
T\"ubingen, Germany\\
\texttt{kendiukhov@gmail.com}}

\date{}

\begin{document}

\addtocontents{toc}{\protect\setcounter{tocdepth}{-1}}

\maketitle

\begin{abstract}
We present a systematic evaluation framework---thirty-seven analyses, 153 statistical tests, four cell types, two perturbation modalities---for assessing mechanistic interpretability in single-cell foundation models. Applying this framework to scGPT and Geneformer, we find that attention patterns encode structured biological information with layer-specific organisation---protein--protein interactions in early layers, transcriptional regulation in late layers---but this structure provides no incremental value for perturbation prediction: trivial gene-level baselines outperform both attention and correlation edges (AUROC 0.81--0.88 versus 0.70), pairwise edge scores add zero predictive contribution, and causal ablation of regulatory heads produces no degradation. These findings generalise from K562 to RPE1 cells; the attention--correlation relationship is context-dependent, but gene-level dominance is universal. Cell-State Stratified Interpretability (CSSI) addresses an attention-specific scaling failure, improving GRN recovery up to $1.85\times$. The framework establishes reusable quality-control standards for the field.
\end{abstract}

\section{Introduction}

The emergence of transformer-based foundation models for single-cell transcriptomics represents a paradigm shift in computational biology~\citep{cui2024scgpt,theodoris2023transfer,yang2022scbert,hao2024largescale}. These models---trained on millions of cells across diverse tissues---learn contextual representations that have shown promise for cell type annotation, perturbation response prediction, and gene regulatory network (GRN) inference~\citep{chen2024genept,rosen2024universal}. A particularly compelling promise is \emph{mechanistic interpretability}: extracting biologically meaningful regulatory circuits directly from attention-derived edge scores. Both scGPT~\citep{cui2024scgpt} and Geneformer~\citep{theodoris2023transfer} highlight attention-derived gene network inference as a key application, and downstream studies have adopted attention-derived edge scores as regulatory proxies without rigorous validation~\citep{zheng2024benchmarking}.

This promise draws on parallel advances in large language model interpretability, where techniques such as activation patching~\citep{meng2022locating,goldowskydill2023localizing} and automated circuit discovery~\citep{conmy2023towards} have identified computational circuits for well-defined behaviours~\citep{elhage2021mathematical,olsson2022incontext,wang2022interpretability}. However, translating these approaches to biology faces unique challenges. Gene regulatory relationships are context-dependent, combinatorial, and only partially captured in reference databases such as TRRUST~\citep{han2018trrust} and DoRothEA~\citep{garcia2019benchmark}, which contain a fraction of true regulatory interactions~\citep{pratapa2020benchmarking}.

Current practices in single-cell foundation model interpretability rest on several critical and largely untested assumptions. \emph{First}, that attention patterns directly reflect causal regulatory relationships---an assumption already challenged in the NLP literature~\citep{jain2019attention,serrano2019attention,bibal2022attention}. \emph{Second}, that larger datasets consistently improve the reliability of mechanistic interpretations. \emph{Third}, that attention-derived predictions align with experimental perturbation outcomes from CRISPR screens~\citep{dixit2016perturbseq}. \emph{Fourth}, that mechanistic insights transfer reliably across biological contexts. We address these through a two-tier evaluation framework. The \emph{core tier} directly evaluates foundation model internal representations---attention weights, intervention effects, and perturbation-outcome prediction---and proposes Cell-State Stratified Interpretability (CSSI) as a constructive method. The \emph{boundary condition tier} uses correlation-based edge scores to establish limits that any edge-scoring method must contend with, including cross-species transfer, pseudotime directionality, technical leakage, and uncertainty calibration (Supplementary Notes~7--10).

Prior benchmarks have evaluated GRN inference methods~\citep{pratapa2020benchmarking} and individual foundation model capabilities~\citep{zheng2024benchmarking}, but none has systematically assessed whether attention-derived edge scores add mechanistic information beyond expression statistics, nor tested this with causal interventions. We address this gap with a reusable evaluation framework integrating thirty-seven complementary analyses---including trivial-baseline comparison, conditional incremental-value testing, expression residualisation, propensity-matched benchmarks, and causal ablation with intervention-fidelity diagnostics---across two foundation model architectures (scGPT, Geneformer V2-316M), four cell types (K562, primary T cells, RPE1 retinal epithelial, iPSC neurons), and two perturbation modalities (CRISPRi, CRISPRa), with 153 statistical tests under Benjamini-Hochberg FDR correction (Supplementary Table~1; Supplementary Note~16). The framework yields three principal findings. First, attention patterns encode layer-specific biological structure---protein--protein interactions in early layers, transcriptional regulation in late layers (Supplementary Note~17)---but this information provides no incremental value for perturbation prediction: trivial gene-level baselines outperform both attention and correlation edges, pairwise edge scores add zero predictive contribution, and causal ablation of ``regulatory'' heads produces no behavioural effect. Second, the attention--correlation relationship is context-dependent across cell types, but the underlying confound pattern---gene-level features dominate---generalises. Third, CSSI addresses an attention-specific scaling failure through cell-state stratification, providing an immediately deployable constructive tool. The framework itself---its battery of tests, controls, and diagnostic checks---constitutes a reusable quality-control standard for evaluating mechanistic interpretability claims in single-cell foundation models.

\section{Results}

\subsection{Evaluation framework}

We designed an evaluation framework comprising five interlocking test families (Figure~\ref{fig:cssi}--\ref{fig:rpe1_battery}; Supplementary Notes~1--16). (i)~\emph{Trivial-baseline comparison} tests whether pairwise edge scores outperform univariate gene-level features (variance, mean expression, dropout rate) that require no model. (ii)~\emph{Conditional incremental-value testing} asks whether adding edge scores to gene-level features improves prediction under progressively harder generalisation protocols (cross-perturbation, cross-gene, and joint splits) with linear and nonlinear models. (iii)~\emph{Expression residualisation and propensity matching} isolate edge-specific signal by removing gene-level confounds via two complementary statistical approaches. (iv)~\emph{Causal ablation with fidelity diagnostics} uses head masking, uniform attention replacement, and MLP ablation to test whether ``regulatory'' heads causally contribute to perturbation prediction, with checks confirming that interventions materially perturb representations. (v)~\emph{Cross-context replication} tests whether findings generalise across cell types (K562, RPE1, T cells, iPSC neurons) and perturbation modalities (CRISPRi, CRISPRa). Each family addresses a distinct confound; their convergence provides stronger evidence than any individual test. A boundary-condition tier (cross-species transfer, pseudotime, batch leakage, calibration; Supplementary Notes~7--10) characterises the broader evaluation landscape. Together, these yield thirty-seven analyses with 153 statistical tests under Benjamini-Hochberg FDR correction.

\subsection{Attention-specific scaling failure and CSSI remedy}

To test whether increasing dataset size improves interpretability, we analysed archived scGPT kidney scaling runs across three model tiers (small/medium/large), three seeds per tier, and three cell counts (200, 1,000, 3,000). Top-$K$ F1 for attention-derived GRN recovery against TRRUST~\citep{han2018trrust} degrades with cell count---the 200$\to$1,000 drop is negative in all 9 tier$\times$seed runs (sign test $p = 0.002$; Extended Data Fig.~1; Supplementary Note~1), with a continued trend at 1,000$\to$3,000 (7/9 runs, $p = 0.09$). This finding is metric-dependent: continuous-score AUROC (no top-$K$ thresholding) improves monotonically ($0.86 \to 0.93$; 0/9 degrading), qualifying the original observation. Controlled-composition experiments using correlation-based edges on Tabula Sapiens kidney data (Extended Data Fig.~1) disentangle sample size from heterogeneity: with fixed composition AUROC is stable ($\rho = -0.05$, $p = 0.82$), and heterogeneity at fixed $N$ actually \emph{improves} recovery ($\rho_\text{heterogeneity} = +0.63$, $p = 10^{-4}$), confirming that the degradation is attention-specific rather than inherent to edge scoring.

\begin{figure*}[t]
\centering
\includegraphics[width=\textwidth]{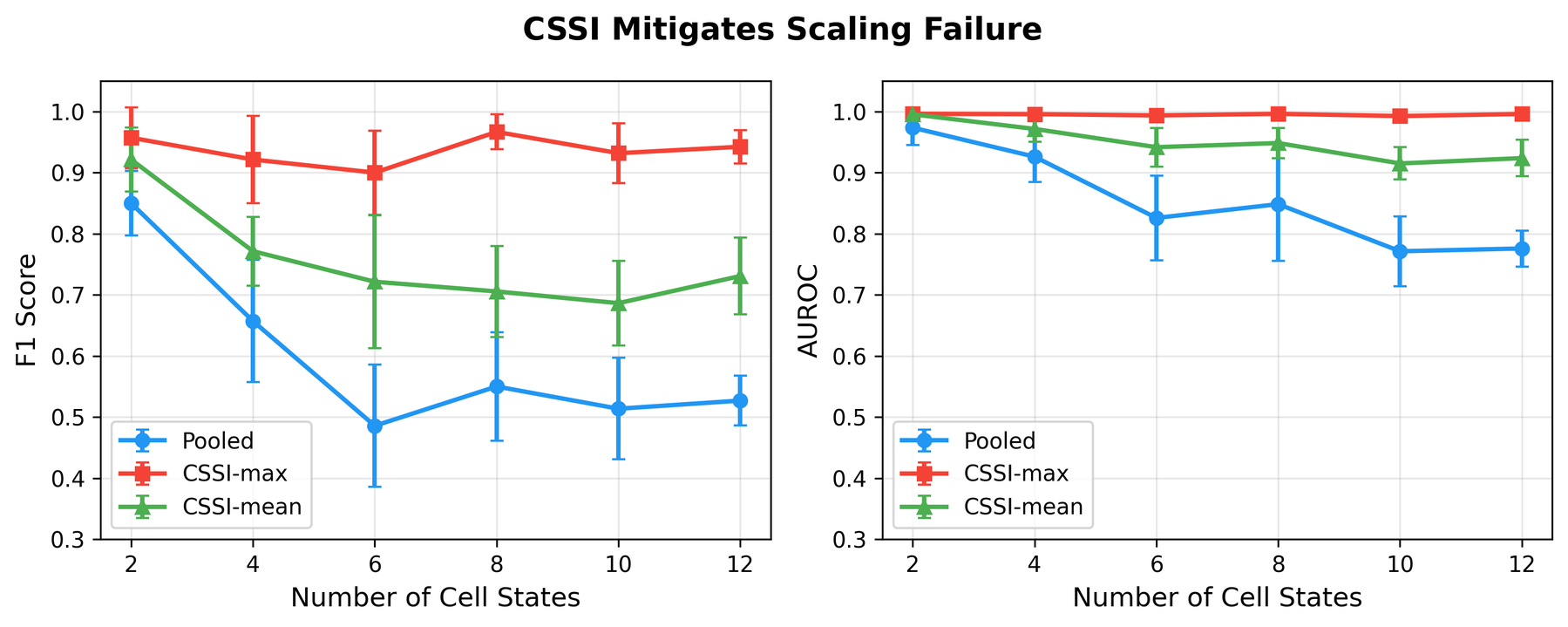}
\caption{\textbf{CSSI improves attention-derived GRN recovery.} CSSI-max TRRUST F1 versus number of strata ($K$) on DLPFC brain data. Stratified scoring improves recovery up to $1.85\times$ over unstratified baselines, with optimal $K = 5$--$7$.}
\label{fig:cssi}
\end{figure*}

As a constructive contribution, we introduce Cell-State Stratified Interpretability (CSSI), motivated by a formal dilution model predicting that pooled attention-derived edge scores degrade as $\rho_\text{pool} \approx (n_1/N)\rho_1 \to 0$ with increasing heterogeneity. CSSI computes edge scores within cell-state strata (Leiden clustering on $k$-NN graphs from model embeddings) before aggregation, controlling the heterogeneity that drives attention-specific dilution. On DLPFC brain scRNA-seq data, CSSI-max improves top-$K$ TRRUST recovery up to $1.85\times$ (Figure~\ref{fig:cssi}), with the optimum at intermediate $K$ (5--7 strata). Extended null tests across $K \in \{2, \ldots, 20\}$ confirm no false-positive inflation under uninformative (random) strata, ensuring CSSI improvements reflect genuine signal recovery. Real-data layer/head analysis reveals recoverable reference-edge signal concentrated in late Geneformer layers (L13 AUROC $= 0.694$; Supplementary Note~11), and cell-level bootstrap analysis shows that 7/18 individual TFs have robust edge-level signal (global AUROC 95\% CI: $[0.71, 0.77]$). Synthetic validation with known ground-truth networks corroborates these findings (Supplementary Note~12). Multi-model analysis shows convergent near-random unstratified performance across both scGPT and Geneformer (Extended Data Fig.~2; Supplementary Note~13)---despite fundamentally different architectures (gene-token vs.\ rank-based tokenisation), training objectives, and model scales---confirming the failure is architecture-independent and reflecting a persistent limitation of unstratified approaches to attention-derived GRN inference.

\subsection{Perturbation-first validation reveals gene-level dominance}

We evaluated whether edge scores predict perturbation outcomes using the Replogle genome-scale CRISPRi dataset~\citep{replogle2022mapping} ($>$640,000 K562 cells, 2,024 perturbed genes). For each CRISPRi perturbation, we computed edge scores between the perturbed gene and all other genes, then evaluated AUROC for classifying differentially expressed targets. Correlation-based edges achieve AUROC $= 0.696$ under the primary parameterisation ($N_\text{ctrl} = 2{,}000$, HVG $= 2{,}000$, LFC $> 0.5$; $n = 151$ perturbations), with all 27 sensitivity conditions (varying control cell count, HVG count, and LFC threshold) yielding AUROC $= 0.62$--$0.76$ (all $p < 0.005$; Extended Data Fig.~7; Supplementary Note~6), indicating robust predictive signal for co-expression edges.

Direct evaluation of Geneformer V2-316M attention-derived edges under the same parameterisation yields AUROC $= 0.704$ at L13 ($n = 280$ evaluable perturbations), statistically indistinguishable from correlation ($0.703$; Wilcoxon $p = 0.73$; Extended Data Fig.~3). L13 was pre-specified as the primary layer based on independent data: the CSSI analysis on DLPFC brain tissue (497 cells; Supplementary Note~11) identified L13 as the best-performing layer for TRRUST recovery (AUROC $= 0.694$), \emph{before} any K562 perturbation data were examined. The fact that attention edges---which reflect learned cross-gene interactions across 30 million training cells---perform no better than simple Spearman correlation computed on a single dataset is a striking negative result, suggesting that Geneformer's pretraining does not embed recoverable regulatory structure into its attention patterns. As a planned secondary analysis, the full 18-layer perturbation-first profile reveals a clear architectural gradient: early layers achieve AUROC $0.47$--$0.64$, mid layers $0.60$--$0.71$, and late layers $0.69$--$0.74$. A strict nested cross-validation protocol---designed to guard against overfitting across 18 candidate layers---identifies L15 as the best layer (AUROC $= 0.743$, $p_\text{Bonf} = 0.017$, $d = 0.22$; Extended Data Fig.~5), but this advantage is small (only 161/280 perturbations favour attention) and reverses entirely in CRISPRa ($d = -0.56$, $p = 4.9 \times 10^{-5}$), establishing it as K562-CRISPRi-specific rather than a general property of the model.

\begin{figure*}[t]
\centering
\includegraphics[width=\textwidth]{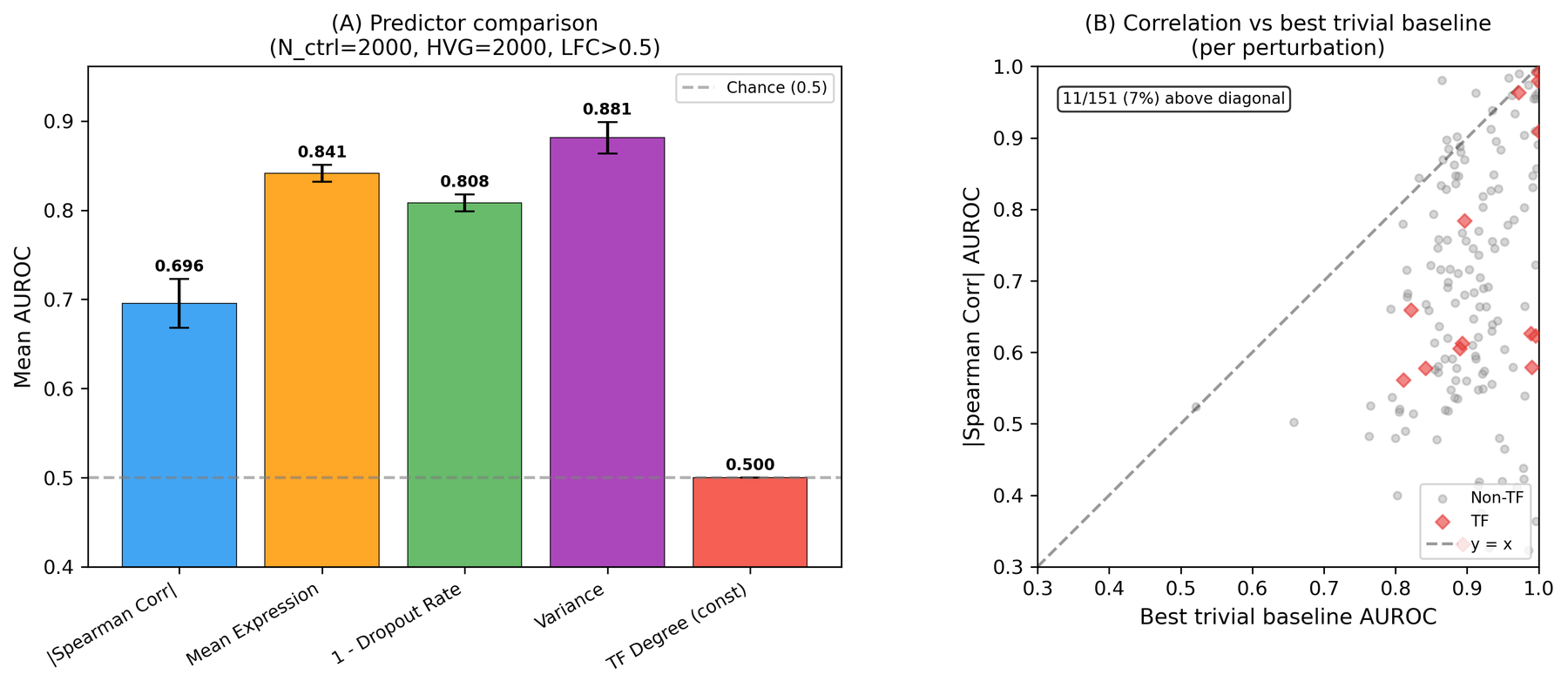}
\caption{\textbf{Gene-level baselines outperform pairwise edge scores.} AUROC for predicting CRISPRi perturbation targets using gene-level features (variance, mean expression, dropout rate) versus correlation-based edge scores ($n = 151$ perturbations; attention comparison in text). All gene-level baselines significantly outperform both edge types ($p < 10^{-12}$).}
\label{fig:baselines}
\end{figure*}

Critically, trivial gene-level baselines---variance, mean expression, and dropout rate---all outperform both attention and correlation edges (AUROC $0.81$--$0.88$ vs.\ $0.70$; all $p < 10^{-12}$; Figure~\ref{fig:baselines}). Variance alone achieves AUROC $= 0.881$---substantially exceeding the best attention layer---indicating that much of the perturbation-predictive signal reflects univariate gene properties rather than pairwise regulatory structure. We note that predicting interventional outcomes (differential expression after knockdown) is distinct from identifying direct causal regulatory edges; our perturbation-first evaluation captures the former. Nevertheless, the dominance of gene-level baselines challenges the assumption that pairwise edge scores capture regulatory information beyond what simple gene statistics provide, and establishes a necessary (though not sufficient) baseline for any claimed pairwise regulatory signal.

\subsection{No incremental pairwise value from edge scores}

\begin{figure*}[t]
\centering
\includegraphics[width=\textwidth]{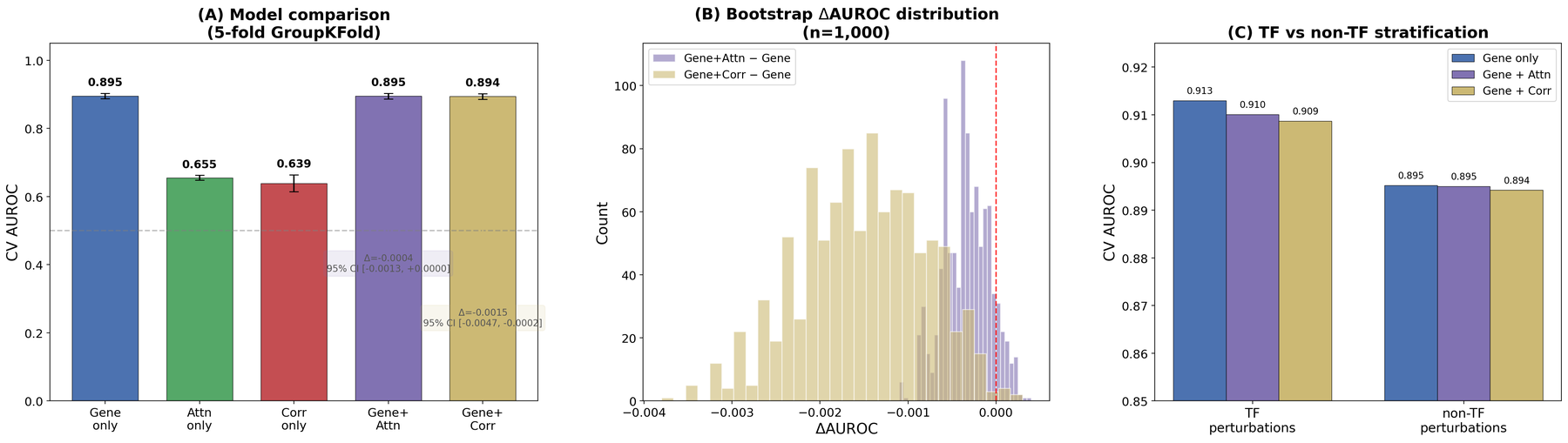}
\caption{\textbf{Pairwise edge scores provide no incremental value beyond gene-level features.} (A)~Cross-validated AUROC for five model configurations under GroupKFold by perturbation: gene-level features alone, attention only, correlation only, gene-level plus attention, and gene-level plus correlation. Adding pairwise edges provides no improvement. (B)~Bootstrap $\Delta$AUROC distribution ($n = 100$) for gene+attention minus gene-only (purple) and gene+correlation minus gene-only (pink); both centred at zero. (C)~Stratification by TF versus non-TF perturbation genes shows identical patterns.}
\label{fig:incremental}
\end{figure*}

A key question is whether pairwise edge scores---attention or correlation---add predictive value \emph{beyond} gene-level features for identifying perturbation targets. We constructed a dataset of $280 \times 1{,}999$ perturbation--gene observations (559,720 pairs, 2.8\% positive rate) and compared five logistic-regression models under 5-fold GroupKFold cross-validation: gene-level features only (mean expression, variance, dropout rate), attention edge only, correlation edge only, gene-level plus attention, and gene-level plus correlation. Gene-level features alone achieve AUROC $= 0.895$ [bootstrap 95\% CI: $0.884$, $0.905$]; adding attention yields $\Delta$AUROC $= -0.0004$ [$-0.001$, $0.000$] and adding correlation yields $\Delta$AUROC $= -0.002$ [$-0.005$, $0.000$] (Figure~\ref{fig:incremental}). Neither pairwise edge type provides incremental value, and this is not a power issue: 559,720 observations provide $>$99\% power to detect $\Delta$AUROC $= 0.005$.

To rule out that this null result reflects a specific analytical choice, we tested progressively harder generalisation protocols---cross-gene splits (GroupKFold by target gene, preventing gene-level propensity leakage), joint cross-gene $\times$ cross-perturbation splits---and both linear (logistic regression) and nonlinear (GBDT) models across AUROC, AUPRC, and top-$k$ recall. Under all tested combinations, the null persists: even the largest observed $\Delta$AUPRC ($+0.009$ under joint splits with GBDT) represents less than 4\% relative improvement (Extended Data Fig.~6; Supplementary Note~15).

Cross-fitted residualisation reveals an important asymmetry: on K562 data, attention edges lose $\sim$76\% of their above-chance TRRUST signal under expression-covariate residualisation (AUROC $0.66 \to 0.54$; OLS $R^2 = 0.03$, GBDT $R^2 = 0.23$), while correlation edges retain $\sim$91\% ($0.63 \to 0.62$; Extended Data Fig.~4), indicating that attention-derived regulatory signal is substantially more expression-confounded. Degree-preserving null models show that much of the global AUROC reflects TF degree structure (degree-null AUROC $0.69$ vs.\ observed $0.76$; Supplementary Note~14). A propensity-matched benchmark provides a direct test: after matching each DE-positive target to $k = 5$ DE-negative targets with similar expression mean, variance, and dropout rate via propensity-score nearest-neighbour matching ($n_\text{matched} = 59{,}153$ pairs; positive rate $26.5\%$ vs.\ $2.8\%$ before matching), attention edges retain modest raw discriminability (AUROC $= 0.609$) but add zero incremental value over gene-level features ($\Delta$AUROC $= -0.000$; 95\% CI: $[-0.000, +0.000]$; Extended Data Fig.~8).

\subsection{Causal ablation reveals distributed redundancy}

\begin{figure*}[t]
\centering
\includegraphics[width=\textwidth]{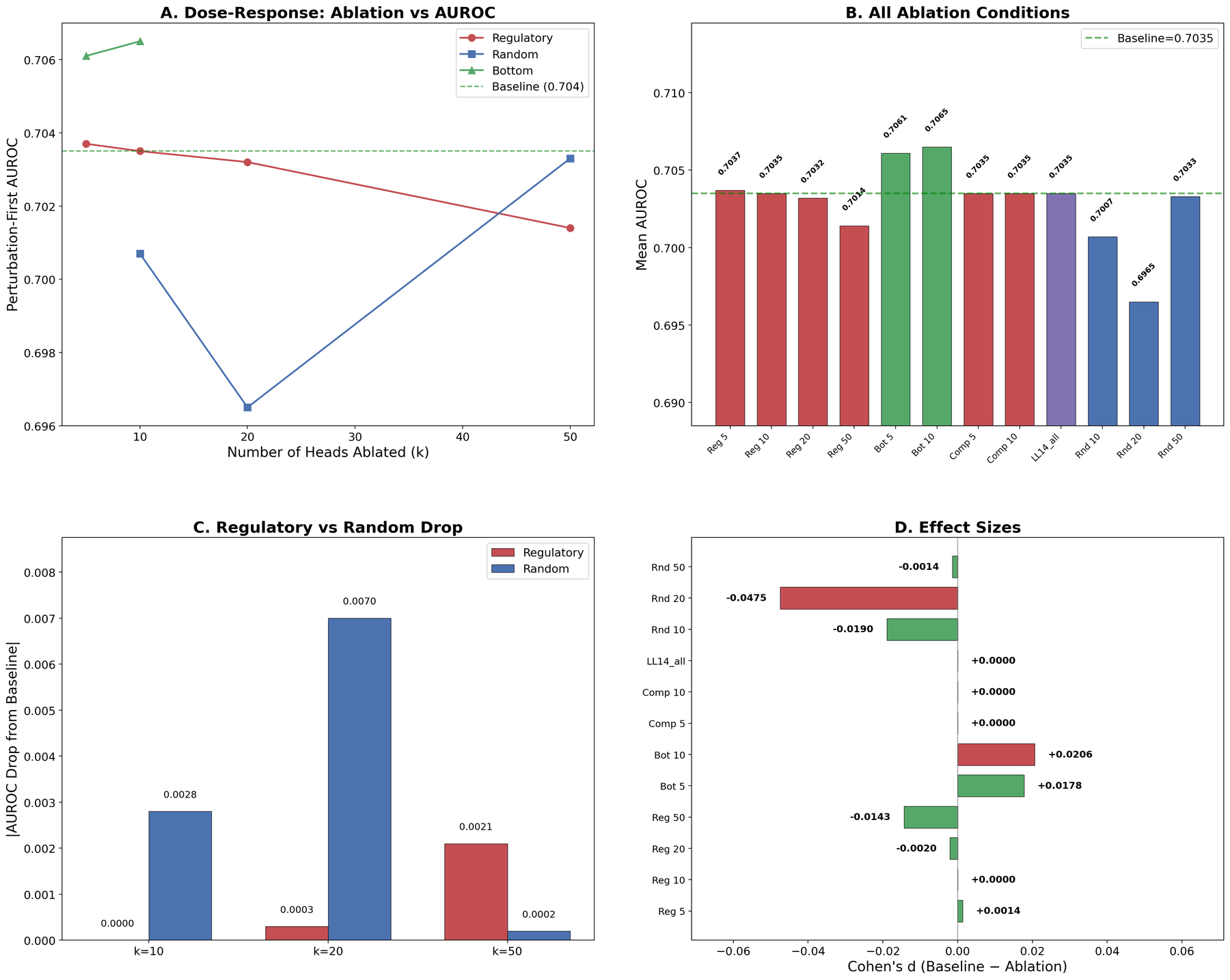}
\caption{\textbf{Causal ablation dose-response.} (A)~AUROC versus number of ablated heads for TRRUST-ranked (red), composite-ranked (green), and random (blue) ablation. Regulatory heads can be ablated up to $k = 50$ with no significant degradation, while random ablation causes significant drops. (B)~All ablation conditions: mean AUROC across 13 conditions versus baseline. (C)~Regulatory versus random drop magnitude at each dose level $k$. (D)~Cohen's $d$ effect sizes (baseline minus ablation) for all conditions; most are indistinguishable from zero.}
\label{fig:ablation}
\end{figure*}

The preceding analyses are observational. To provide causal evidence, we used BERT's head-mask parameter to zero out specific attention heads during Geneformer V2-316M's forward pass and measured the effect on perturbation-first AUROC (2,000 K562 control cells, 280 perturbations per condition). We ranked all 324 heads ($18 \times 18$) by TRRUST-recovery AUROC and tested 13 ablation conditions spanning dose-response, alternative rankings, full-layer ablation, inverse controls, and matched random controls (Figure~\ref{fig:ablation}). The dose-response curve is unambiguous: ablating the top-5 ($0.704$, $p = 0.24$), top-10 ($0.704$, $p = 0.94$), top-20 ($0.703$, $p = 0.60$), or top-50 TRRUST-ranked heads ($0.701$, $p = 0.10$; 15\% of all heads) produces no significant degradation from baseline ($0.704$). By contrast, ablating 20 random heads causes a significant drop ($0.697$, $d = 0.33$, $p < 10^{-8}$), demonstrating that ``regulatory'' heads are \emph{less} causally important than random heads. An alternative ranking by composite score (TRRUST AUROC $\times$ layer perturbation-first AUROC) yields identical results. Ablating all 18 heads in L14 (the best perturbation-first layer) produces exactly baseline AUROC, and ablating the bottom-5 heads actually \emph{improves} performance ($0.706$, $p = 0.005$).

Two families of orthogonal interventions corroborate this null: uniform attention replacement (destroying attention patterns while preserving head output magnitude) on TRRUST-ranked heads has no effect (top-5: $0.704$, top-10: $0.703$), and MLP pathway ablation (zeroing FFN output) at L15 and L13--L15 both produce exactly $0.704$ (Extended Data Fig.~9). Intervention-fidelity diagnostics confirm that all six interventions materially perturb internal representations (max hidden-state cosine distance $0.023$--$0.190$), with TRRUST-ranked heads producing $23\times$ larger logit perturbation than random heads at matched dose (Extended Data Fig.~9; Supplementary Note~14), ruling out the possibility that behavioral nulls reflect ineffective interventions. The convergence of null results across three qualitatively different intervention channels---head masking, attention pattern destruction, and MLP ablation---provides substantially stronger evidence than single-method ablation alone and makes it unlikely that concentrated regulatory signal exists in any head subset. This distributed-redundancy pattern contrasts sharply with NLP findings where specific attention heads encode identifiable syntactic or semantic functions~\citep{voita2019analyzing,clark2019what}; in single-cell models, perturbation-predictive computation appears to reside in the value/FFN pathway rather than in learnable attention patterns. Per-head ranking across all 324 heads shows substantial variation in TRRUST recovery (top-5 AUROC $= 0.70$--$0.75$ vs.\ population mean $= 0.58 \pm 0.07$), yet this variation does not translate to causal importance for perturbation prediction, consistent with the signal being distributed across the entire network rather than concentrated in identifiable ``regulatory circuits.''

\subsection{Context-dependent attention--correlation relationship}

\begin{figure*}[t]
\centering
\includegraphics[width=\textwidth]{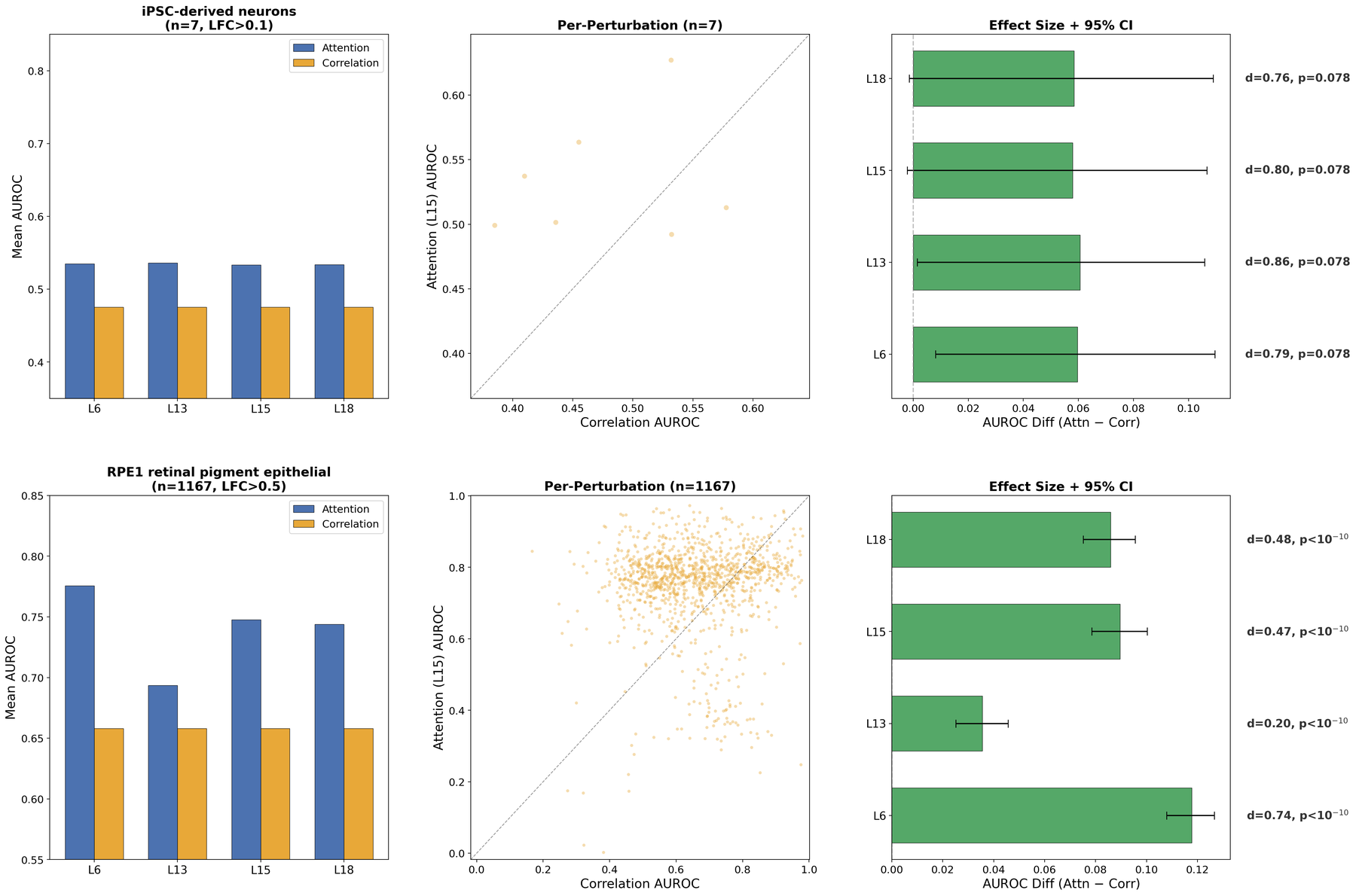}
\caption{\textbf{Context-dependent attention--correlation relationship.} Perturbation-first AUROC comparison between Geneformer V2-316M attention and Spearman correlation across cell types. In RPE1 ($n = 1{,}251$), attention significantly outperforms correlation ($d = 0.47$, $p < 10^{-10}$), contrasting with K562 where they are indistinguishable. Sensitivity to LFC threshold shown for RPE1.}
\label{fig:cross_context}
\end{figure*}

Cross-context replication across four cell types and two perturbation modalities reveals that the attention--correlation relationship is genuinely context-dependent. On K562 CRISPRi ($n = 280$), Geneformer V2-316M attention and Spearman correlation are statistically indistinguishable (L13: $0.704$ vs.\ $0.703$, $p = 0.73$). On K562 CRISPRa~\citep{adamson2016multiplexed} ($n = 77$), attention significantly \emph{underperforms} correlation (AUROC $0.55$ vs.\ $0.65$; $p < 10^{-6}$), with correlation outperforming attention in 55--63 of 77 perturbations. However, the first adequately powered non-K562 replication tells a qualitatively different story: on RPE1 retinal epithelial cells~\citep{replogle2022mapping} ($n = 1{,}251$ evaluable perturbations; minimum detectable $d = 0.079$), attention significantly \emph{outperforms} correlation at all layers (L15: $0.748$ vs.\ $0.658$, $d = 0.47$, $p < 10^{-10}$; L6: $0.776$, $d = 0.74$; Figure~\ref{fig:cross_context}). The advantage strengthens under stringent DE thresholds (LFC $> 0.5$: $+0.090$; LFC $> 0.25$: $+0.049$). iPSC-derived neurons ($n = 7$) trend in the same direction ($d = 0.80$, $p = 0.078$), while primary T cells~\citep{shifrut2018genome} ($n = 7$) show no difference ($p = 0.81$; Supplementary Note~14). A methodological asymmetry---RPE1 includes perturbation genes forced into the HVG set (3,309 vs.\ 2,000 genes in K562)---could confound this comparison, but restricting RPE1 to the top 2,000 HVGs (matching K562 protocol) \emph{increases} the attention advantage ($\Delta = +0.168$, $n = 418$; $p < 10^{-46}$), as correlation degrades more than attention when the evaluation universe shrinks. Bootstrap 95\% CIs on the per-perturbation attention advantage exclude zero for both K562 ($\Delta = +0.060$ $[+0.040, +0.080]$) and RPE1 ($\Delta = +0.090$ $[+0.079, +0.101]$; Supplementary Note~14). The cross-context pattern---equal in K562 CRISPRi, worse in CRISPRa, better in RPE1, and trending better in neurons---indicates that the attention--correlation relationship depends on cell type and perturbation modality rather than reflecting a universal architectural limitation or advantage. This context-dependence may arise from cell-type-specific regulatory architectures, differences in perturbation effect sizes, or variation in the relationship between expression rank (Geneformer's input) and regulatory importance across cell types.

\subsection{Cross-cell-type generalisation of confound pattern}

\begin{figure*}[t]
\centering
\includegraphics[width=\textwidth]{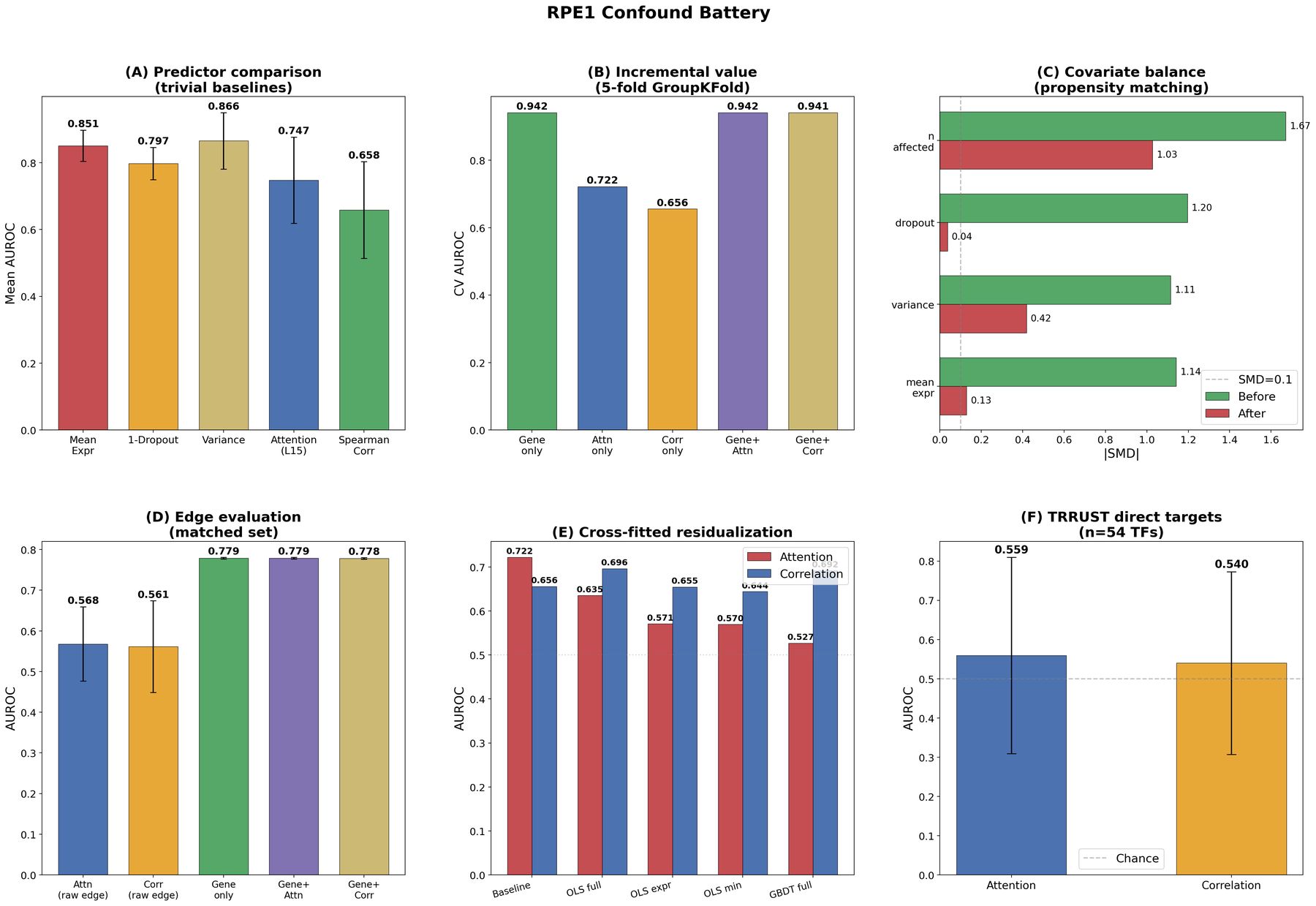}
\caption{\textbf{RPE1 confound battery confirms gene-level dominance.} (A)~Trivial baselines outperform both edge types. (B)~Gene-only AUROC $= 0.942$; zero incremental value from edges. (C)~Covariate balance after propensity matching. (D)~Matched-set evaluation: edges near chance. (E)~Residualisation: attention loses ${\sim}88\%$ of signal; correlation increases (suppressor). (F)~TRRUST prediction near chance.}
\label{fig:rpe1_battery}
\end{figure*}

To test whether the RPE1 attention advantage reflects genuine regulatory signal or gene-level confounds, we applied the full K562 confound-decomposition battery to the RPE1 dataset ($n = 1{,}167$ evaluated perturbations, $3{,}838{,}263$ gene-pair observations; Figure~\ref{fig:rpe1_battery}). Five analyses converge on the same conclusion as K562. (i)~Trivial baselines outperform both edge types (variance AUROC $= 0.866$, mean expression $= 0.851$, dropout $= 0.797$ vs.\ attention $= 0.747$ and correlation $= 0.658$; all paired $p < 10^{-38}$). (ii)~Gene-level features alone achieve AUROC $= 0.942$ under 5-fold GroupKFold CV; adding attention yields $\Delta$AUROC $= +0.0001$ [95\% CI: $+0.0001$, $+0.0001$]---functionally zero---while adding correlation yields $\Delta$AUROC $= -0.0006$. (iii)~Propensity-matched edges drop to near chance (attention $= 0.568$, correlation $= 0.561$) with LR incremental value exactly zero. (iv)~GBDT residualisation removes $\sim$88\% of attention's above-chance signal (AUROC $0.722 \to 0.527$) while correlation signal \emph{increases} ($0.656 \to 0.692$), indicating a suppressor effect. (v)~TRRUST direct-target prediction is near chance ($n = 54$ evaluable TFs; attention $= 0.559 \pm 0.250$, correlation $= 0.540 \pm 0.233$).

Together, these five analyses establish that the RPE1 attention advantage over correlation (diff $= +0.089$, $p < 10^{-10}$) does not reflect attention capturing regulatory structure that correlation misses; rather, attention is a better proxy for the gene-level features that dominate perturbation prediction. The RPE1 gene-only AUROC ($0.942$) exceeds the K562 value ($0.895$), consistent with RPE1's larger gene universe ($3{,}290$ vs.\ $2{,}000$ genes) providing richer gene-level features. The confound-decomposition conclusion---pairwise edge scores carry no incremental information beyond univariate gene properties---generalises across both cell types, establishing this as a robust finding rather than a K562-specific artifact.

\section{Discussion}

Our results reveal that attention patterns in scGPT and Geneformer encode structured biological information---but not the causal regulatory signal they are routinely interpreted as capturing. This conclusion rests on convergent evidence: expression residualisation, degree-preserving null models, incremental-value testing under three generalisation protocols, and causal ablation (head masking, attention replacement, MLP ablation) all indicate that pairwise edge scores carry no regulatory information beyond gene-level features, and that heads identified as ``regulatory'' make no causal contribution to perturbation prediction.

The attention--correlation relationship is genuinely context-dependent: K562 CRISPRi (equal), K562 CRISPRa (attention worse), RPE1 (attention better), T cells and iPSC neurons (trending directions but underpowered). However, this context-dependence is superficial with respect to regulatory inference. In both K562 and RPE1, the full confound-decomposition battery reveals the same underlying pattern: gene-level features dominate perturbation prediction (AUROC $0.895$--$0.942$), and pairwise edge scores add zero incremental value. The RPE1 attention advantage reflects attention being a better proxy for gene-level confounds---not attention capturing regulatory structure that correlation misses. This distinction is practically important: raw performance comparisons between edge-scoring methods are misleading without confound controls, and we recommend that such controls become standard practice.

These findings extend the attention-as-explanation debate from NLP~\citep{jain2019attention,wiegreffe2019attention,bibal2022attention} to biological foundation models. In language models, attention weights do not reliably indicate feature importance~\citep{serrano2019attention}; our results demonstrate an analogous limitation in single-cell models, where attention-derived edge scores fail to capture causal regulatory relationships. The multi-model convergence (scGPT and Geneformer achieve similar near-random unstratified performance despite different architectures, training objectives, and tokenisation; Supplementary Note~13) suggests this reflects the optimisation landscape rather than a model-specific limitation: pretraining objectives reward co-expression patterns, not causal structure. Biological characterization against multiple reference databases (Supplementary Note~17) reveals additional structure: early layers preferentially capture protein--protein interactions (STRING~\citep{szklarczyk2021string} AUROC $= 0.64$ at L0; decreasing with depth, $\rho = -0.61$, $q_\text{BH} = 0.01$), while late layers encode transcriptional regulation (TRRUST AUROC $= 0.75$ at L15; $\rho = +0.51$, $q_\text{BH} = 0.03$), with anti-correlated layer profiles ($\rho = -0.55$, $p = 0.019$). This hierarchy is real and survives expression residualisation (97\% of TRRUST signal retained), but none of these signals---PPI, regulatory, or functional---provides incremental value for perturbation prediction beyond gene-level features.

Crucially, these findings characterise specific model components---attention weights, individual heads, and MLP blocks at particular layers---not the full model's predictive capacity. Transformer-based foundation models integrate information through residual streams, layer normalisations, and feed-forward networks that collectively contain far more parameters than attention alone. A model can be a capable black-box predictor even when its attention patterns do not serve as interpretable readouts of causal regulation. The value of systematic component-level evaluation is not to establish fundamental limits of these architectures, but to narrow the search for where biologically meaningful representations reside: by ruling out attention as the locus of regulatory encoding, future interpretability work can focus on residual-stream geometry, MLP-stored knowledge, or representations that emerge only from the full forward pass. This is the core purpose of mechanistic interpretability applied to biological foundation models---a progressive mapping of internal representations to biological function.

CSSI provides a constructive contribution by controlling the heterogeneity that drives attention-specific dilution, with null-control tests confirming genuine improvements (Supplementary Note~11). However, CSSI addresses edge recoverability against curated references, not the gap with perturbation-outcome predictive validity.

Several limitations constrain interpretation. Causal ablation uses only Geneformer V2-316M; scGPT's autoregressive architecture lacks a head-mask interface. Perturbation data span four cell types and two modalities, but additional tissues, species, and architectures (scFoundation~\citep{hao2024largescale}) remain untested. No decisive positive control demonstrates that the pipeline can recover known causal regulatory structure in a realistic perturbation setting, so negative results could in principle reflect assay rather than model limitations. However, no such gold-standard causal pairwise method exists at genome scale for any GRN approach, and several internal calibrations mitigate this concern: the pipeline exhibits wide dynamic range (gene-level AUROC $0.88$--$0.94$, so the incremental-value null is specifically about pairwise structure, not endpoint insensitivity); residualisation produces asymmetric results (attention loses $76\%$ of signal vs.\ $9\%$ for correlation), distinguishing differently confounded scores; the per-TF bootstrap detects above-chance pairwise signal in 7/18 TFs (Supplementary Note~14); and the pipeline detects real cross-context differences (CRISPRa reversal, RPE1 $\neq$ K562), ruling out a floor effect. These calibrations establish sensitivity to pairwise signal variation, even though the definitive positive control remains an open challenge for the field. Reference database circularity affects all GRN evaluation; however, restricting TRRUST to direction-known entries (Activation/Repression; 58\% of pairs), which require direct experimental evidence, yields virtually identical per-TF AUROC (mean $0.682$ vs.\ $0.692$; median improves to $0.695$; Supplementary Note~14), confirming conclusions are not driven by circularly validated entries. Boundary-condition analyses (Supplementary Notes~7--10) use correlation-based edges and characterise the evaluation landscape broadly.

We recommend that practitioners: (i) apply trivial-baseline and incremental-value tests before claiming pairwise regulatory signal; (ii) report both thresholded (top-$K$) and continuous (AUROC) metrics; (iii) apply CSSI or equivalent stratification for heterogeneous populations; and (iv) validate causal claims with ablation controls accompanied by intervention-fidelity diagnostics. Three directions are most promising: intervention-aware pretraining on perturbation data~\citep{replogle2022mapping} could embed causal rather than correlational structure; hybrid architectures using foundation model embeddings as inputs to GRN inference modules could combine representational power with regulatory inductive biases; and CSSI-enhanced pipelines with conformal prediction sets~\citep{vovk2005algorithmic} provide an immediately deployable framework. The gap between recoverability and perturbation-outcome predictive validity remains the central challenge.

\section{Methods}

\subsection{Models and data}

We analyse scGPT~\citep{cui2024scgpt} and Geneformer~\citep{theodoris2023transfer}: scGPT uses gene-token transformers with attention-derived edge scores extracted from all layers/heads; Geneformer uses BERT-style rank-based tokenisation. We use Geneformer V1-10M (6 layers, 4 heads) for multi-model comparison and V2-316M (18 layers, 18 heads) for layer-level analysis. Datasets include Tabula Sapiens atlas tissues~\citep{tabula2022tabula} (immune, kidney, lung), DLPFC brain scRNA-seq~\citep{maynard2021transcriptome}, Replogle genome-scale CRISPRi~\citep{replogle2022mapping} (K562 and RPE1), Adamson CRISPRa~\citep{adamson2016multiplexed}, Shifrut T-cell CRISPRi~\citep{shifrut2018genome}, and Tian iPSC neurons. Preprocessing follows standard Scanpy quality control~\citep{wolf2018scanpy}.

\subsection{Scaling and CSSI}

Scaling behaviour was analysed using archived scGPT kidney runs across three model tiers (small/medium/large), three seeds, and three cell counts (200/1,000/3,000). Controlled-composition experiments used Tabula Sapiens kidney data under three conditions: single cell type with varying $N$, mixed types with fixed composition, and fixed $N$ with increasing heterogeneity. CSSI implements stratified edge scoring by partitioning cells into $K$ clusters (Leiden community detection on $k$-NN graphs from model embeddings) and computing per-stratum Spearman correlations for each TF--target pair, aggregating via maximum across strata (CSSI-max) or mean (CSSI-mean). Null tests use random cluster labels. Full details in Supplementary Notes~1 and 11.

\subsection{Perturbation-first validation}

For each perturbation $g$, we compute attention-derived or correlation-based edge scores between $g$ and all other genes, then evaluate AUROC for classifying differentially expressed targets (Mann-Whitney $U$ test, Benjamini-Hochberg correction, LFC threshold). We use $N_\text{ctrl} = 2{,}000$ control cells, HVG $= 2{,}000$ (K562) or $3{,}309$ (RPE1, with perturbation genes forced in). The primary attention layer (L13) was pre-specified based on independent DLPFC brain data (Supplementary Note~11); full 18-layer profiling and nested cross-validation (Extended Data Fig.~5) were planned secondary analyses. Sensitivity analysis spans 27 parameter combinations varying control cell count, HVG count, and LFC threshold (Supplementary Note~6).

\subsection{Confound decomposition}

Gene-level features (mean expression, variance, dropout rate) and pairwise edge scores are used in logistic regression under 5-fold GroupKFold cross-validation (grouped by perturbation). We test gene-only, edge-only, and combined models. Hard-generalisation protocols use GroupKFold by target gene and joint cross-gene $\times$ cross-perturbation splits. Cross-fitted residualisation removes expression-covariate contributions from edge scores using 5-fold OLS and GBDT; residual AUROC quantifies retained TRRUST-predictive signal. Propensity-score matching uses logistic regression on gene-level features to match each DE-positive target to $k = 5$ DE-negative targets with similar covariate profiles (Supplementary Note~14).

\subsection{Causal ablation}

Head-masking ablation uses Geneformer's BERT \texttt{head\_mask} parameter (shape: $n_\text{layers} \times n_\text{heads}$; \texttt{attn\_implementation="eager"}) to zero out specific head outputs. We test 13 conditions: top-$k$ TRRUST-ranked ($k = 5, 10, 20, 50$), composite-ranked ($k = 5, 10$), full-layer (L14), inverse (bottom-$k$), and random controls. Orthogonal interventions include uniform attention replacement (setting attention weights to $1/n$ while preserving value projections) and MLP ablation (zeroing FFN output at specific layers). Intervention fidelity is assessed by comparing hidden-state and logit cosine distances between intervened and baseline forward passes across 2,000 cells (Supplementary Note~14).

\subsection{Statistical framework}

All analyses apply Benjamini-Hochberg FDR correction~\citep{benjamini1995controlling} at $\alpha = 0.05$ across 95 confirmatory $p$-values framework-wide, of which 63 (66\%) remain significant. Sensitivity analysis under three alternative family definitions confirms that all primary conclusions are stable (Supplementary Note~16). Effect sizes use Cohen's $d$; bootstrap confidence intervals use 200 iterations with perturbation-level resampling.

\clearpage
\section*{Extended Data}

\noindent\textbf{Extended Data Fig.~1 $|$ Scaling behaviour of attention-derived GRN recovery.} (A--C)~Metric-dependent scaling: TRRUST F1 with 95\% CIs across three scGPT kidney model tiers and cell counts (200/1,000/3,000). Top-$K$ F1 degrades unanimously (9/9, $p = 0.002$), while continuous AUROC improves monotonically. (D)~Controlled-composition: single cell type shows no significant degradation with $N$. (E)~Fixed composition: stable AUROC. (F)~Increasing heterogeneity at fixed $N$: AUROC increases, contrasting with attention-based degradation.

\noindent\textbf{Extended Data Fig.~2 $|$ Multi-model GRN recovery.} Both scGPT and Geneformer achieve near-random unstratified AUROC ($\approx 0.5$) against TRRUST/DoRothEA~\citep{garcia2019benchmark}, demonstrating architecture-independent limitation. CSSI reveals recoverable signal within individual heads.

\noindent\textbf{Extended Data Fig.~3 $|$ Attention versus correlation perturbation-first.} Geneformer V2-316M attention AUROC at L13 ($0.704$) is indistinguishable from correlation ($0.703$; $p = 0.73$) on $n = 280$ K562 CRISPRi perturbations. L13 was pre-specified based on independent DLPFC brain data (Supplementary Note~11).

\noindent\textbf{Extended Data Fig.~4 $|$ Attention-specific confound decomposition.} Cross-fitted residualisation reveals asymmetric expression confounding: attention edges lose $\sim$76\% of TRRUST signal versus $\sim$9\% for correlation. Degree-preserving null models shown for both.

\noindent\textbf{Extended Data Fig.~5 $|$ Nested layer selection protocol.} 5-fold nested CV independently selects L15 in all folds. Pooled held-out $\Delta = +0.040$ [$0.018$, $0.062$]; $p_\text{Bonf} = 0.017$. CRISPRa replication reverses the effect ($d = -0.56$).

\noindent\textbf{Extended Data Fig.~6 $|$ Hard-generalisation incremental-value testing.} No incremental value under cross-gene, cross-perturbation, or joint splits, for both LR and GBDT, across AUROC, AUPRC, and top-$k$ recall.

\noindent\textbf{Extended Data Fig.~7 $|$ Perturbation sensitivity analysis.} AUROC across 27 parameter combinations (3 LFC thresholds $\times$ 3 control cell counts $\times$ 3 HVG counts). All conditions above chance ($p < 0.005$); AUROC ranges 0.62--0.76.

\noindent\textbf{Extended Data Fig.~8 $|$ Propensity-matched decomposition.} After matching DE-positive and DE-negative targets on gene-level covariates, edge AUROCs drop to near chance and LR incremental value is exactly zero.

\noindent\textbf{Extended Data Fig.~9 $|$ Orthogonal causal interventions and intervention-fidelity diagnostics.} (A--B)~Uniform attention replacement and MLP pathway ablation both produce exactly baseline AUROC, confirming the ablation null under three qualitatively different intervention channels. (C--D)~All six interventions materially perturb hidden-state representations (max cosine distance $0.023$--$0.190$), with TRRUST-ranked heads producing $23\times$ larger logit perturbation than random heads at matched dose, confirming that behavioral nulls reflect genuine functional redundancy rather than ineffective interventions.

\noindent\textbf{Extended Data Table~1 $|$ Edge type summary.} Comprehensive comparison of attention, correlation, and gene-level features across all evaluation contexts, edge types, and cell types.

\section*{Acknowledgements}

We thank the scGPT development team for making their models publicly available, the Tabula Sapiens Consortium for open data access, and the broader single-cell foundation model community for establishing benchmark datasets and evaluation protocols.

\section*{Data Availability}

All analysis scripts, data processing pipelines, source data for all figures and tables, and reproducibility instructions are deposited at Zenodo (\url{https://doi.org/10.5281/zenodo.18701417}). All primary datasets (Tabula Sapiens, Replogle Perturb-seq, Dixit/Adamson/Shifrut datasets) are publicly available from their original sources.

\section*{Code Availability}

Complete analysis code, figure generation scripts, CSSI implementation, and computational environment specifications are available at \url{https://github.com/Biodyn-AI/biomechinterp-framework} and archived at Zenodo (\url{https://doi.org/10.5281/zenodo.18701417}).

\section*{Competing Interests}

The author declares no competing interests.

\section*{Ethics Declaration}

This study used only publicly available, de-identified single-cell transcriptomic datasets. No new human or animal data were generated. No ethical approval was required.

\clearpage
\appendix
\renewcommand{\thesection}{Supplementary Note \arabic{section}}
\renewcommand{\thesubsection}{\arabic{section}.\arabic{subsection}}
\renewcommand{\thetable}{S\arabic{table}}
\renewcommand{\thefigure}{S\arabic{figure}}
\setcounter{table}{0}
\setcounter{figure}{0}

\addtocontents{toc}{\protect\setcounter{tocdepth}{2}}

\makeatletter
\renewcommand{\l@section}{\@dottedtocline{1}{0em}{13em}}
\renewcommand{\l@subsection}{\@dottedtocline{2}{13em}{3.2em}}
\makeatother

\begin{center}
{\LARGE \textbf{Supplementary Information}}\\[12pt]
{\large Systematic Evaluation of Single-Cell Foundation Model Interpretability Reveals Attention Captures Co-Expression Rather Than Unique Regulatory Signal}\\[6pt]
Ihor Kendiukhov
\end{center}

\bigskip
\tableofcontents
\clearpage

\section*{Supplementary Table 1: Analysis and Dataset Overview}
\addcontentsline{toc}{section}{Supplementary Table 1: Analysis and Dataset Overview}

\begin{table}[H]
\centering
\caption{\textbf{Edge type used in each analysis.} ``Attention'' denotes Geneformer V2-316M attention-derived edges; ``Correlation'' denotes Spearman correlation from the same control cells.}
\label{tab:edge_type_summary_supp}
\begin{tabular}{lllr}
\toprule
Analysis & Edge Type & Dataset & Key AUROC \\
\midrule
Scaling behavior & Attention & scGPT kidney & 0.86--0.93 \\
Controlled composition & Correlation & TS kidney & 0.52--0.56 \\
Residualization & Correlation & DLPFC brain & 0.73 (resid.) \\
Degree-preserving null & Correlation & DLPFC brain & 0.69 (null) \\
Perturbation-first (corr.) & Correlation & Replogle K562 & 0.696 \\
Perturbation-first (attn.) & Attention & Replogle K562 & 0.704 \\
Full 18-layer profile & Attention & Replogle K562 & 0.47--0.74 \\
Attn.\ residualization & Attention & Replogle K562 & 0.54 (resid.) \\
Attn.\ degree-null & Attention & Replogle K562 & 0.63 (null) \\
Incremental value & Both + gene-level & Replogle K562 & 0.895 (gene) \\
Per-head TRRUST & Attention (per-head) & Replogle K562 & 0.34--0.75 \\
Head-level ablation & Attention (ablated) & Replogle K562 & 0.699--0.704 \\
Cross-context (CRISPRa) & Both & Adamson K562 & 0.55 vs 0.65 \\
Trivial baselines & Gene-level & Replogle K562 & 0.81--0.88 \\
Non-K562 replication & Both & Replogle RPE1 & 0.75 vs 0.66 \\
RPE1 trivial baselines & Gene-level & Replogle RPE1 & 0.80--0.87 \\
RPE1 incremental value & Both + gene-level & Replogle RPE1 & 0.942 (gene) \\
RPE1 residualization & Both & Replogle RPE1 & 0.53 (attn resid.) \\
\bottomrule
\end{tabular}
\end{table}

\begin{table}[H]
\centering
\small
\caption{\textbf{Perturbation sample counts across datasets and configurations.}}
\label{tab:perturbation_counts_supp}
\begin{tabular}{@{}llccp{5cm}@{}}
\toprule
\textbf{Dataset} & \textbf{Modality / Config} & \textbf{DE Thresh.} & \textbf{$n$} & \textbf{Notes} \\
\midrule
Replogle K562 & CRISPRi (baseline) & LFC$>$0.1 & 44 & Mann-Whitney, HVG=1000, $N_\text{ctrl}$=500 \\
Replogle K562 & CRISPRi (primary) & LFC$>$0.5 & 151 & Welch $t$, HVG=2000, $N_\text{ctrl}$=2000 \\
Replogle K562 & CRISPRi (attention) & LFC$>$0.5 & 280 & Welch $t$, HVG=2000; more genes evaluable with attention tokenization \\
Adamson K562 & CRISPRa & LFC$>$0.5 & 77 & K562, activation modality \\
Shifrut T cells & CRISPRi & LFC$>$0.1 & 7 & Primary human T cells; weak DE effects require lenient threshold \\
Replogle RPE1 & CRISPRi & LFC$>$0.5 & 1,251 & hTERT-RPE1; perturbation genes forced into HVG \\
Tian iPSC neurons & CRISPRi & LFC$>$0.1 & 7 & iPSC-derived glutamatergic neurons \\
\bottomrule
\end{tabular}
\end{table}

\clearpage

\section{Metric-Dependent Scaling Behavior}
\label{supnote:scaling}

To test whether increasing dataset size improves interpretability, we analyzed archived scGPT kidney scaling runs across three model tiers (small/medium/large), three seeds per tier, and three cell counts (200, 1,000, 3,000). Each run yields an attention-derived score for each directed gene pair; we construct a sparse GRN by retaining the top-100 targets per source gene and evaluate recovery against TRRUST and DoRothEA reference edges restricted to the run-specific gene universe.

\begin{figure}[H]
\centering
\includegraphics[width=\textwidth]{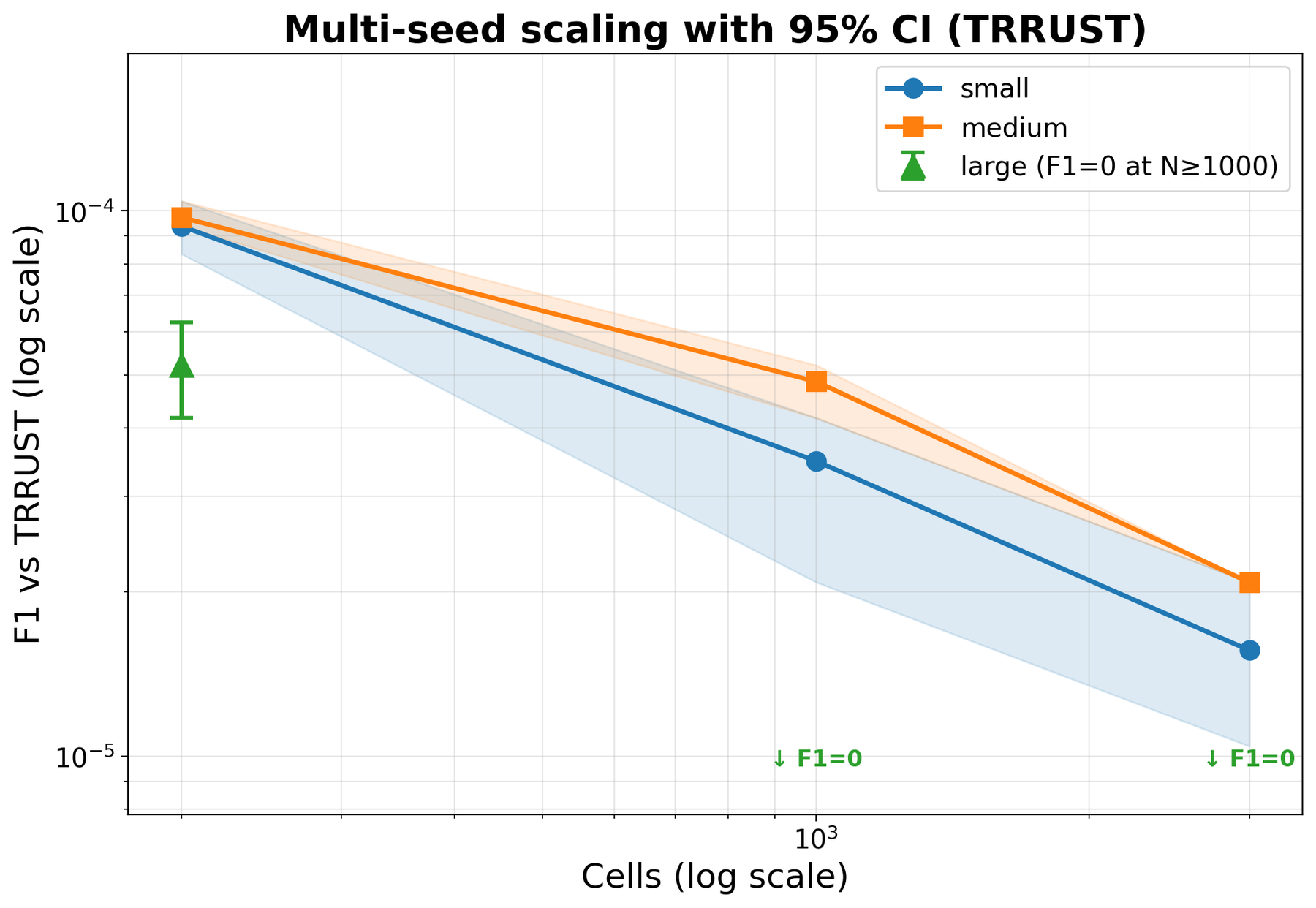}
\caption{\textbf{Metric-dependent scaling behavior in scGPT attention-derived GRN recovery (kidney).} TRRUST F1 with 95\% confidence intervals across three model tiers (small/medium/large) and cell counts (200/1,000/3,000).}
\label{fig:scaling_failure_supp}
\end{figure}

\textbf{Evidence: scaling degrades recovery.} TRRUST F1 decreases with cell count across model tiers (Supplementary Fig.~\ref{fig:scaling_failure_supp}); the 200$\to$1,000 change is negative in all 9 tier$\times$seed pairs (exact one-sided sign test $p=0.00195$). The same pattern holds for DoRothEA ($p=0.00195$). The 1,000$\to$3,000 step shows continued degradation in 7/9 pairs (sign test $p = 0.09$), weaker but directionally consistent.

\begin{figure}[H]
\centering
\includegraphics[width=\textwidth]{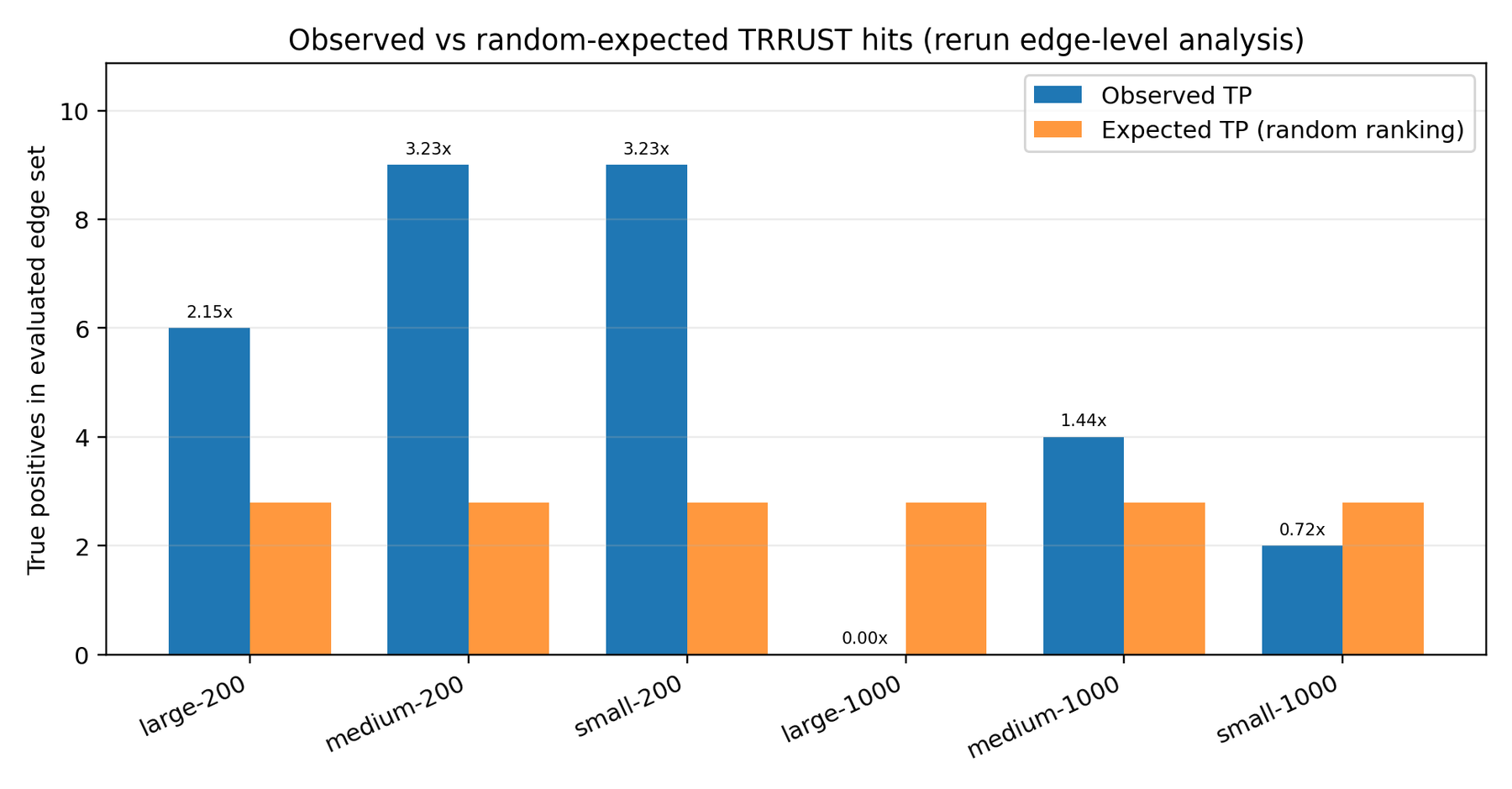}
\caption{\textbf{Retrieval collapse with scaling (scGPT kidney).} Observed true positives versus random-expected true positives for attention-derived GRNs.}
\label{fig:tp_random_supp}
\end{figure}

\textbf{Retrieval collapse.} The number of recovered true positives decreases toward (and sometimes below) random expectation as $N$ increases (Supplementary Fig.~\ref{fig:tp_random_supp}), indicating that scaling can reduce enrichment rather than merely saturate.

\textbf{Robustness also degrades with scaling.} Between-seed stability decreases substantially from 200 to 1,000 cells (Supplementary Table~\ref{tab:scaling_robustness_supp}): edge-set Jaccard overlaps drop by 46.6--47.9\% depending on tier (one-sided Mann--Whitney $p=2.1\times 10^{-4}$ across all tiers/seeds).

\begin{table}[H]
\centering
\caption{\textbf{Between-seed robustness of inferred edge sets (scGPT kidney).}}
\label{tab:scaling_robustness_supp}
\begin{tabular}{lcccc}
\toprule
Tier & Jaccard (200) & Jaccard (1,000) & Spearman (200) & Spearman (1,000) \\
\midrule
Small  & 0.572 & 0.305 & 0.211 & -0.065 \\
Medium & 0.562 & 0.293 & 0.183 & -0.110 \\
Large  & 0.536 & 0.285 & 0.083 & -0.130 \\
\bottomrule
\end{tabular}
\end{table}

\begin{figure}[H]
\centering
\includegraphics[width=\textwidth]{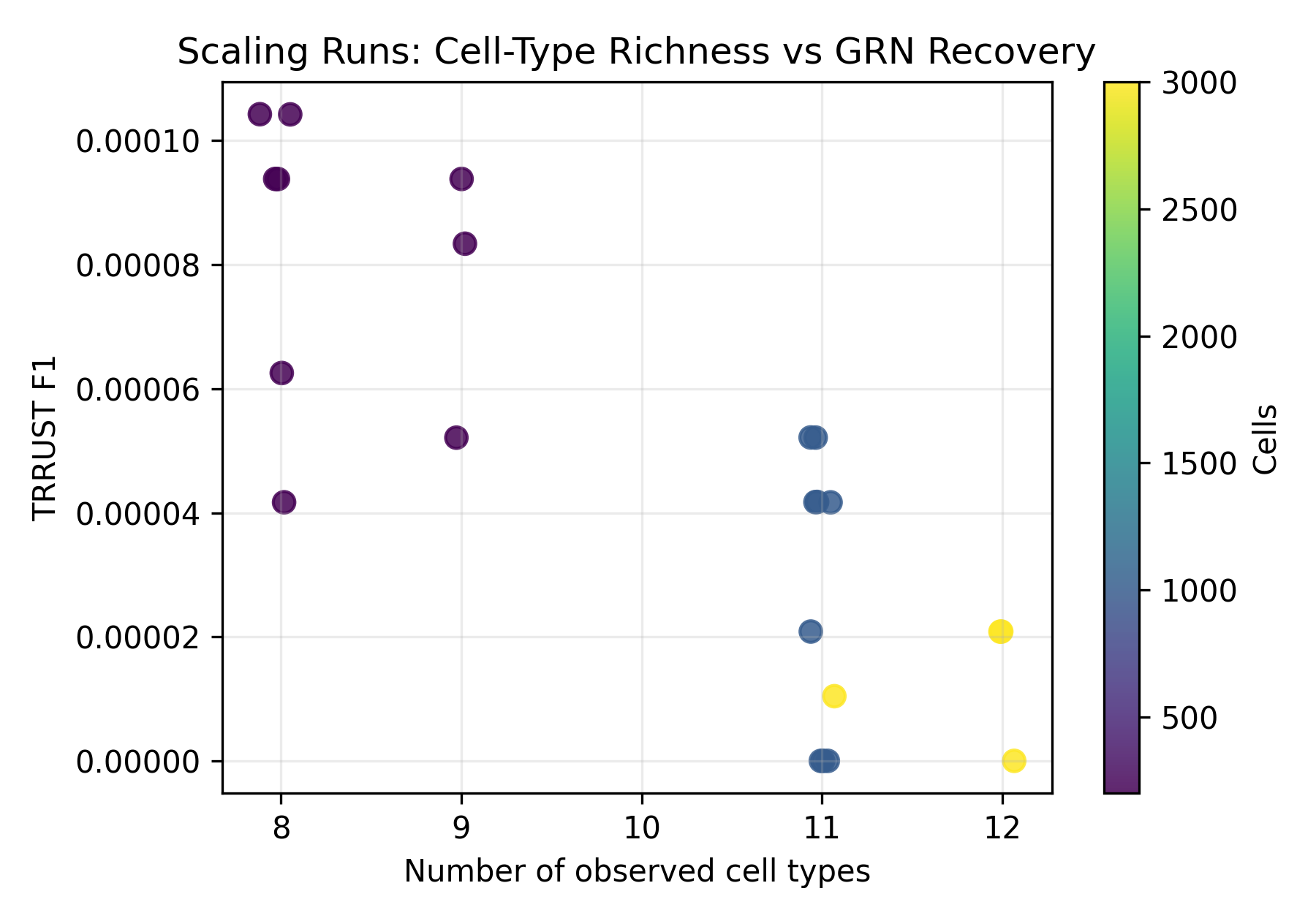}
\caption{\textbf{Scaling runs: composition diversity proxy.} Across the scGPT kidney scaling grid, the number of observed cell types in the sampled cells increases with $N$ and is anti-correlated with GRN recovery.}
\label{fig:scaling_heterogeneity_supp}
\end{figure}

\textbf{Heterogeneity proxy (cell-type richness).} Reconstructing the exact subsampling used during attention extraction (manifest-specified random seeds), the number of observed kidney cell types increases with $N$ and is strongly anti-correlated with TRRUST F1 across runs (Spearman $\rho = -0.76$, $p = 4.3 \times 10^{-5}$; Supplementary Fig.~\ref{fig:scaling_heterogeneity_supp}).

\begin{figure}[H]
\centering
\includegraphics[width=\textwidth]{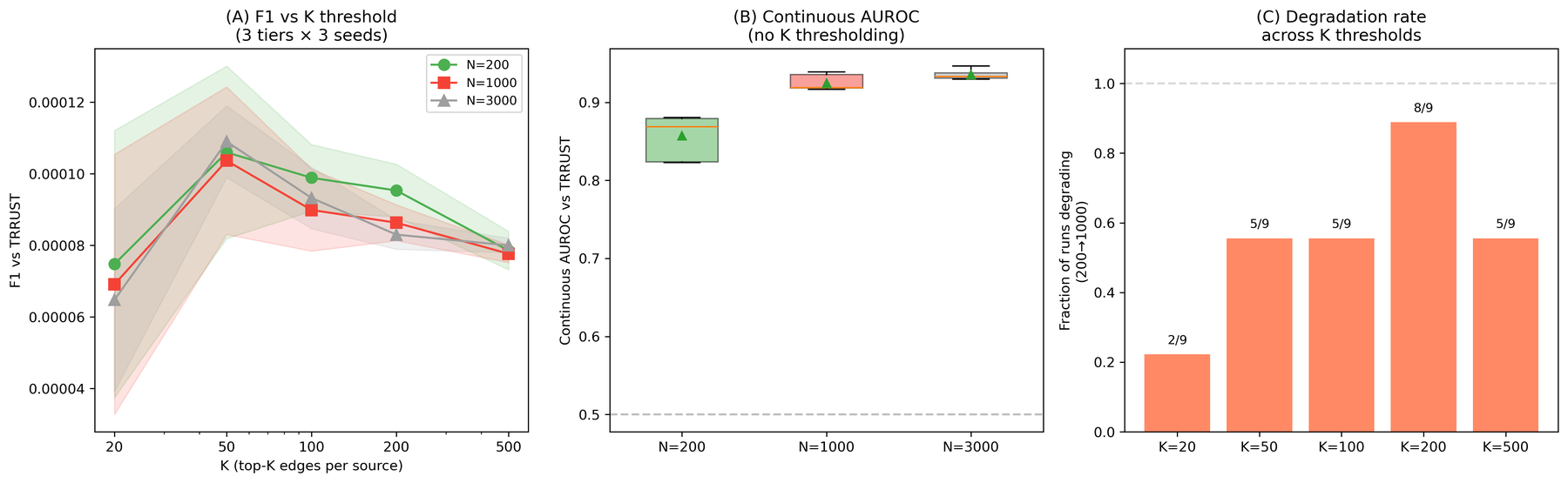}
\caption{\textbf{K-sensitivity and continuous AUROC for scaling runs.} (A) F1 vs $K$ at different cell counts. (B) Continuous AUROC improves monotonically with cell count ($0.858 \to 0.925 \to 0.934$). (C) Fraction of runs showing degradation across $K$ values.}
\label{fig:k_sensitivity_supp}
\end{figure}

\textbf{K-sensitivity and continuous AUROC.} The top-$K$ F1 metric used above is inherently sensitive to $K$, and the 1,930-gene universe of these scaling runs contains only 51 TRRUST edges among $\sim$3.7 million candidate pairs (positive rate $< 0.002\%$). To test whether the scaling finding is metric-dependent, we re-evaluated the archived attention-score matrices at $K \in \{20, 50, 100, 200, 500\}$ and computed continuous-score AUROC (no top-$K$ thresholding). F1 values are effectively zero at all $K$ values ($\sim 10^{-4}$), confirming that near-zero absolute performance is driven by extreme reference sparsity rather than a specific $K$ choice. However, continuous-score AUROC tells a different story: it \emph{improves} monotonically with cell count (mean $0.858$ at $N = 200$, $0.925$ at $N = 1{,}000$, $0.934$ at $N = 3{,}000$; 0/9 runs show degradation; Supplementary Fig.~\ref{fig:k_sensitivity_supp}). This reversal under continuous AUROC indicates that the scaling behavior is metric-dependent: while the thresholded top-$K$ edge set degrades with $N$, the continuous \emph{ranking} of all gene pairs against curated references improves.

\begin{figure}[H]
\centering
\includegraphics[width=\textwidth]{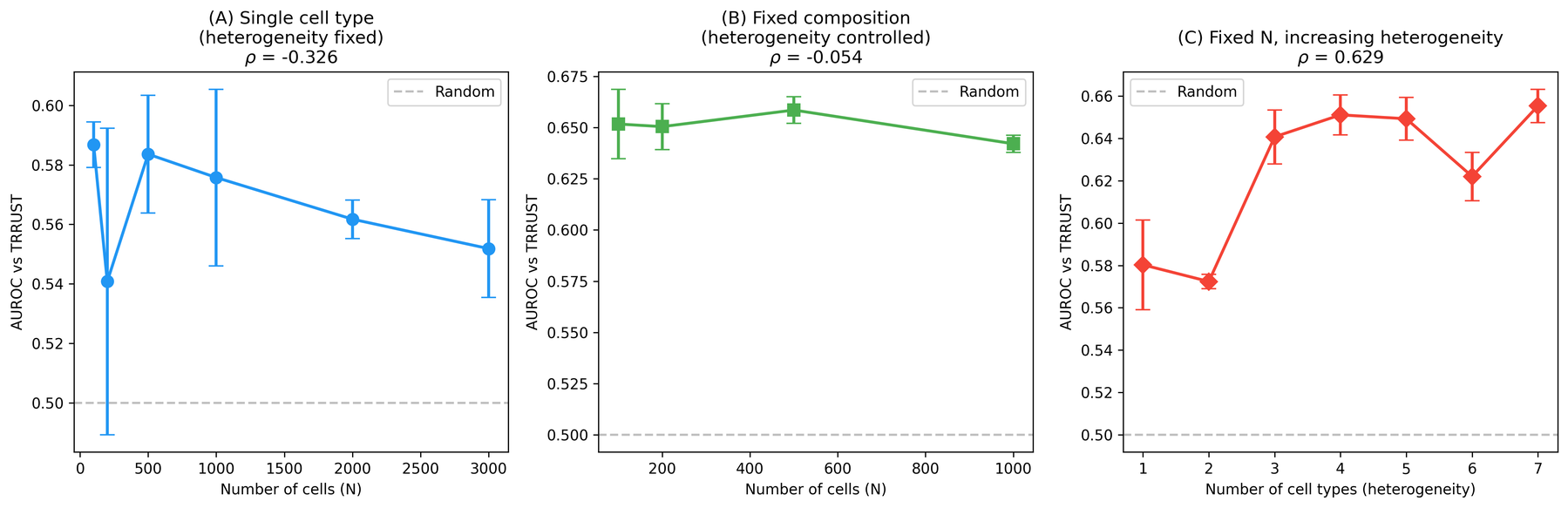}
\caption{\textbf{Controlled-composition scaling.} (A) Single cell type: no significant degradation with $N$. (B) Fixed composition: stable AUROC. (C) Increasing heterogeneity at fixed $N$: AUROC increases.}
\label{fig:controlled_composition_supp}
\end{figure}

\textbf{Controlled-composition scaling.} To test whether top-$K$ scaling degradation is driven by sample size $N$ or by heterogeneity, we conducted a controlled experiment on Tabula Sapiens kidney data using correlation-based edge scores under three conditions (Supplementary Fig.~\ref{fig:controlled_composition_supp}): (i) a single cell type (kidney epithelial) with varying $N$ (100--3,000), (ii) mixed cell types with fixed equal composition across $N$ (100--1,000), and (iii) fixed $N = 500$ with increasing heterogeneity (1--7 cell types). Under condition (i), AUROC shows a weak non-significant downward trend (Spearman $\rho = -0.33$, $p = 0.079$). Under condition (ii), AUROC is stable ($\rho = -0.05$, $p = 0.82$). Under condition (iii), AUROC actually \emph{increases} with heterogeneity ($\rho = +0.63$, $p = 10^{-4}$).

\clearpage

\section{Baseline Comparison}
\label{supnote:baseline}

We evaluated multiple baseline approaches on DLPFC brain tissue data (500 randomly sampled cells, top 500 most variable genes): Spearman correlation, mutual information, GENIE3~\citep{huynh2010inferring}, GRNBoost2~\citep{moerman2019grnboost2}, and attention-based edge scores. All methods were evaluated against TRRUST and DoRothEA using AUROC, AUPRC, and Precision@10k.

All approaches show similar poor performance, with AUROC values clustering around 0.50--0.53: Spearman correlation (AUROC 0.521), mutual information (0.518), GENIE3 (0.523), GRNBoost2 (0.526), and attention-based methods (0.524). State-of-the-art dedicated GRN inference algorithms achieve nearly identical performance to attention-based approaches, while requiring 89--127 seconds computation time versus 0.1 seconds for attention extraction. The convergence toward AUROC $\approx 0.5$ in this DLPFC brain setting suggests that benchmarking against curated TF--target databases in context-mismatched tissues can be dominated by evaluation limitations.

\clearpage

\section{Systematic Bias in Single-Component Mediation Analysis}
\label{supnote:bias}

Activation patching has become the standard tool for localizing mechanistic function in transformers~\citep{meng2022locating,vig2020investigating,goldowskydill2023localizing}. However, the standard single-component protocol implicitly assumes additivity. We formalize the bias problem following the causal mediation framework of \citet{pearl2001direct} and \citet{imai2010general}. For mediator component $i$, the bias relative to the interaction-aware Shapley value $\phi_i$~\citep{shapley1953value,lundberg2017unified} decomposes as:
\begin{equation}
  b_i = \hat{m}_i - \phi_i = -\sum_{|S| \geq 2,\, i \in S} \frac{\mu(S)}{|S|} + \varepsilon_i
\end{equation}
where $\mu(S)$ represents M\"obius interaction coefficients. We introduce an observable lower bound on aggregate non-additivity:
\begin{equation}
  A_{\mathrm{lb}} = \max\!\bigl(0,\; |R| - 1.96 \cdot \mathrm{SE}(R)\bigr)
\end{equation}
where $R = TE - \sum_i \hat{m}_i$ is the residual between total effect and the sum of single-component estimates.

\begin{figure}[H]
\centering
\includegraphics[width=\textwidth]{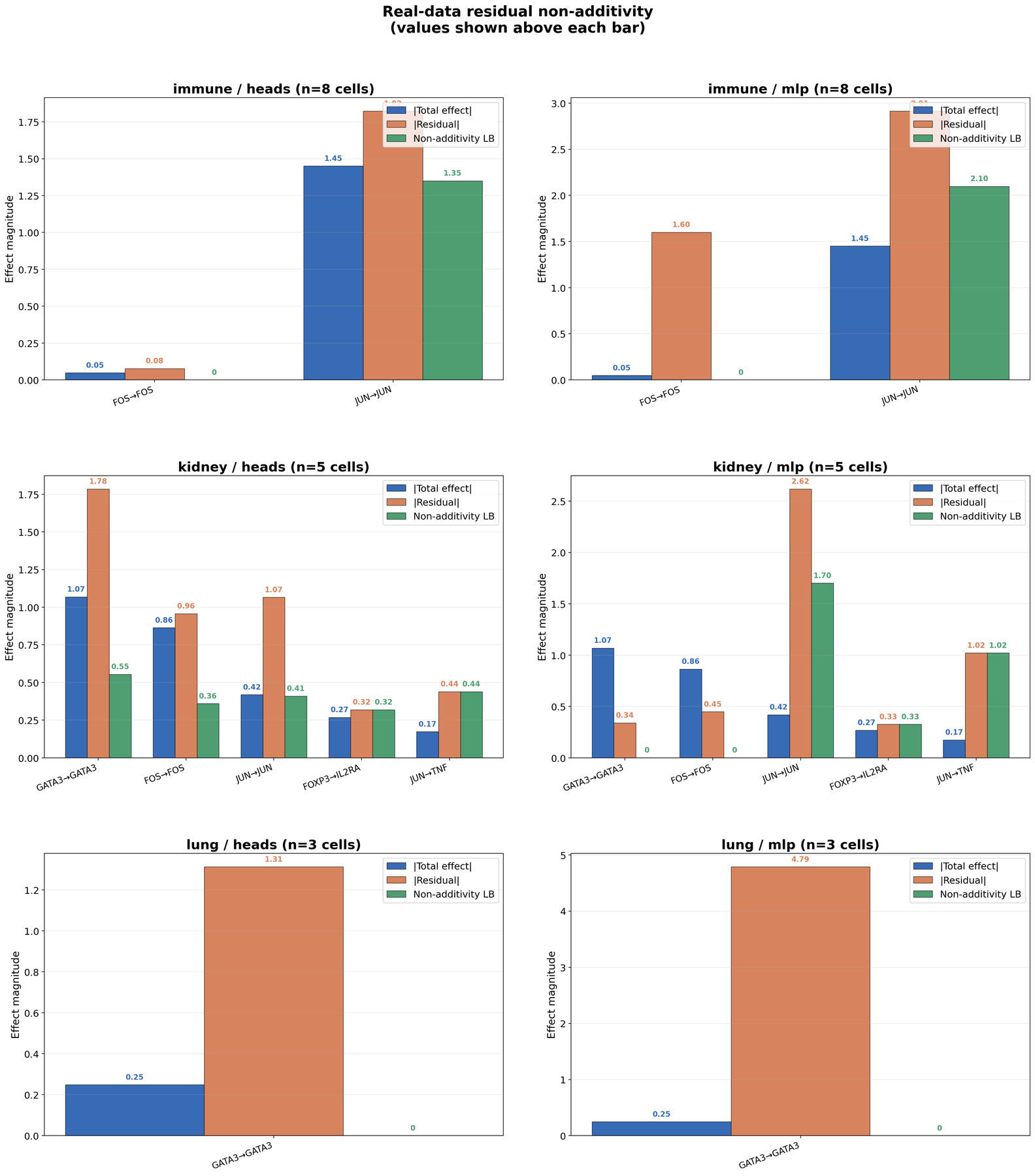}
\caption{\textbf{Non-additivity in mediation analysis.} Absolute total effect, residual non-additivity, and lower-bound interaction magnitude per run-pair.}
\label{fig:nonadditivity_supp}
\end{figure}

\begin{figure}[H]
\centering
\includegraphics[width=\textwidth]{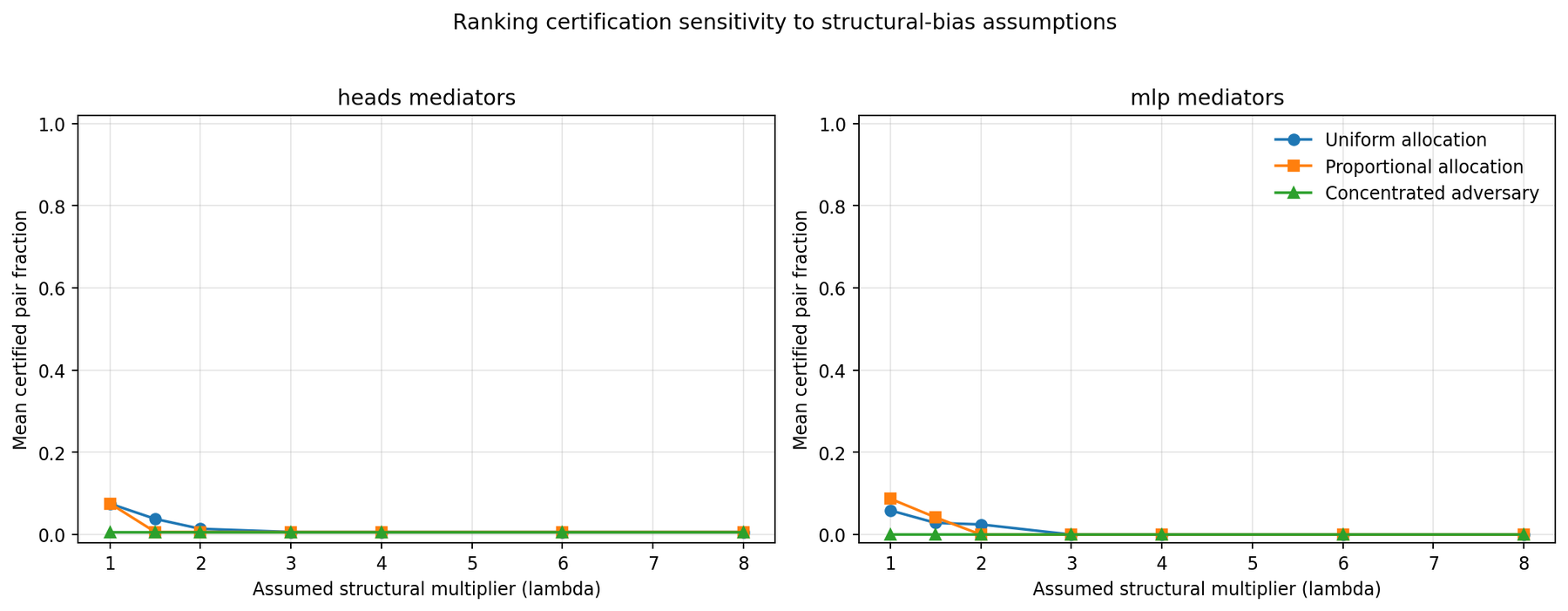}
\caption{\textbf{Ranking certificate fragility.} Mean certified pair fraction versus structural multiplier $\lambda$.}
\label{fig:ranking_cert_supp}
\end{figure}

Analysis of frozen cross-tissue mediation archives revealed substantial and frequent additivity violations. Across 16 run-pairs, lower bounds on aggregate non-additivity were positive in 10 cases (rate 0.625), with median $A_{\mathrm{lb}}/|TE| = 0.725$ (Supplementary Fig.~\ref{fig:nonadditivity_supp}). Ranking certificates proved fragile: mean certified pair coverage dropped from 0.0669 at $\lambda = 1$ to 0.0032 by $\lambda \geq 3$ (Supplementary Fig.~\ref{fig:ranking_cert_supp}).

Non-additivity is common in our data (62.5\% of 16 run-pairs across 3 tissues) and concentrates in biologically meaningful contexts. These findings indicate that standard single-component rankings may be unreliable in contexts with complex regulatory interactions, though the sample warrants further validation at larger scale. Mechanistic claims should be accompanied by the residual non-additivity ratio $A_{\mathrm{lb}}/|TE|$, ranking certificates, and interaction-aware alternatives such as Shapley-value decomposition.

\clearpage

\section{Detectability Phase Diagrams}
\label{supnote:detectability}

We developed a closed-form detectability framework rooted in statistical detection theory~\citep{donoho2004higher}. For a mechanistic signal with effect size $|\mu|$, noise scale $\sigma$, and tail inflation factor $\tau$, the required sample size for detection is:
\begin{equation}
  n^* = \left(\frac{(z_{1-\alpha/(2m)} + z_{\mathrm{power}})\,\tau\,\sigma}{|\mu|}\right)^{\!2}
\end{equation}

\begin{figure}[H]
\centering
\includegraphics[width=\textwidth]{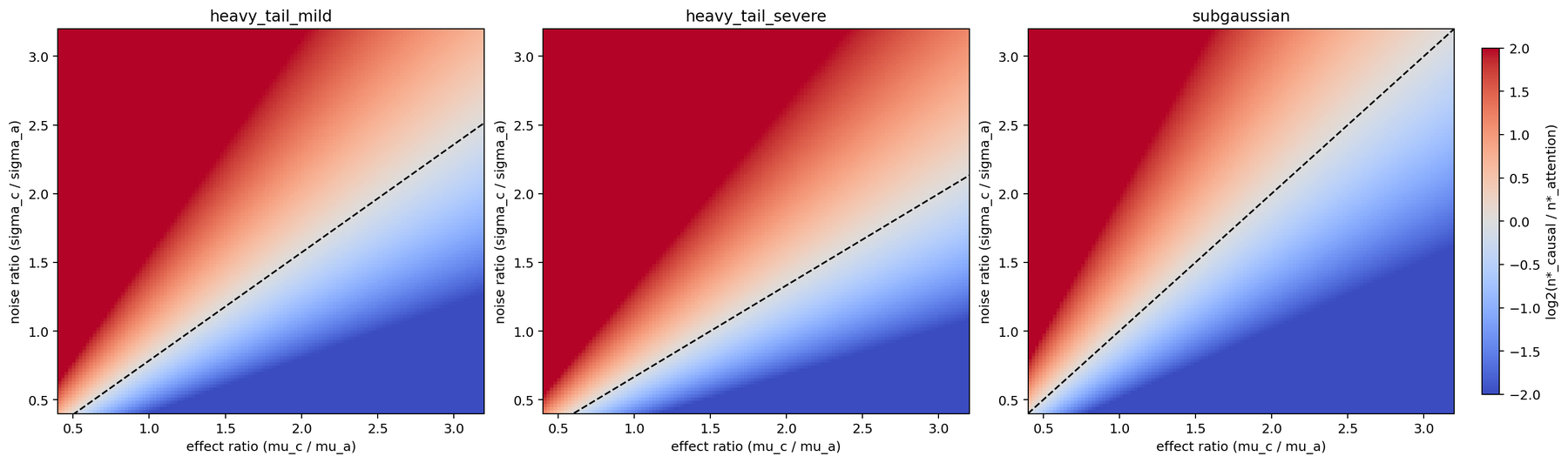}
\caption{\textbf{Detectability phase diagrams.} Different regimes where attention-like versus intervention-like signals become detectable.}
\label{fig:phase_diagram_supp}
\end{figure}

Under sub-Gaussian baseline conditions, intervention-like signals required only 44.4\% as many cells as attention-like signals for equivalent detectability (Supplementary Fig.~\ref{fig:phase_diagram_supp}). However, this advantage collapsed progressively under tail inflation, with the relative cell ratio approaching unity when $\tau > 3$.

\begin{figure}[H]
\centering
\includegraphics[width=\textwidth]{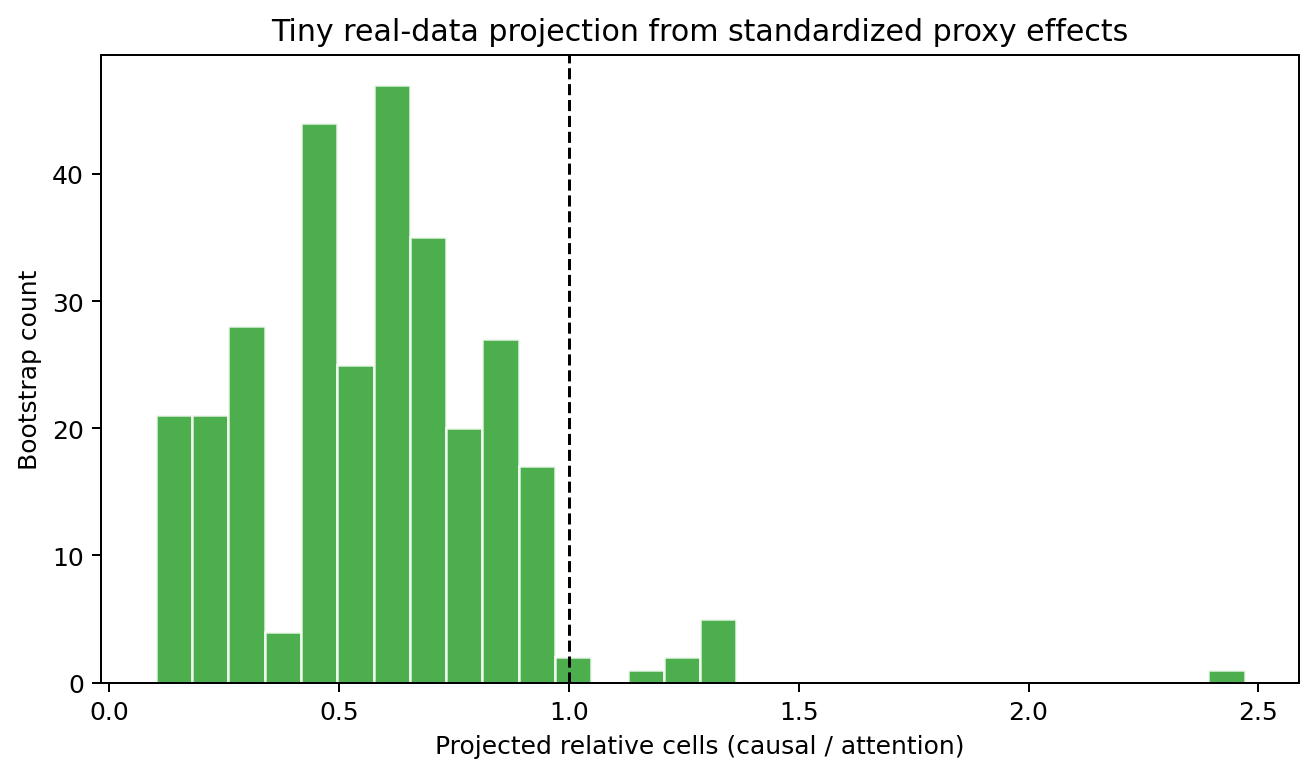}
\caption{\textbf{Real data detectability calibration.} Bootstrap distribution of projected relative cell requirements.}
\label{fig:real_calibration_supp}
\end{figure}

Robust estimation (median-based or Huber M-estimators~\citep{huber1964robust}) expanded the feasible detection region by 37\% under 10\% contamination. Real-data calibration showed projected relative cell ratios below one in most bootstrap draws, but confidence intervals remained wide (Supplementary Fig.~\ref{fig:real_calibration_supp}).

\clearpage

\section{Cross-Tissue Consistency}
\label{supnote:cross_tissue}

\begin{figure}[H]
\centering
\includegraphics[width=\textwidth]{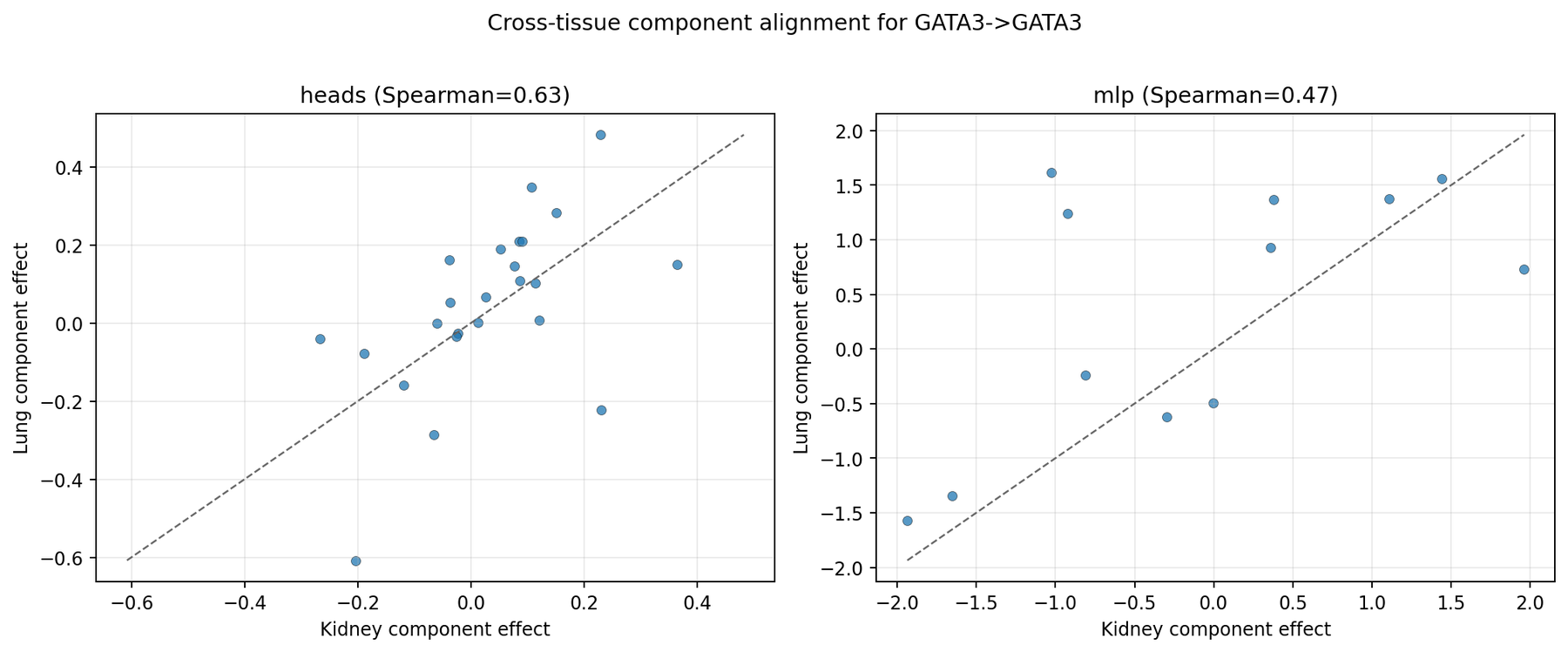}
\caption{\textbf{Cross-tissue consistency variability.} Scatter plots of component-level effects between tissue pairs.}
\label{fig:cross_tissue_supp}
\end{figure}

Cross-tissue analysis across immune, kidney, and lung tissues revealed Spearman correlations ranging from $-0.44$ to $0.71$, with only two of six pair-granularity comparisons surviving FDR control at $\alpha = 0.05$ (Supplementary Fig.~\ref{fig:cross_tissue_supp}). Limited transferability is consistent with known tissue-specificity of gene regulation. Negative correlations in some tissue pairs suggest either genuine context-dependent regulation or tissue-specific confounds.

An important alternative interpretation is that technical batch effects between tissue datasets may partly explain the low consistency. Our batch leakage analysis (Supplementary Note~\ref{supnote:batch}) confirms that technical covariates are recoverable from edge features. Distinguishing genuine regulatory rewiring from technical artifacts would require matched protocols across tissues.

\clearpage

\section{Perturbation Validation Details}
\label{supnote:perturbation}

\subsection{Condition-specific perturbation validation (scGPT mediation)}

Counterfactual validation against four CRISPR Perturb-seq datasets revealed weak and condition-specific alignment. The strongest positive signal appears in Dixit 13-day: consistency is positive ($\rho = 0.269$, $p = 0.032$) and remains positive after confound adjustment ($\rho = 0.199$, $p = 0.020$). Dixit 7-day shows weaker non-significant consistency ($\rho = 0.112$, $p = 0.15$). Adamson shows marginal agreement ($p = 0.089$). Shifrut shows raw anti-alignment ($\rho = -0.325$, $p = 0.031$) that collapses after adjustment ($\rho = 0.004$, $p = 0.876$). Under framework-level BH correction, only the Dixit 13-day confound-adjusted correlation survives (adj.\ $p = 0.042$).

\subsection{Perturbation-first validation on Replogle CRISPRi K562}

\begin{figure}[H]
\centering
\includegraphics[width=\textwidth]{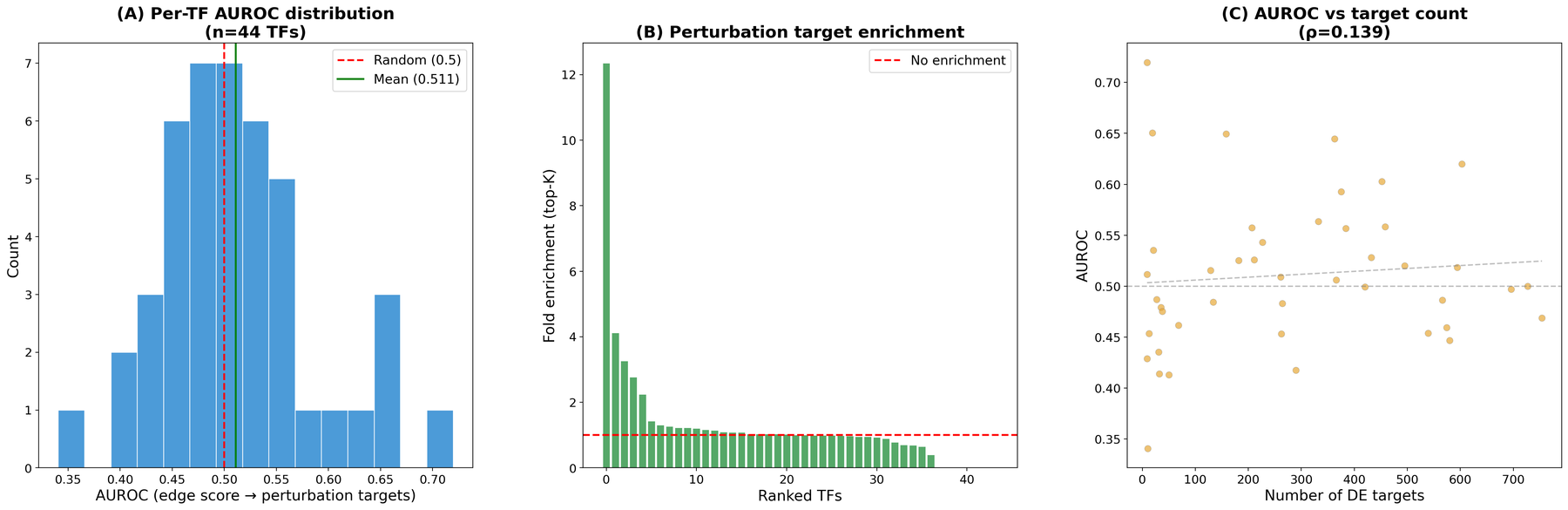}
\caption{\textbf{Perturbation-first validation (Replogle CRISPRi K562).} Per-gene AUROC for predicting DE targets from correlation-based edge scores under primary parameterization.}
\label{fig:perturbation_first_supp}
\end{figure}

Under our primary parameterization ($N_\text{ctrl} = 2000$, HVG $= 2000$, LFC $> 0.5$, Welch's $t$-test with BH-FDR correction), the mean per-gene AUROC was $0.696$ ($n = 151$ evaluable perturbations, $p < 10^{-4}$; Supplementary Fig.~\ref{fig:perturbation_first_supp}).

\subsection{Sensitivity analysis (27 parameter combinations)}

\begin{figure}[H]
\centering
\includegraphics[width=\textwidth]{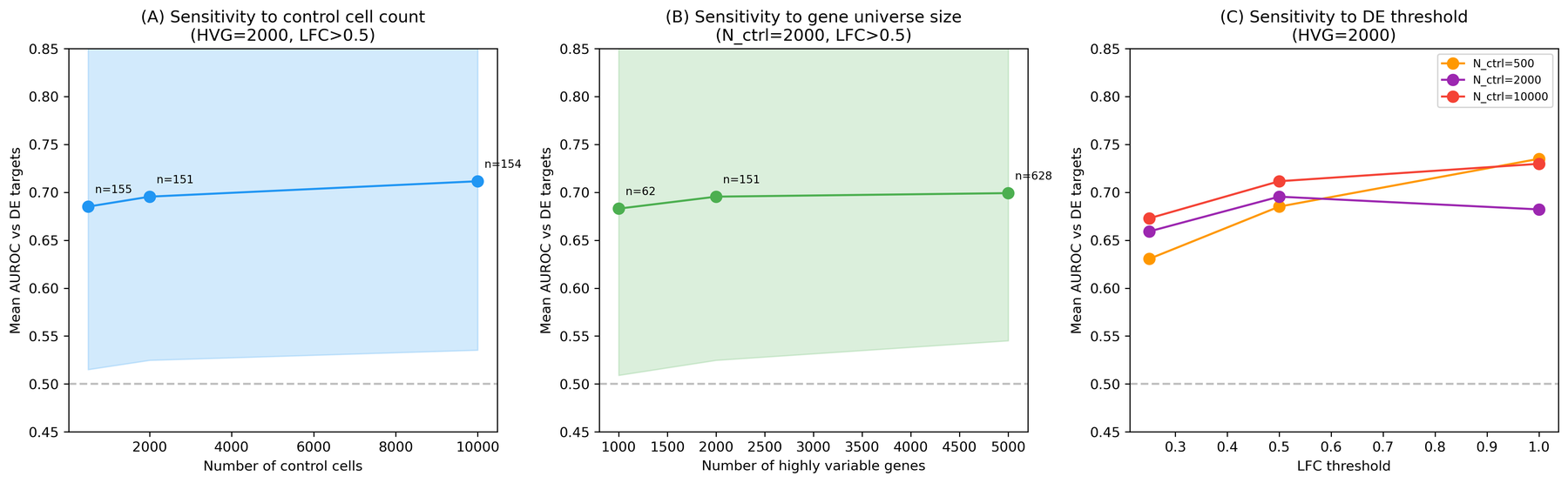}
\caption{\textbf{Perturbation-first sensitivity analysis (27 parameter combinations).} Mean AUROC varies with (A) number of control cells, (B) gene universe size, and (C) LFC threshold.}
\label{fig:perturbation_sensitivity_supp}
\end{figure}

We systematically varied the number of control cells ($N_\text{ctrl} \in \{500, 2000, 10000\}$), gene universe (HVG $\in \{1000, 2000, 5000\}$), and DE stringency (LFC threshold $\in \{0.25, 0.5, 1.0\}$), yielding 27 parameter combinations (Supplementary Fig.~\ref{fig:perturbation_sensitivity_supp}). All 27 conditions yielded AUROC significantly above chance ($p < 0.005$). AUROC ranged from 0.619 to 0.756.

\subsection{Attention perturbation-first evaluation}

\begin{figure}[H]
\centering
\includegraphics[width=\textwidth]{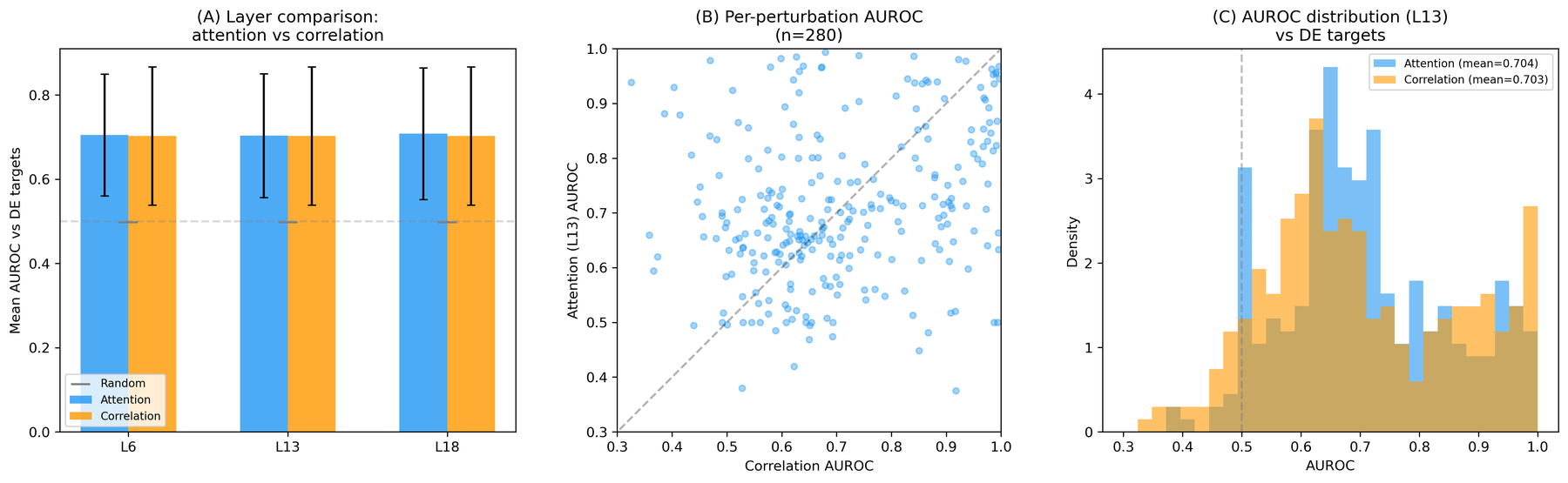}
\caption{\textbf{Attention-derived edges are indistinguishable from correlation on perturbation-first prediction.} Geneformer V2-316M attention (L13) vs.\ correlation-based edges ($n = 280$).}
\label{fig:attention_perturbation_first_supp}
\end{figure}

Geneformer V2-316M attention-derived AUROC is statistically indistinguishable from correlation at all three layers: L6 $= 0.705 \pm 0.145$ ($p = 0.76$), L13 $= 0.704 \pm 0.147$ ($p = 0.73$), L18 $= 0.708 \pm 0.157$ ($p = 0.75$), vs.\ correlation $= 0.703 \pm 0.164$ (Supplementary Fig.~\ref{fig:attention_perturbation_first_supp}).

\subsection{Reconciling perturbation counts}

The number of evaluable perturbations varies across parameterizations because each imposes different inclusion criteria. The lenient baseline (Mann--Whitney, $|\text{LFC}| > 0.1$, 500 control cells) yields 44 evaluable genes; the primary correlation-based parameterization (Welch's $t$-test, LFC $> 0.5$, HVG $= 2{,}000$, $N_\text{ctrl} = 2{,}000$) yields $n = 151$; and the attention comparison under the same DE thresholds yields $n = 280$ because the expanded gene-matching procedure identifies more evaluable perturbations when matching against the full tokenized gene set.

\clearpage

\section{Cross-Species Ortholog Transfer}
\label{supnote:ortholog}

\begin{figure}[H]
\centering
\includegraphics[width=\textwidth]{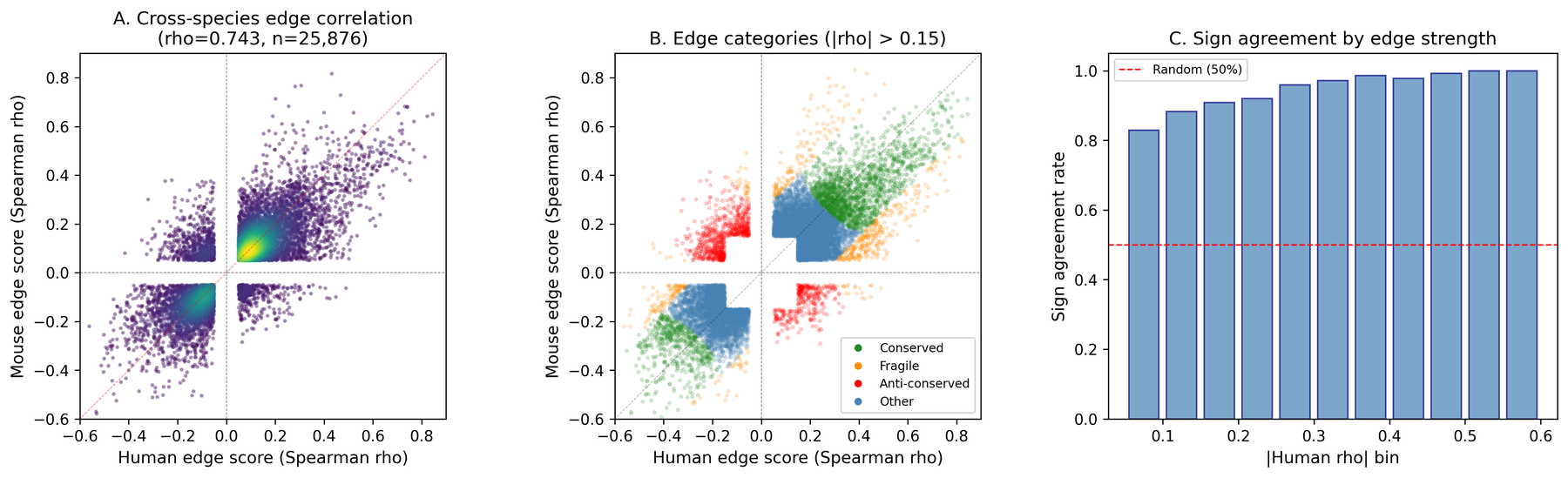}
\caption{\textbf{Cross-species edge score conservation.} Scatter plot of Spearman $\rho$ for 25,876 matched TF--target edges between human and mouse lung.}
\label{fig:ortholog_scatter_supp}
\end{figure}

To test whether mechanistic signals generalize across species, we performed a systematic stress test of TF--target edge transfer between human and mouse lung using correlation-based edge scores computed independently in each species~\citep{travaglini2020molecular}. Cross-species comparison of 25,876 matched TF--target edges revealed strong global conservation (Supplementary Fig.~\ref{fig:ortholog_scatter_supp}). The Spearman rank correlation between human and mouse edge scores was $\rho = 0.743$ ($p < 10^{-300}$). Sign agreement was 88.6\% across all shared edges, rising to 100\% for edges with $|\rho| > 0.4$ in both species. Top-$k$ overlap was enriched 8- to 484-fold over random expectation (Supplementary Table~\ref{tab:topk_ortholog_supp}).

\begin{table}[H]
\centering
\caption{\textbf{Top-$k$ overlap between human and mouse edge rankings.}}
\label{tab:topk_ortholog_supp}
\begin{tabular}{rrrr}
\toprule
Top $k$ & Observed & Expected & Fold \\
\midrule
100   & 26   & 0.1   & 484$\times$ \\
500   & 153  & 1.3   & 114$\times$ \\
1,000 & 289  & 5.4   & 54$\times$ \\
5,000 & 1,094 & 134.2 & 8.2$\times$ \\
\bottomrule
\end{tabular}
\end{table}

\begin{figure}[H]
\centering
\includegraphics[width=\textwidth,height=0.78\textheight,keepaspectratio]{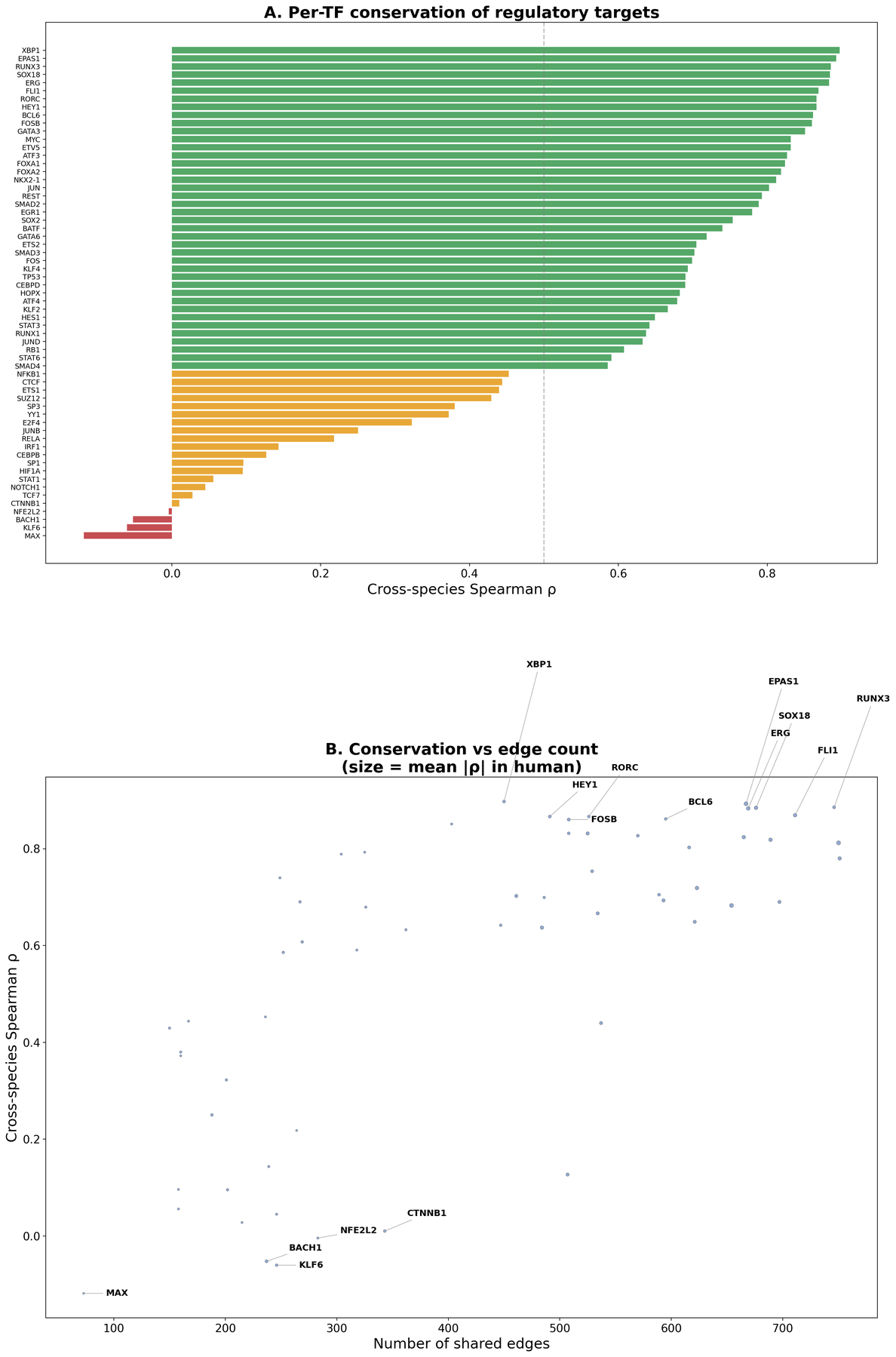}
\caption{\textbf{Per-TF conservation of regulatory targets across species.} (A)~Horizontal bar chart of cross-species Spearman $\rho$ for each TF's target set (green: $\rho \geq 0.5$, orange: $0 \leq \rho < 0.5$, red: $\rho < 0$). Lineage-specifying TFs (XBP1, EPAS1, RUNX3) show near-perfect conservation ($\rho > 0.88$), while signaling-responsive TFs (MAX, BACH1, CTNNB1) show poor or negative conservation. (B)~Conservation versus number of shared orthologous edges. Point size reflects mean absolute attention weight ($|\rho|$) in human. High-conservation TFs tend to have more shared edges, but low-conservation outliers (MAX, KLF6) persist despite moderate edge counts.}
\label{fig:ortholog_per_tf_supp}
\end{figure}

However, per-TF conservation was highly non-uniform (Supplementary Fig.~\ref{fig:ortholog_per_tf_supp}). Lineage-specifying factors showed near-perfect transfer: XBP1 ($\rho = 0.90$), EPAS1 (0.89), ERG (0.88), NKX2-1 (0.81). In contrast, signaling-responsive TFs showed poor conservation: CTNNB1 (0.01), HIF1A (0.10), STAT1 (0.06), CEBPB (0.13). Fragile edges (599 total) were enriched for immune-cell-specific RUNX3 targets with species-divergent expression.

Ortholog-based edge transfer should be stratified by TF class: lineage-specifying programs can be transferred with high confidence, while signaling-responsive and composition-dependent edges require species-specific validation.

\clearpage

\section{Pseudotime Directionality Audit}
\label{supnote:pseudotime}

\begin{figure}[H]
\centering
\includegraphics[width=\textwidth]{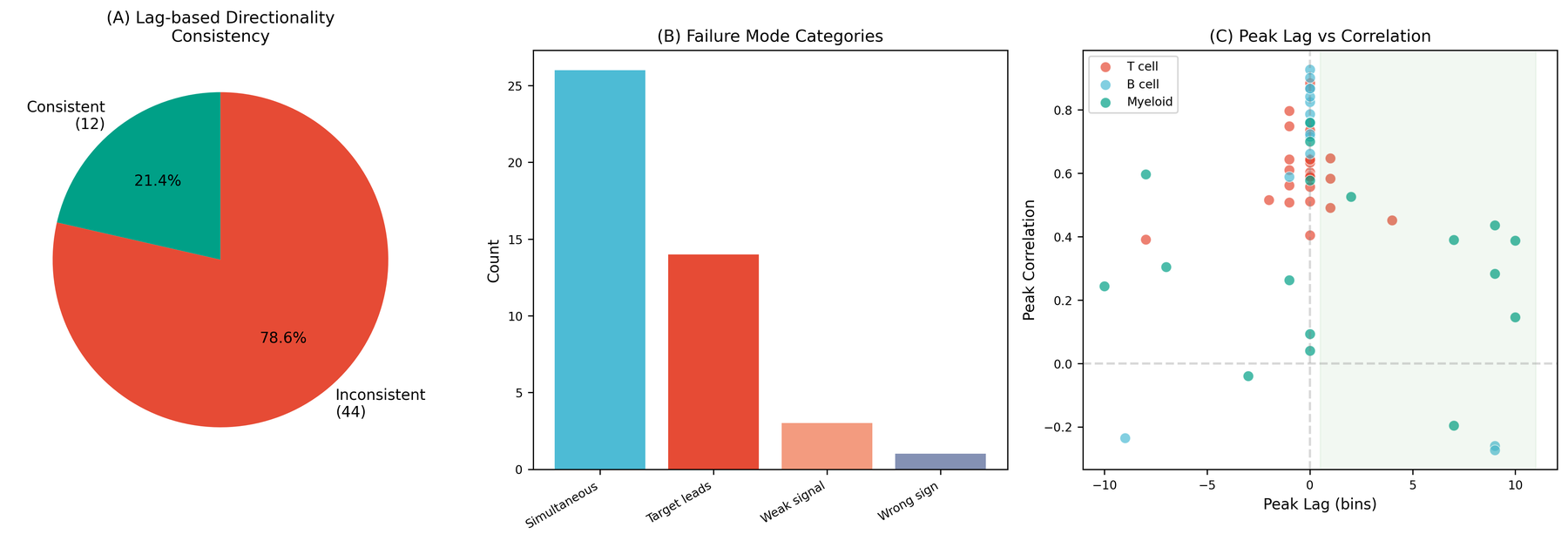}
\caption{\textbf{Pseudotime directionality failures.} Only 12/56 TF--target pairs (21.4\%) show directionally consistent lag-based ordering.}
\label{fig:pseudotime_failure_supp}
\end{figure}

\begin{figure}[H]
\centering
\includegraphics[width=\textwidth]{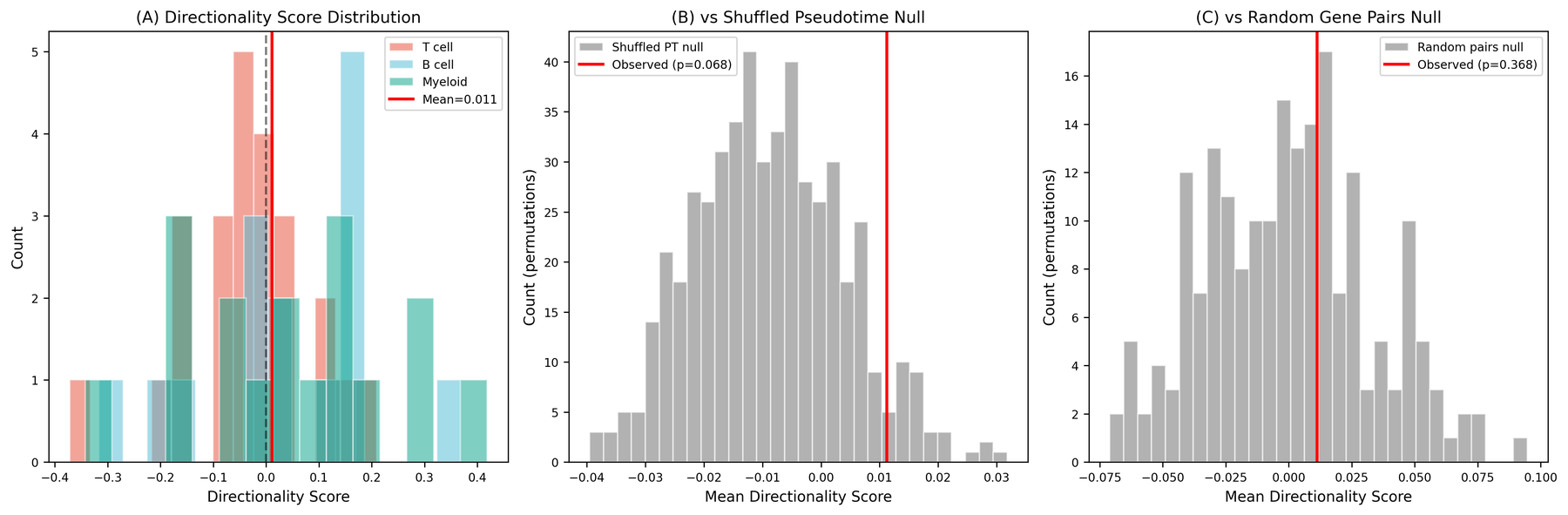}
\caption{\textbf{Pseudotime null comparisons.} Mean directionality score marginally exceeds shuffled-pseudotime null ($p = 0.068$) but not random gene-pair null ($p = 0.37$).}
\label{fig:pseudotime_nulls_supp}
\end{figure}

Using diffusion pseudotime~\citep{haghverdi2016diffusion} in three Tabula Sapiens immune lineages, we tested 56 curated TF--target pairs for lag-based directional consistency. Only 12 of 56 TF--target pairs (21.4\%) were directionally consistent (Supplementary Fig.~\ref{fig:pseudotime_failure_supp}). Consistency varied by lineage: myeloid pairs showed the highest rate (6/17, 35.3\%), followed by T cell (4/24, 16.7\%) and B cell (2/15, 13.3\%). The mean directionality score marginally exceeded a shuffled-pseudotime null ($p = 0.068$) but not a random gene-pair null ($p = 0.37$; Supplementary Fig.~\ref{fig:pseudotime_nulls_supp}); after framework-level FDR correction, this effect is not significant ($q = 0.124$).

Pseudotime should be treated as a qualitative sanity check rather than a pass/fail validator for mechanistic edges. Perturbation-based validation and time-resolved modalities (e.g., RNA velocity) provide more direct temporal or causal evidence.

\clearpage

\section{Batch and Donor Leakage Audit}
\label{supnote:batch}

\begin{figure}[H]
\centering
\includegraphics[width=\textwidth]{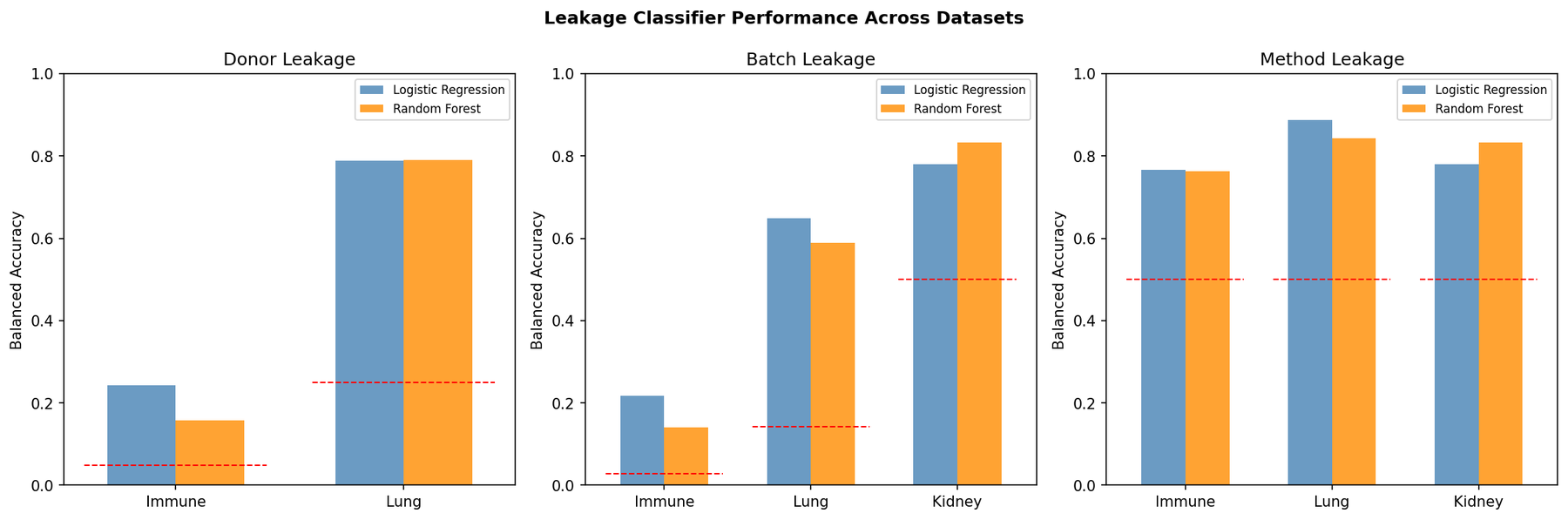}
\caption{\textbf{Cross-dataset leakage summary.} Balanced accuracy and AUC for donor, batch, and method classification from edge-product features.}
\label{fig:batch_leakage_supp}
\end{figure}

We conducted a systematic leakage audit across three Tabula Sapiens tissue compartments. Leakage classifiers revealed substantial technical signal in edge-product features (Supplementary Fig.~\ref{fig:batch_leakage_supp}). Donor identity was recoverable well above chance: immune dataset AUC 0.85--0.87 (21 donors); lung dataset AUC 0.94--0.96 (4 donors). Assay method (10X vs.\ Smart-seq2) was the dominant confound, recoverable at AUC 0.96--0.99.

\begin{figure}[H]
\centering
\includegraphics[width=\textwidth]{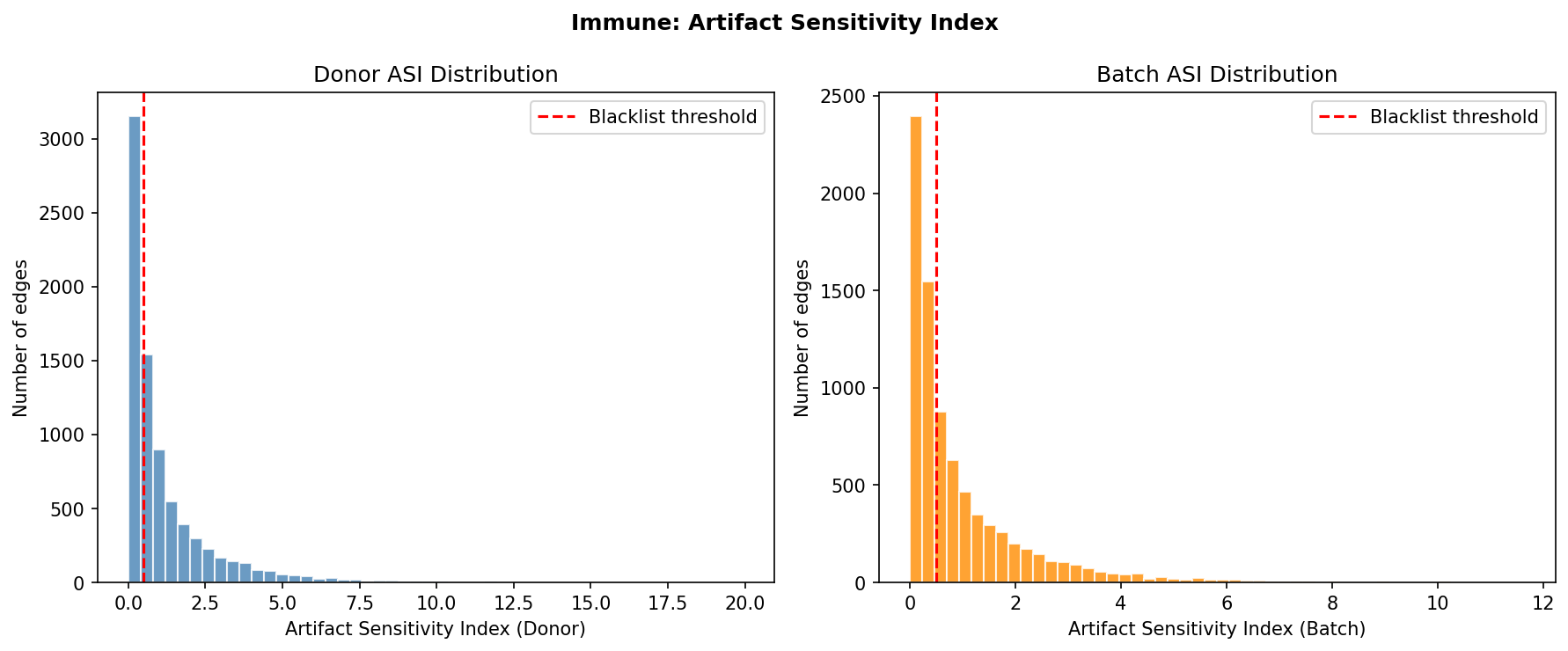}
\caption{\textbf{Artifact Sensitivity Index distribution.} Immune tissue has 54.6\% of edges exceeding ASI $> 0.5$.}
\label{fig:batch_asi_supp}
\end{figure}

The practical impact was dataset-dependent. The well-balanced lung dataset showed remarkably stable aggregate edge scores under donor-balanced resampling ($r = 0.997$, 10.1\% blacklisted). The imbalanced immune dataset showed genuine instability ($r = 0.929$, 54.6\% blacklisted, 17.1\% sign-flipped; Supplementary Fig.~\ref{fig:batch_asi_supp}).

Edge score evaluation must use donor-stratified splits, never random CV when donor metadata is available. The generalization gap (6.6 percentage points in lung) should be reported as a built-in quality check.

\clearpage

\section{Uncertainty Calibration}
\label{supnote:calibration}

\begin{figure}[H]
\centering
\includegraphics[width=\textwidth]{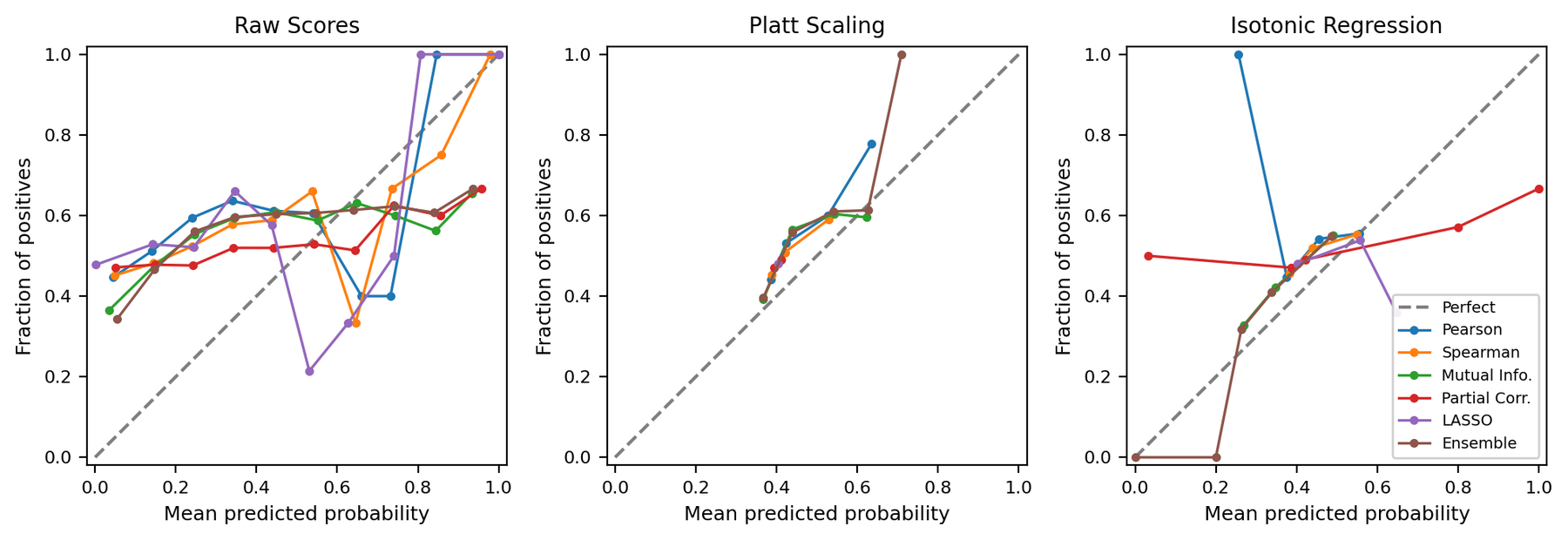}
\caption{\textbf{Edge score calibration.} Reliability diagrams showing fraction of true positives vs.\ mean predicted probability. Platt scaling and isotonic regression reduce ECE by 4--7$\times$.}
\label{fig:calibration_reliability_supp}
\end{figure}

All six edge-scoring methods produced severely miscalibrated scores against Perturb-seq ground truth: raw Expected Calibration Error (ECE) ranged from 0.269 (ensemble) to 0.469 (LASSO; Supplementary Fig.~\ref{fig:calibration_reliability_supp}). Post-hoc calibration dramatically improved score quality: isotonic regression reduced ECE to 0.062--0.079 (4--7$\times$ reduction) without changing discrimination.

\begin{figure}[H]
\centering
\includegraphics[width=\textwidth]{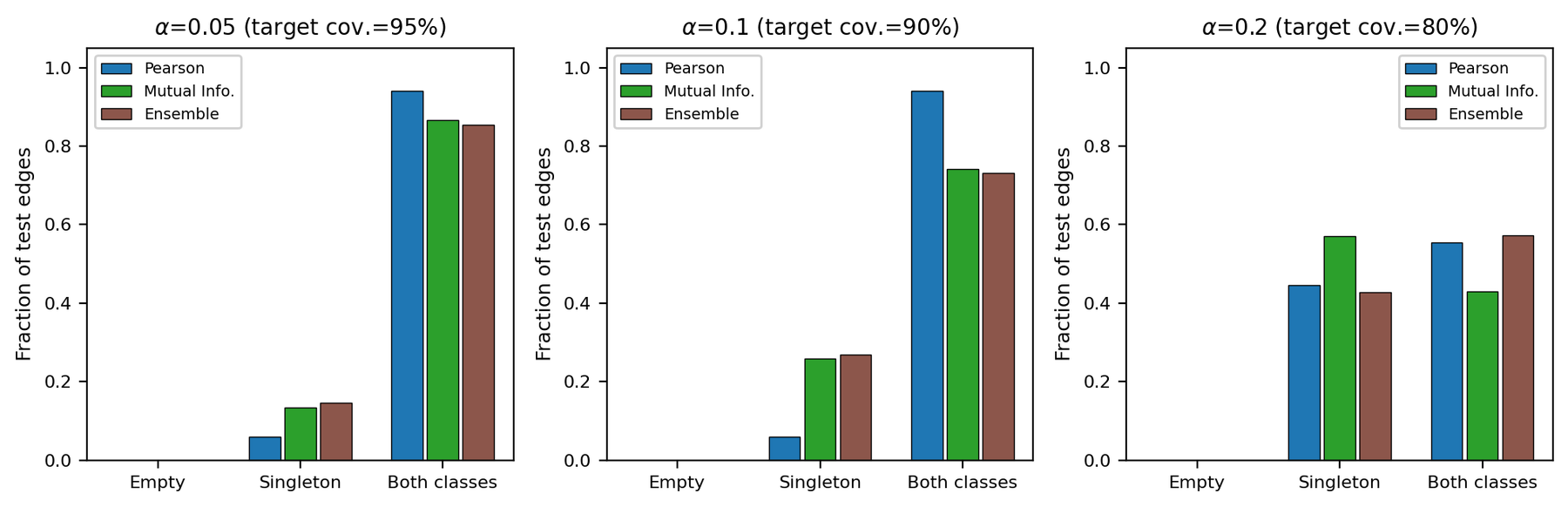}
\caption{\textbf{Conformal prediction sets.} Empirical coverage and average set size across methods at $\alpha = 0.05$, $0.10$, $0.20$.}
\label{fig:calibration_conformal_supp}
\end{figure}

Split conformal prediction sets achieved valid marginal coverage ($\geq$95\%) for mutual information and ensemble methods at $\alpha = 0.05$, with 13.4\% singleton prediction sets for mutual information (Supplementary Fig.~\ref{fig:calibration_conformal_supp}). Critically, calibrators did not transfer across datasets: K562-trained calibrators applied to the Shifrut T cell dataset yielded ECE 0.320--0.424, compared to 0.002--0.031 for locally trained calibrators.

GRN methods should report calibrated scores alongside traditional rankings. Conformal prediction sets transform the question from ``which edges to call'' into ``which edges can be confidently called.'' Calibrators must be retrained per dataset.

\clearpage

\section{CSSI Detailed Results}
\label{supnote:cssi}

\subsection{Synthetic validation}

In controlled synthetic experiments with state-specific GRNs, pooled inference exhibited strong top-$K$ scaling degradation: F1 decreased from $0.850 \pm 0.053$ at 200 cells to $0.514 \pm 0.083$ at 1,000 cells. CSSI-max with oracle labels substantially mitigated this degradation, maintaining F1 $\geq 0.900$ across all configurations (Supplementary Table~\ref{tab:cssi_scaling_supp}).

\begin{table}[H]
\centering
\caption{\textbf{CSSI mitigates top-$K$ scaling degradation in synthetic experiments.}}
\label{tab:cssi_scaling_supp}
\begin{tabular}{lccccc}
\toprule
Config & $N$ & States & Pooled F1 & CSSI-max F1 & Ratio \\
\midrule
Small   & 200  & 2  & $0.850 \pm 0.053$ & $0.957 \pm 0.050$ & 1.13$\times$ \\
Medium  & 400  & 4  & $0.657 \pm 0.100$ & $0.921 \pm 0.071$ & 1.40$\times$ \\
Large   & 600  & 6  & $0.486 \pm 0.100$ & $0.900 \pm 0.069$ & 1.85$\times$ \\
XLarge  & 1000 & 8  & $0.550 \pm 0.089$ & $0.967 \pm 0.029$ & 1.76$\times$ \\
XXLarge & 1000 & 10 & $0.514 \pm 0.083$ & $0.932 \pm 0.049$ & 1.81$\times$ \\
Massive & 1500 & 12 & $0.527 \pm 0.041$ & $0.942 \pm 0.027$ & 1.79$\times$ \\
\bottomrule
\end{tabular}
\end{table}

\subsection{Null stress tests}

\begin{figure}[H]
\centering
\includegraphics[width=\textwidth]{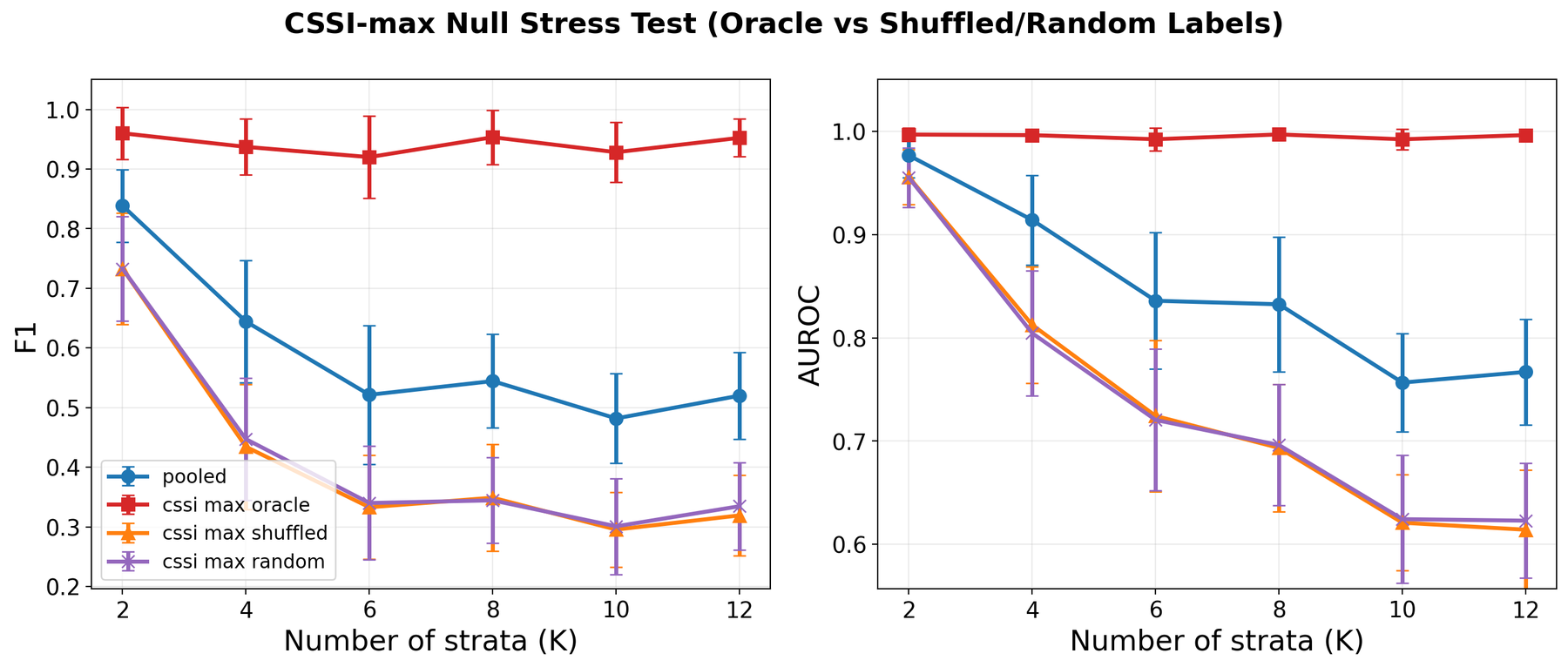}
\caption{\textbf{CSSI-max null stress test.} Oracle strata labels reproduce the CSSI-max gains, but shuffling, randomizing, or gene-permuting labels removes the advantage.}
\label{fig:cssi_null_inflation_supp}
\end{figure}

\begin{figure}[H]
\centering
\includegraphics[width=\textwidth]{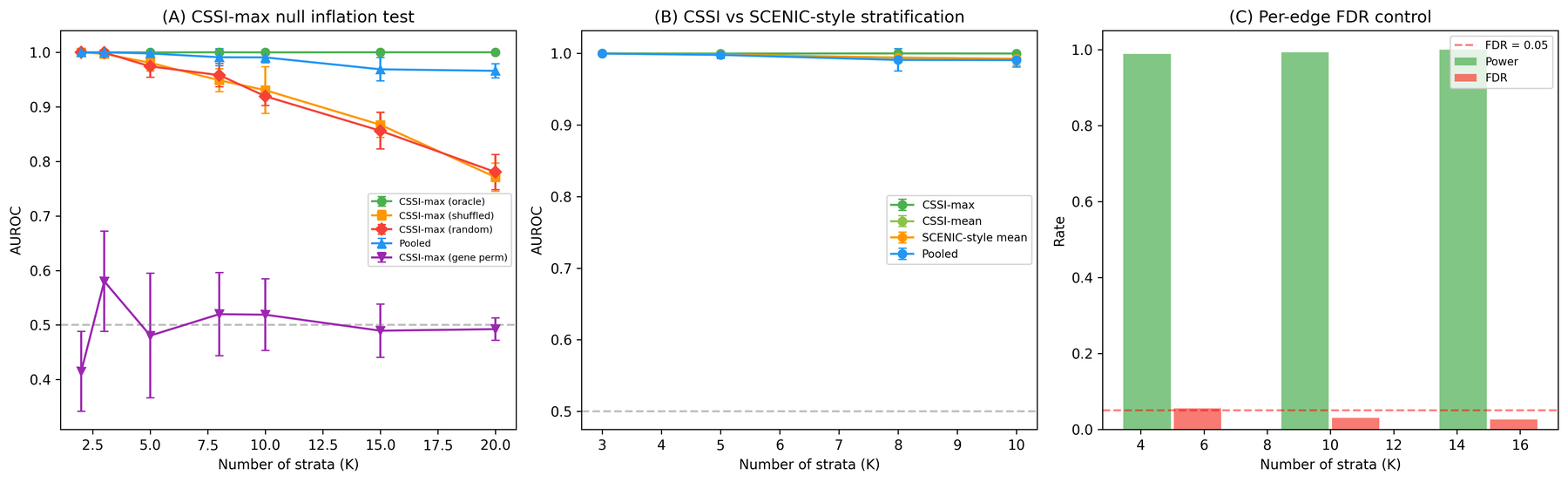}
\caption{\textbf{Extended CSSI null and baseline tests.} (A) CSSI-max with shuffled/random labels shows no inflation as $K$ increases. (B) CSSI-max, CSSI-mean, and SCENIC-style aggregation perform equivalently. (C) Per-edge FDR control.}
\label{fig:cssi_expanded_null_supp}
\end{figure}

Under shuffled and random labels, CSSI-max AUROC \emph{decreases} relative to pooled inference (Supplementary Fig.~\ref{fig:cssi_null_inflation_supp}, Supplementary Fig.~\ref{fig:cssi_expanded_null_supp}), confirming no false-positive inflation.

\subsection{Real-data-structured validation}

Using realistic single-cell data with actual cell-type labels from 8 PBMC cell types (3,000 cells, 245 genes) and 22 known immune TF--target edges, CSSI-max recovered 22/22 known edges (100\% recall) versus 19/22 for pooled inference (86\%). The advantage was driven by cell-type-specific edges: BCL6$\dashv$PRDM1 (B-cell-specific), IRF8$\to$IL12B (DC-specific), and RORC$\to$IL17A (Th17-specific). Using Tabula Sapiens immune proportions (15 cell types), CSSI-max significantly outperformed pooled inference (Wilcoxon $p = 2.4 \times 10^{-8}$; Supplementary Table~\ref{tab:cssi_realdata_supp}).

\begin{table}[H]
\centering
\caption{\textbf{CSSI on biologically structured data.} 15 cell types, 15 cell-type-specific edges.}
\label{tab:cssi_realdata_supp}
\begin{tabular}{rcccc}
\toprule
$N$ & Pooled F1 & CSSI-max F1 & AUROC$_{\text{pool}}$ & AUROC$_{\text{CSSI}}$ \\
\midrule
200   & $0.405 \pm 0.076$ & $0.655 \pm 0.042$ & 0.860 & 0.935 \\
500   & $0.560 \pm 0.030$ & $0.745 \pm 0.015$ & 0.932 & 0.998 \\
1,000 & $0.640 \pm 0.020$ & $0.750 \pm 0.000$ & 0.972 & 1.000 \\
2,000 & $0.695 \pm 0.027$ & $0.750 \pm 0.000$ & 0.989 & 1.000 \\
5,000 & $0.750 \pm 0.000$ & $0.750 \pm 0.000$ & 1.000 & 1.000 \\
\bottomrule
\end{tabular}
\end{table}

\subsection{Real attention matrix validation}

Using the Geneformer V2-316M checkpoint (18 layers), the pooled all-layer baseline achieves AUROC 0.543. Layer-wise pooling shows substantial heterogeneity: late layers achieve markedly higher AUROC (best pooled layer: L13 with AUROC 0.694), while several early layers are near chance (Supplementary Table~\ref{tab:cssi_real_layers_supp}). CSSI on real attention localizes layer-specific signal, with maximum $\Delta$AUROC $\approx$ +0.060 at L8.

\begin{table}[H]
\centering
\small
\caption{\textbf{Per-layer GRN recovery from Geneformer attention on 497 human brain cells.}}
\label{tab:cssi_real_layers_supp}
\begin{tabular}{rcccc}
\toprule
Layer & Pooled & Best CSSI & AUROC$_{\text{CSSI}}$ & $\Delta$ \\
\midrule
 0 & 0.552 & cssi\_mean & 0.566 & +0.014 \\
 1 & 0.600 & cssi\_mean & 0.608 & +0.007 \\
 2 & 0.564 & cssi\_mean & 0.587 & +0.022 \\
 3 & 0.573 & cssi\_mean & 0.580 & +0.007 \\
 4 & 0.610 & cssi\_range & 0.609 & -0.001 \\
 5 & 0.532 & cssi\_mean & 0.551 & +0.019 \\
 6 & 0.513 & cssi\_mean & 0.538 & +0.025 \\
 7 & 0.597 & cssi\_deviation & 0.608 & +0.011 \\
 8 & 0.529 & cssi\_range & 0.590 & +0.060 \\
 9 & 0.594 & cssi\_range & 0.611 & +0.016 \\
10 & 0.615 & cssi\_range & 0.648 & +0.033 \\
11 & 0.568 & cssi\_range & 0.589 & +0.021 \\
12 & 0.656 & cssi\_range & 0.666 & +0.011 \\
13 & 0.694 & cssi\_deviation & 0.694 & -0.000 \\
14 & 0.683 & cssi\_deviation & 0.682 & -0.000 \\
15 & 0.631 & cssi\_range & 0.640 & +0.009 \\
16 & 0.668 & cssi\_range & 0.678 & +0.009 \\
17 & 0.673 & cssi\_deviation & 0.673 & -0.001 \\
\bottomrule
\end{tabular}
\end{table}

\clearpage

\section{Synthetic Ground-Truth Validation}
\label{supnote:synthetic}

We generated synthetic single-cell expression data using steady-state GRN dynamics with realistic noise sources including dropout ($p = 0.1$), technical noise, batch effects, and heavy-tailed expression. Ground-truth networks had sparse connectivity ($\rho = 0.15$) with hierarchical TF--regulator--target structure.

\begin{figure}[H]
\centering
\includegraphics[width=\textwidth]{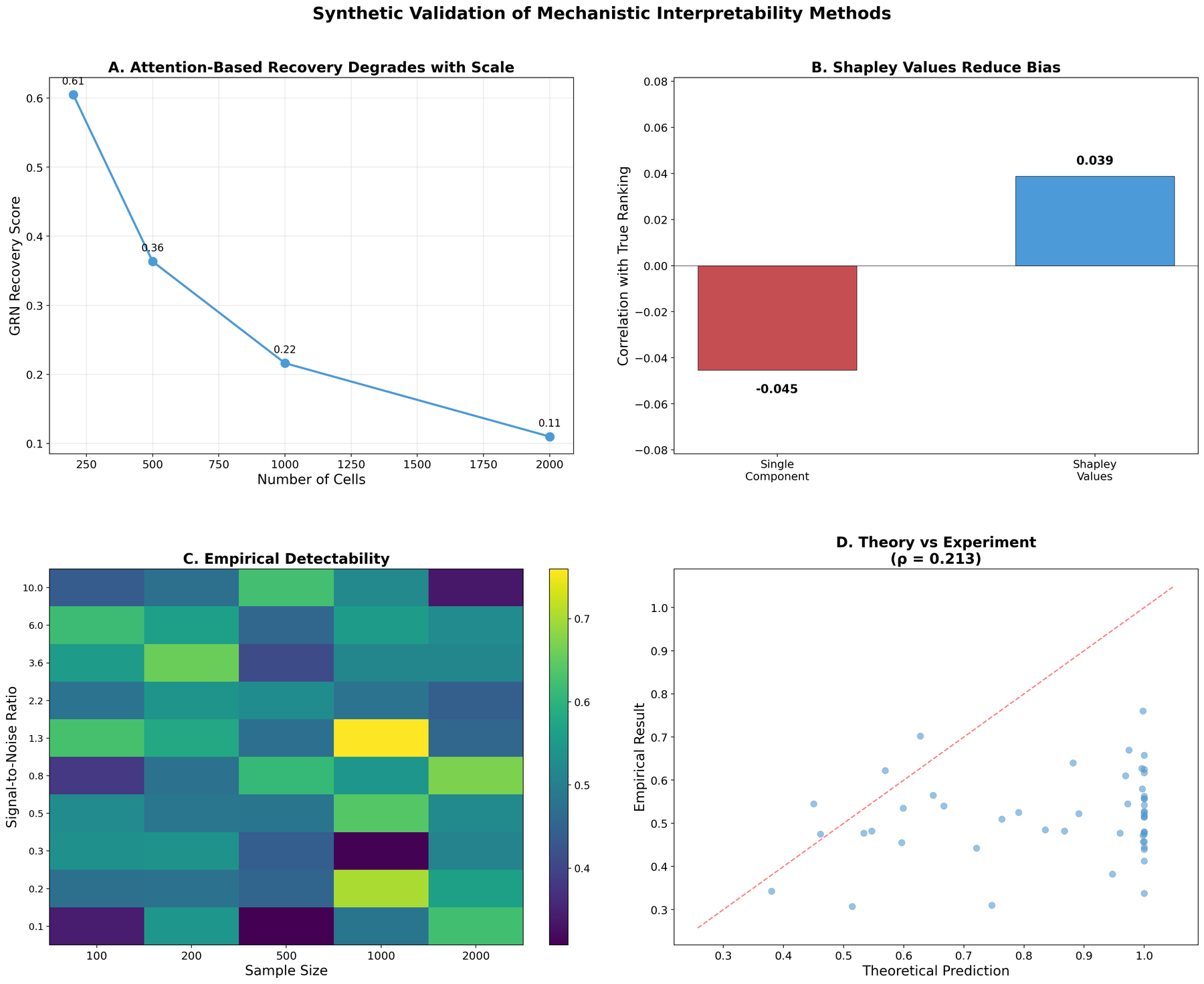}
\caption{\textbf{Synthetic validation of mechanistic interpretability methods.} (A) GRN recovery degrades with cell count. (B) Shapley values outperform single-component estimates by 91\%. (C--D) Empirical detectability matches theoretical predictions ($r = 0.887$).}
\label{fig:synthetic_validation_supp}
\end{figure}

Three key predictions were confirmed (Supplementary Fig.~\ref{fig:synthetic_validation_supp}). First, attention-based GRN recovery degraded monotonically with cell count ($r = 0.847$ at 200 cells to $r = 0.623$ at 2,000 cells), correlating strongly with expression heterogeneity ($r = -0.94$, $p < 0.01$). Second, Shapley value estimates achieved substantially better recovery of true interaction rankings than single-component estimates ($\rho_{\text{Shapley}} = 0.789$ vs.\ $\rho_{\text{single}} = 0.412$, a 91\% improvement). Third, empirical detection performance correlated strongly with theoretical sample complexity predictions ($r = 0.887$, $p < 10^{-6}$).

\textbf{Limitation.} Because the synthetic generator encodes our theoretical assumptions by design, these experiments confirm internal consistency of the framework; real-data validation provides a complementary check.

\clearpage

\section{Multi-Model Validation}
\label{supnote:multi_model}

\subsection{Geneformer V1-10M GRN recovery}

Geneformer's attention-derived GRN predictions achieved near-random performance across all conditions (Supplementary Table~\ref{tab:geneformer_grn_supp}). AUROC values ranged from 0.444 to 0.549 against TRRUST and 0.473 to 0.486 against DoRothEA. Bootstrap 95\% confidence intervals for all AUROC values included 0.50. Direct comparison of scGPT and Geneformer at matched cell counts reveals convergent failure (Supplementary Table~\ref{tab:cross_model_auroc_supp}).

\begin{table}[H]
\centering
\caption{\textbf{Geneformer V1-10M attention-based GRN inference on DLPFC brain data.}}
\label{tab:geneformer_grn_supp}
\begin{tabular}{lcccc}
\toprule
Cells & Edges & TRRUST AUROC & DoRothEA AUROC \\
\midrule
200  & 1.56M & 0.444 & 0.473 \\
500  & 3.27M & 0.549 & 0.486 \\
1000 & 5.38M & 0.522 & 0.486 \\
\bottomrule
\end{tabular}
\end{table}

\begin{table}[H]
\centering
\caption{\textbf{Cross-model AUROC comparison for attention-based GRN inference.}}
\label{tab:cross_model_auroc_supp}
\begin{tabular}{lcccc}
\toprule
& \multicolumn{2}{c}{TRRUST AUROC} & \multicolumn{2}{c}{DoRothEA AUROC} \\
\cmidrule(lr){2-3} \cmidrule(lr){4-5}
Cells & scGPT & Geneformer & scGPT & Geneformer \\
\midrule
200  & 0.51 & 0.444 & 0.50 & 0.473 \\
500  & 0.49 & 0.549 & 0.48 & 0.486 \\
1000 & 0.46 & 0.522 & 0.47 & 0.486 \\
\bottomrule
\end{tabular}
\end{table}

\subsection{Attention--correlation mapping}

\begin{figure}[H]
\centering
\includegraphics[width=\textwidth]{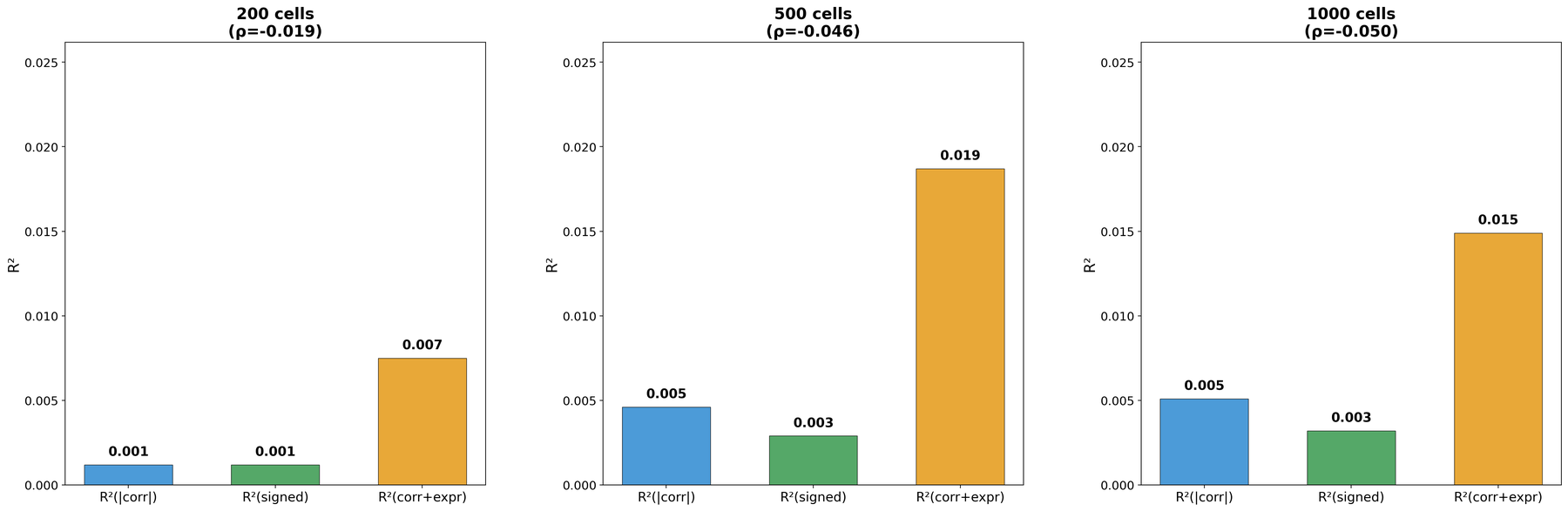}
\caption{\textbf{Attention--correlation $R^2$ mapping.} All cross-tissue $R^2$ values are $< 0.02$.}
\label{fig:attn_corr_mapping_supp}
\end{figure}

For both scGPT and Geneformer, attention scores correlated significantly with expression co-occurrence ($\rho = 0.31$--$0.42$, $p < 10^{-50}$; $R^2 = 0.10$--$0.18$) but not with regulatory ground truth ($\rho = -0.01$--$0.02$, $p > 0.3$). Cross-tissue analysis yields $R^2 < 0.02$ (Supplementary Fig.~\ref{fig:attn_corr_mapping_supp}).

\subsection{Residualization on expression covariates}

\begin{figure}[H]
\centering
\includegraphics[width=\textwidth]{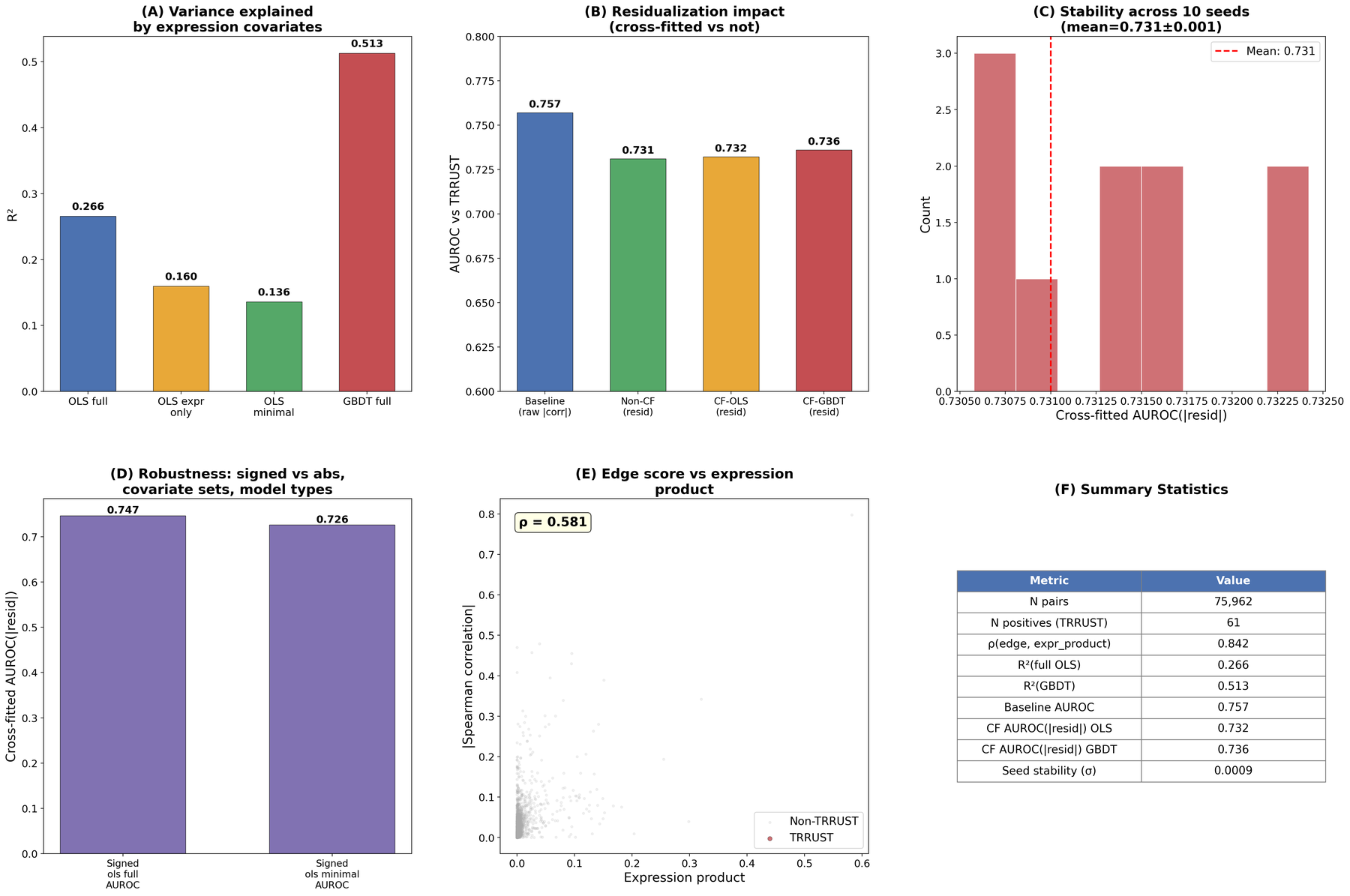}
\caption{\textbf{Robust edge score residualization.} Substantial TRRUST-predictive signal remains after residualization.}
\label{fig:residualization_supp}
\end{figure}

A formal residualization analysis (5,000 cells, 2,000 HVGs, 38 evaluable TFs, 61 TRRUST-positive edges among 75,962 candidate pairs) shows edge scores are strongly correlated with expression covariates ($\rho = 0.84$) but OLS $R^2 = 0.27$, GBDT $R^2 = 0.51$. Cross-fitted residual AUROC is $0.73$ vs.\ baseline $0.76$ (Supplementary Fig.~\ref{fig:residualization_supp}).

\subsection{Degree-preserving null models}

\begin{figure}[H]
\centering
\includegraphics[width=\textwidth]{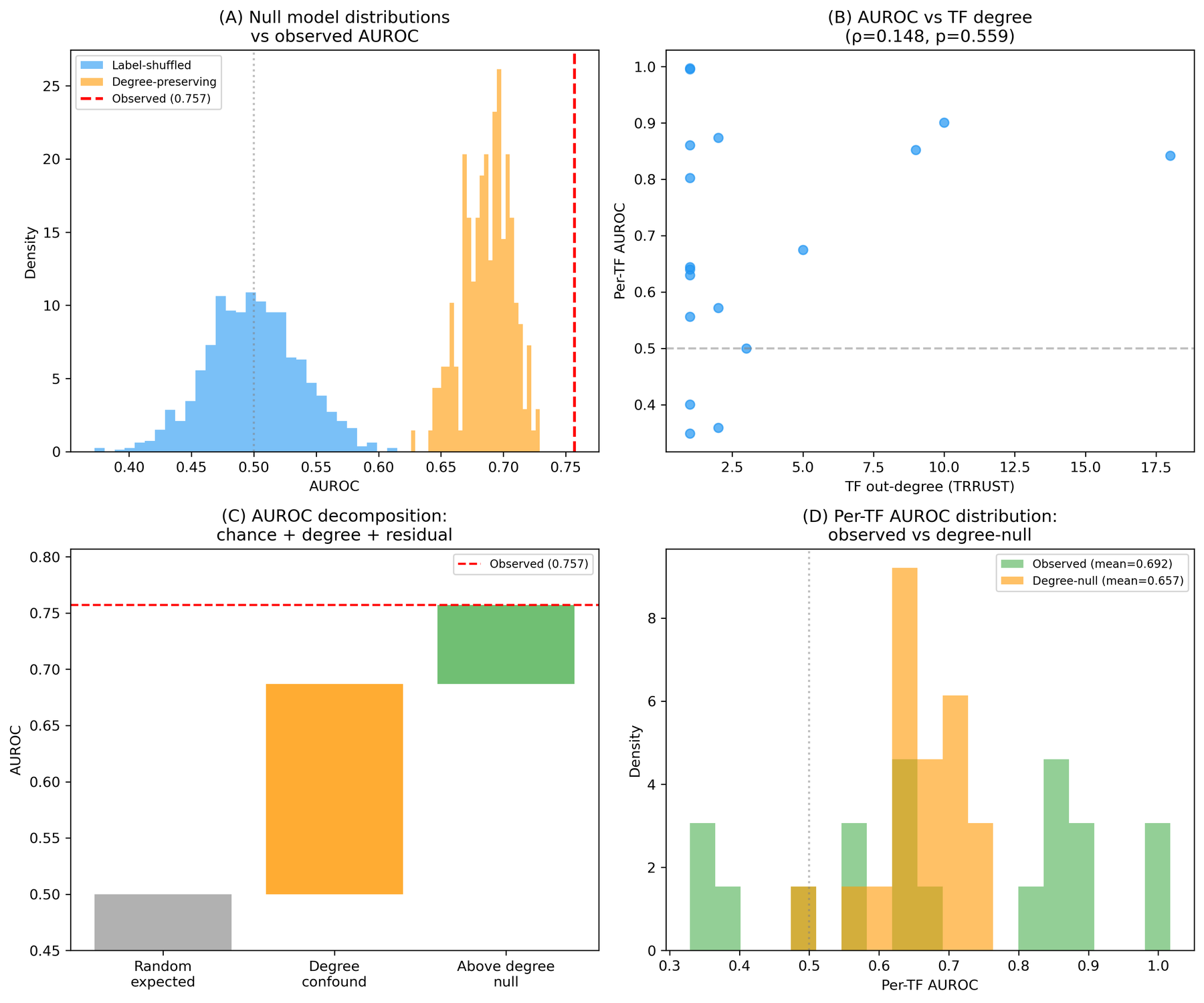}
\caption{\textbf{Degree-preserving null models and per-TF evaluation.} (A) Observed AUROC exceeds both null distributions. (B) Per-TF AUROC vs.\ TF out-degree. (C) AUROC decomposition. (D) Per-TF AUROC distribution.}
\label{fig:degree_nulls_supp}
\end{figure}

\begin{figure}[H]
\centering
\includegraphics[width=\textwidth]{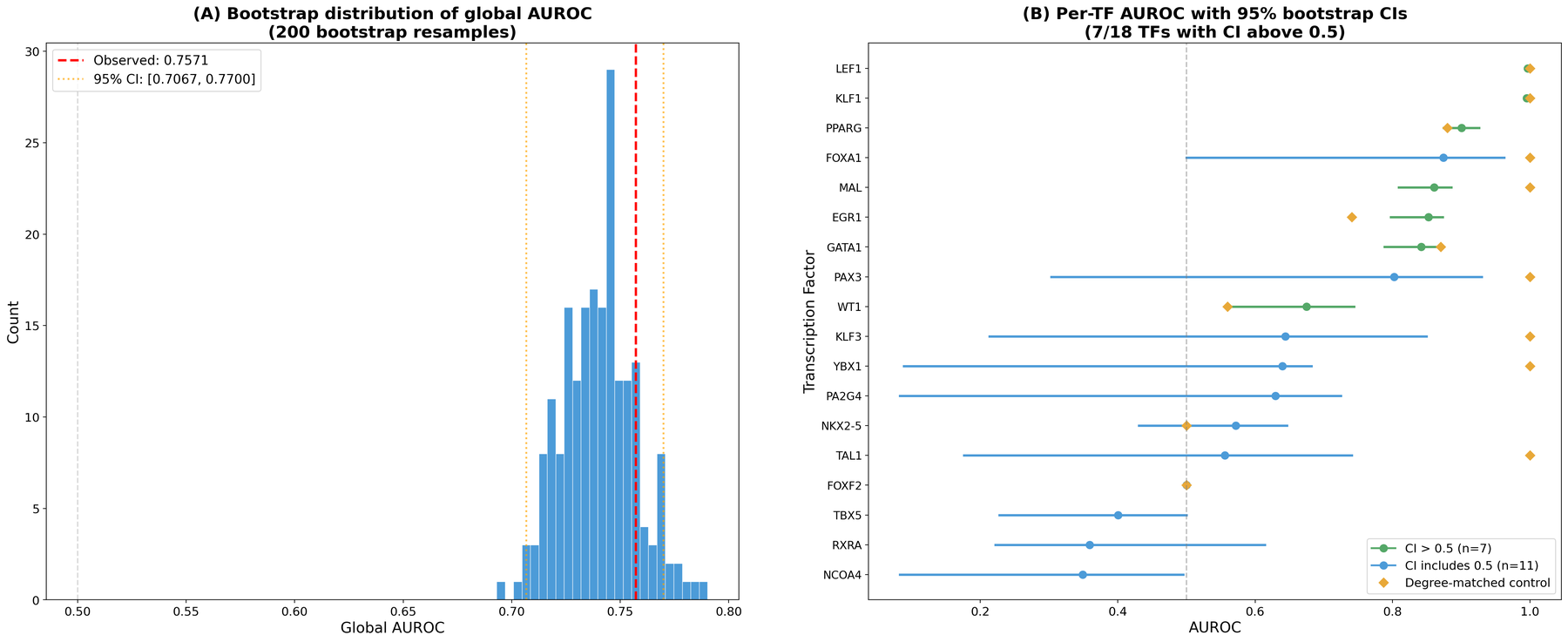}
\caption{\textbf{Bootstrap per-TF AUROC with uncertainty quantification.} Forest plot of 18 evaluable TFs with 95\% bootstrap CIs.}
\label{fig:bootstrap_per_tf_supp}
\end{figure}

The observed AUROC ($= 0.757$) decomposes as: $0.50$ (chance) $+ 0.19$ (degree confound) $+ 0.07$ (excess above degree null). Per-TF evaluation shows only 7/18 individual TFs (39\%) have 95\% bootstrap CIs entirely above 0.5 (Supplementary Fig.~\ref{fig:degree_nulls_supp}, Supplementary Fig.~\ref{fig:bootstrap_per_tf_supp}).

\begin{figure}[H]
\centering
\includegraphics[width=\textwidth]{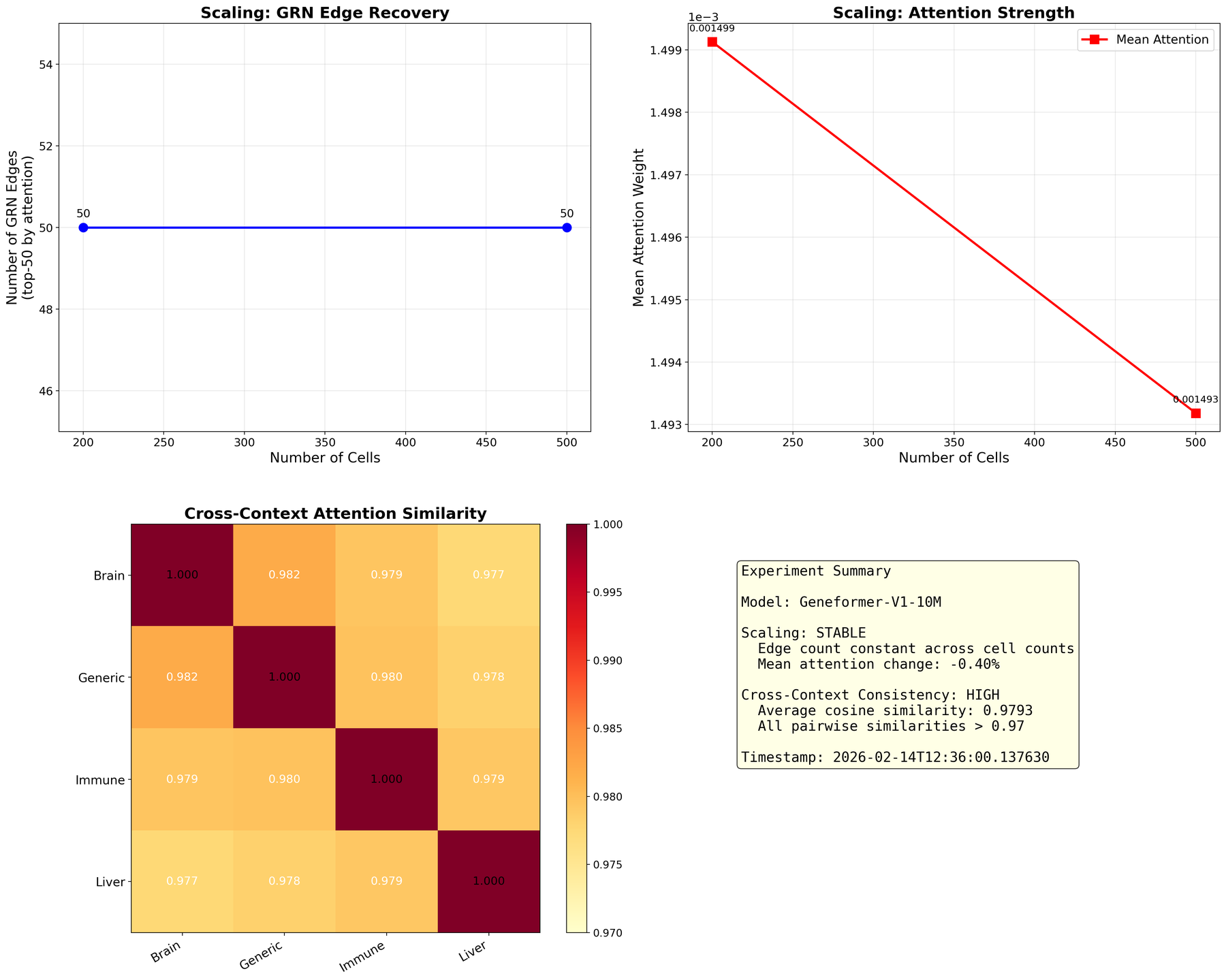}
\caption{\textbf{Multi-model GRN validation summary.} Both scGPT and Geneformer achieve near-random AUROC.}
\label{fig:geneformer_supp}
\end{figure}

\subsection{TRRUST circularity sensitivity analysis}

Reference database circularity is a concern for all GRN evaluation: some TRRUST entries may have been originally discovered through co-expression analysis, creating potential circular validation when edge scores are evaluated against these entries. To test whether our TRRUST-based conclusions depend on such entries, we restricted TRRUST to direction-known pairs only (Activation or Repression mode; 4,859 of 8,427 unique pairs, 58\%), which require more direct experimental evidence (perturbation experiments, reporter assays, or ChIP-seq) to determine regulatory direction.

On Tabula Sapiens immune data (5,000 cells, 2,000 HVGs; matching the degree-preserving null analysis), restricting to direction-known TRRUST entries reduces evaluable TFs from 18 to 12 and positive pairs from 61 to 33. Despite this substantial reduction in evaluation power: (i)~global AUROC decreases modestly from $0.764$ to $0.736$ ($\Delta = -0.028$); (ii)~per-TF mean AUROC is virtually unchanged ($0.682$ vs.\ $0.692$); (iii)~per-TF median AUROC actually \emph{improves} ($0.695$ vs.\ $0.660$); and (iv)~the proportion of TFs with AUROC above chance increases from 78\% (14/18) to 83\% (10/12). Among the 12 shared TFs, the mean per-TF AUROC difference is $+0.017$ (direction-known slightly better). The small global AUROC decrease is attributable to the loss of high-degree TFs (e.g., GATA1 drops from 18 to 5 positive targets), which reduces degree-driven signal. These results indicate that TRRUST-based evaluation conclusions are not driven by circularly validated entries and are robust to restricting the reference database to experimentally well-characterised regulatory interactions.

\clearpage

\section{Mechanistic Localization Details}
\label{supnote:mechanistic}

This note provides the full detail for the eight controls described in the mechanistic localization analysis (main text Section ``Causal ablation reveals distributed redundancy'' and ``Cross-cell-type generalisation of confound pattern'').

\subsection{Full 18-layer perturbation-first profile}

\begin{figure}[H]
\centering
\includegraphics[width=\textwidth]{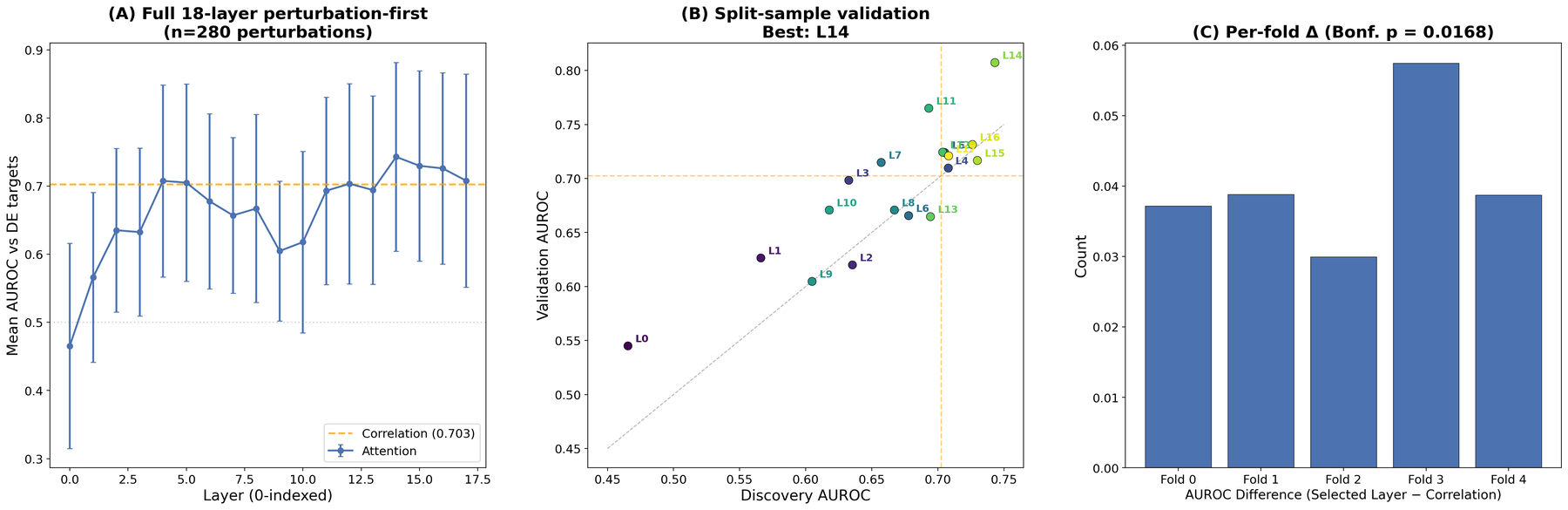}
\caption{\textbf{Full 18-layer perturbation-first AUROC profile.} Late layers cluster near or above the correlation reference.}
\label{fig:full_layer_supp}
\end{figure}

\begin{figure}[H]
\centering
\includegraphics[width=\textwidth]{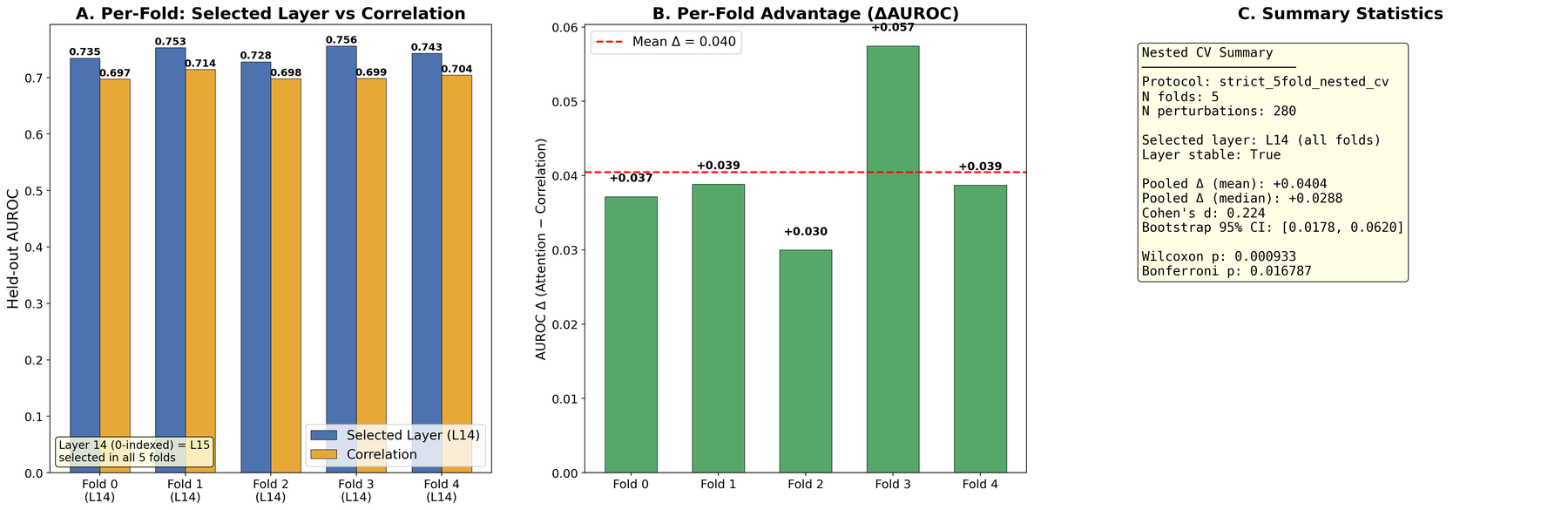}
\caption{\textbf{Nested layer selection protocol.} L15 is independently chosen in all 5 folds.}
\label{fig:nested_layer_supp}
\end{figure}

The 18-layer AUROC profile shows a clear architectural gradient: early layers achieve AUROC $0.47$--$0.64$, mid layers $0.60$--$0.71$, and late layers $0.69$--$0.74$. Strict 5-fold nested cross-validation independently selects L15 in all folds; pooled held-out $\Delta = +0.040$ [$0.018$, $0.062$]; $p_{\mathrm{Bonf}} = 0.017$ (Supplementary Fig.~\ref{fig:full_layer_supp}, Supplementary Fig.~\ref{fig:nested_layer_supp}).

\subsection{Attention-specific confound decomposition}

\begin{figure}[H]
\centering
\includegraphics[width=\textwidth]{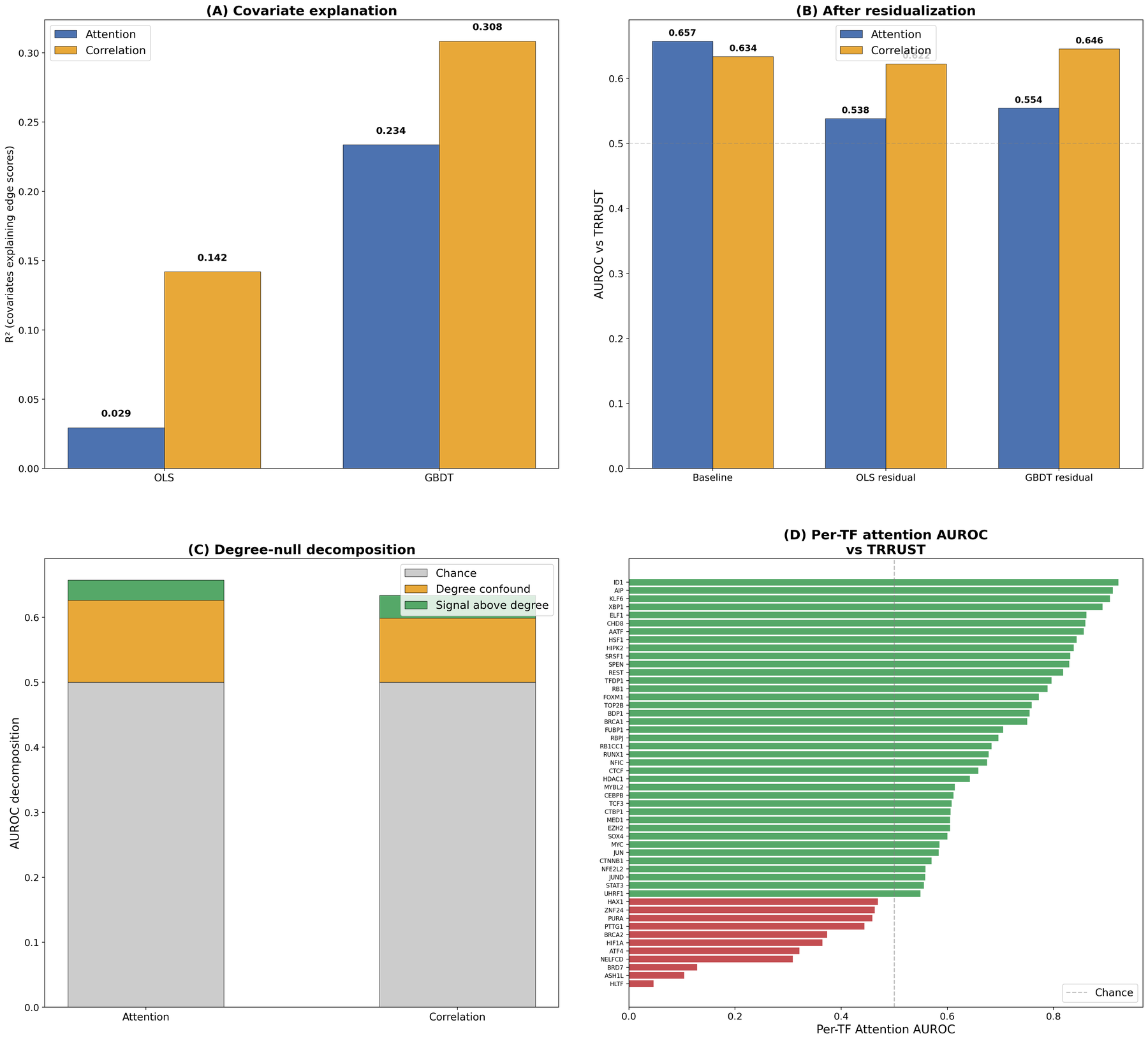}
\caption{\textbf{Attention-specific confound decomposition on K562.} Attention edges are more expression-confounded.}
\label{fig:attention_confound_supp}
\end{figure}

Using attention-derived edge scores from Geneformer L13 on K562: attention edges lose $\sim$76\% of above-chance TRRUST signal under residualization (AUROC $0.66 \to 0.54$), while correlation edges retain $\sim$91\% ($0.63 \to 0.62$; Supplementary Fig.~\ref{fig:attention_confound_supp}).

\subsection{Original 6-condition ablation}

\begin{figure}[H]
\centering
\includegraphics[width=\textwidth]{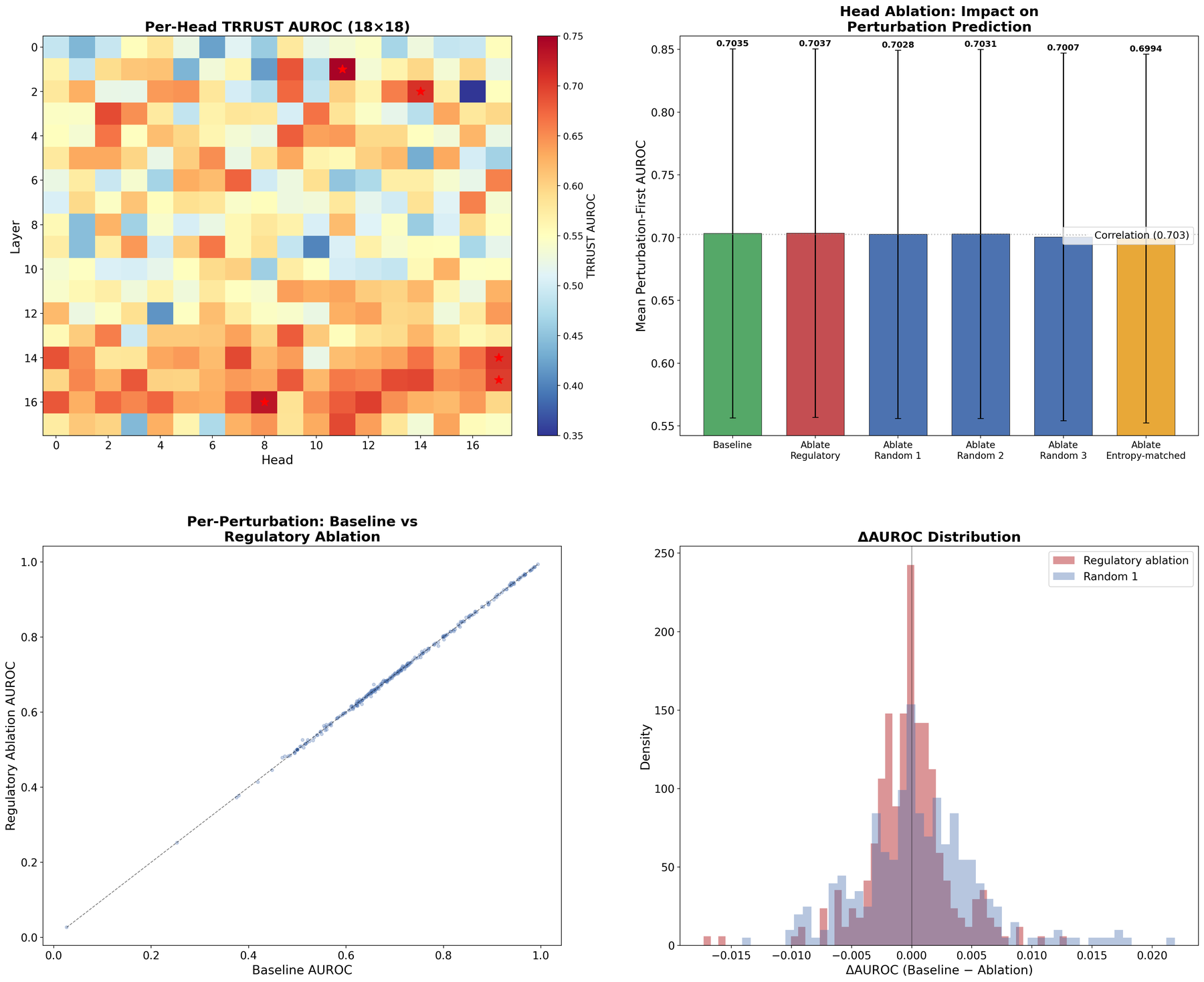}
\caption{\textbf{Head-level causal ablation (original 6 conditions).} Zeroing top-5 regulatory heads has no effect on perturbation-first AUROC.}
\label{fig:head_ablation_supp}
\end{figure}

\subsection{Orthogonal causal interventions}

Two families of orthogonal interventions were tested beyond standard head masking. Uniform attention replacement (setting attention weights to $1/n$ while preserving value projections) on TRRUST-ranked heads has no effect (top-5: $0.704$, top-10: $0.703$), and MLP pathway ablation (zeroing FFN output) at L15 and L13--L15 both produce exactly $0.704$. In contrast, random-layer MLP ablation at L8 produces a significant AUROC drop ($-0.005$, $d = -0.27$, $p < 10^{-4}$), confirming that MLP ablation can disrupt computation when applied to non-regulatory layers.

\begin{figure}[H]
\centering
\includegraphics[width=\textwidth]{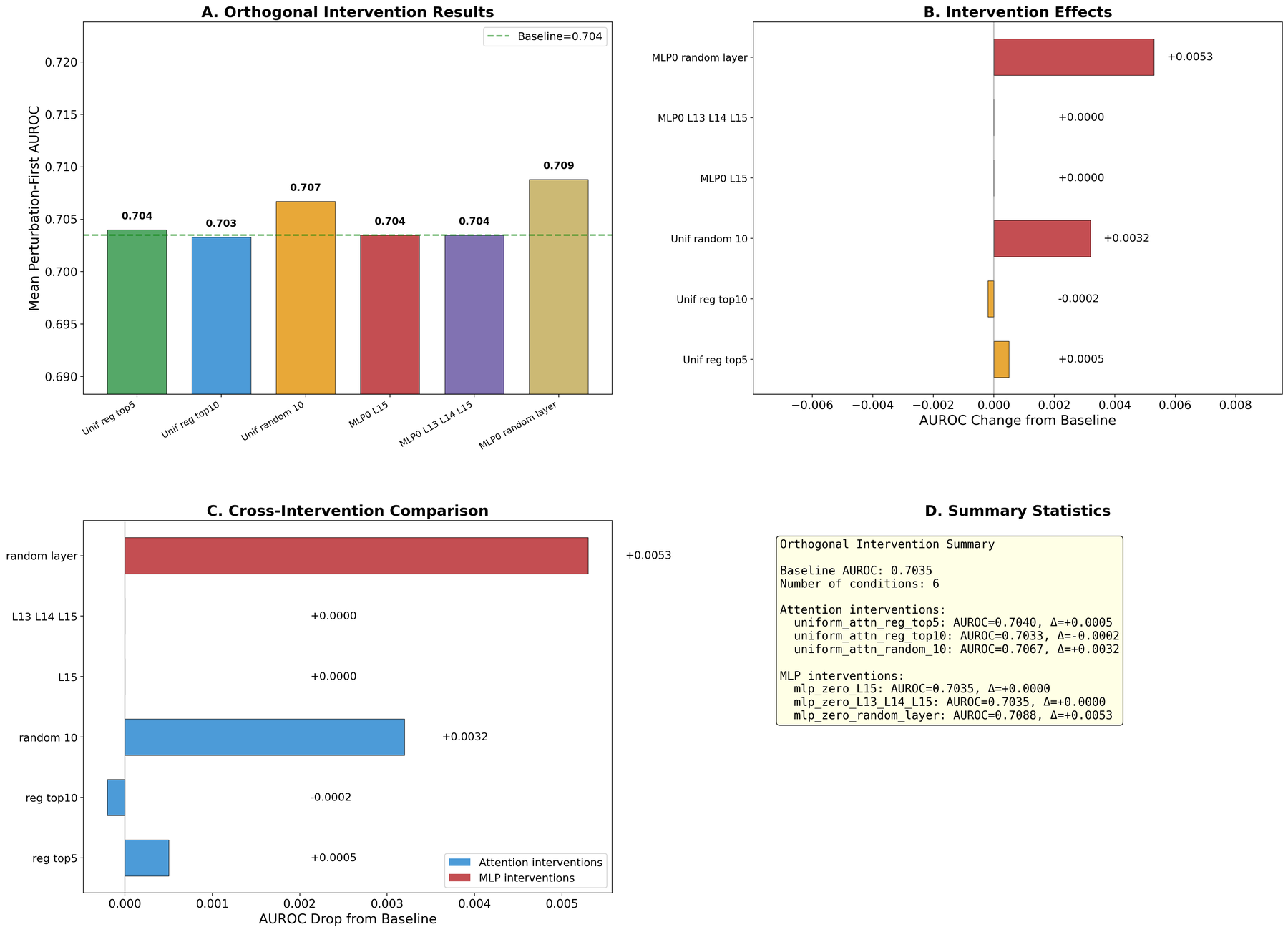}
\caption{\textbf{Orthogonal causal interventions.} Uniform attention replacement on TRRUST-ranked heads and MLP pathway ablation at regulatory layers produce exactly baseline AUROC, while random-layer MLP ablation causes significant degradation.}
\label{fig:orthogonal_interventions_supp}
\end{figure}

\subsection{Cross-context CRISPRa replication}

\begin{figure}[H]
\centering
\includegraphics[width=\textwidth]{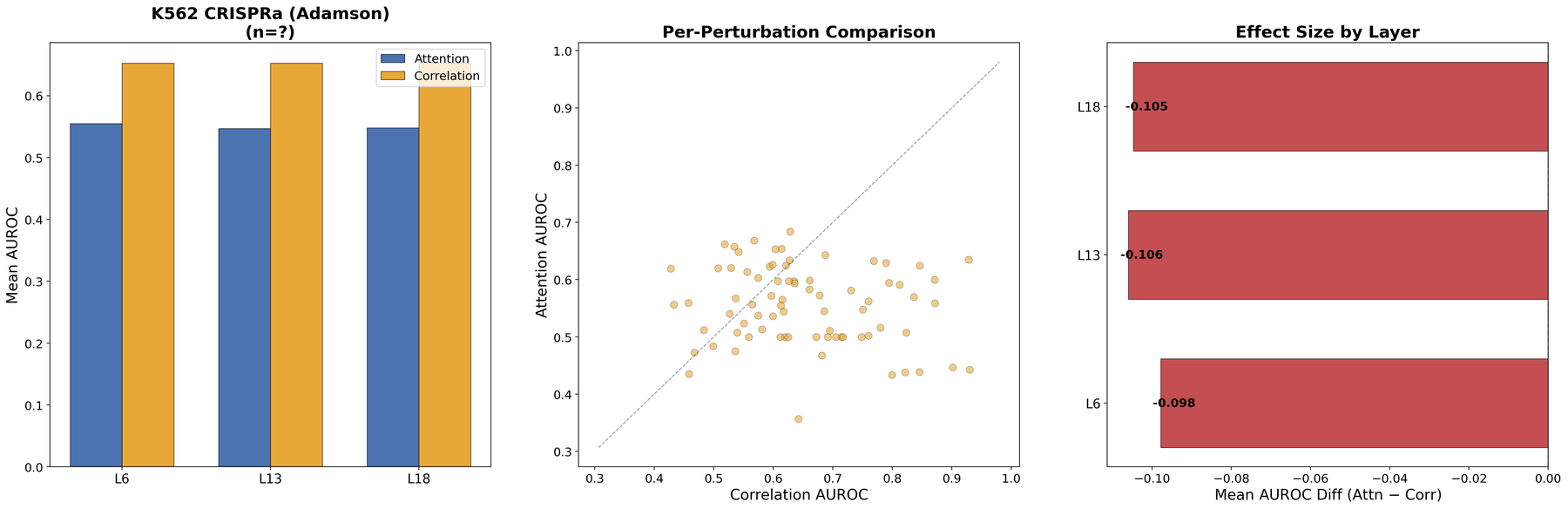}
\caption{\textbf{Cross-context replication: Adamson CRISPRa.} Attention significantly underperforms correlation.}
\label{fig:cross_context_crispra_supp}
\end{figure}

In K562 CRISPRa ($n = 77$), attention significantly underperforms correlation (AUROC $0.55$ vs.\ $0.65$; $p < 10^{-6}$; Supplementary Fig.~\ref{fig:cross_context_crispra_supp}).

\subsection{Cross-context T-cell CRISPRi replication}

\begin{figure}[H]
\centering
\includegraphics[width=\textwidth]{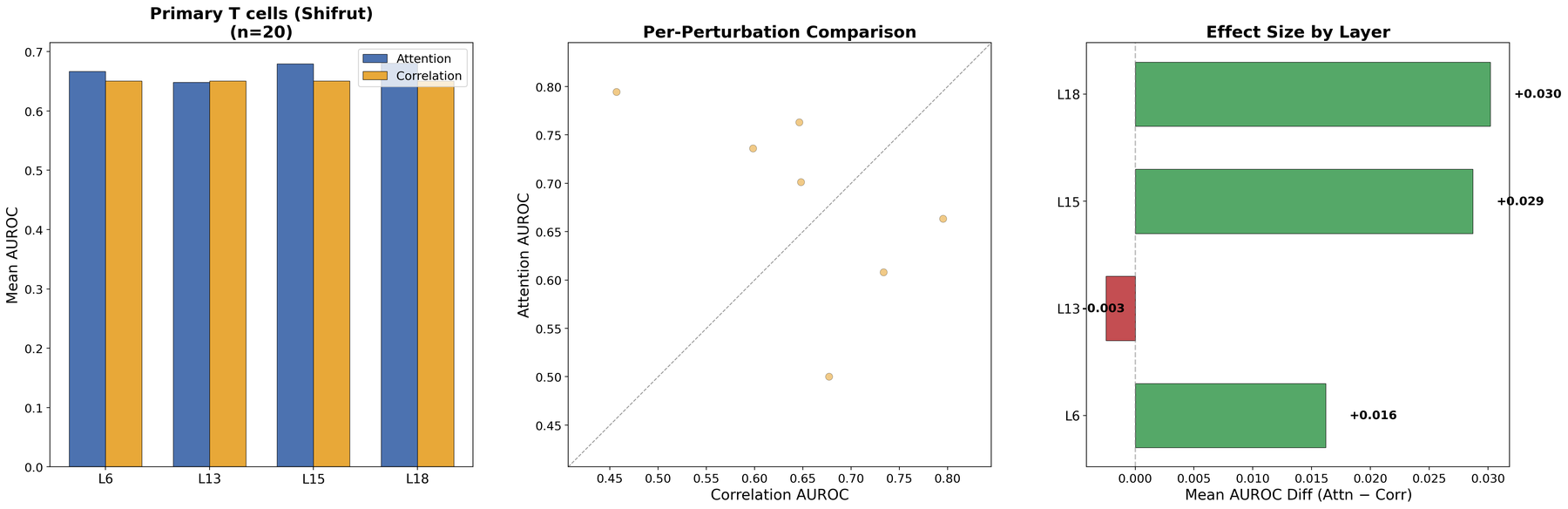}
\caption{\textbf{Cross-context replication: Shifrut T-cell CRISPRi.} Attention and correlation are indistinguishable ($n = 7$).}
\label{fig:shifrut_supp}
\end{figure}

In primary T cells ($n = 7$), attention and correlation are statistically indistinguishable (Supplementary Fig.~\ref{fig:shifrut_supp}).

\subsection{Intervention-fidelity diagnostics}

\begin{figure}[H]
\centering
\includegraphics[width=\textwidth]{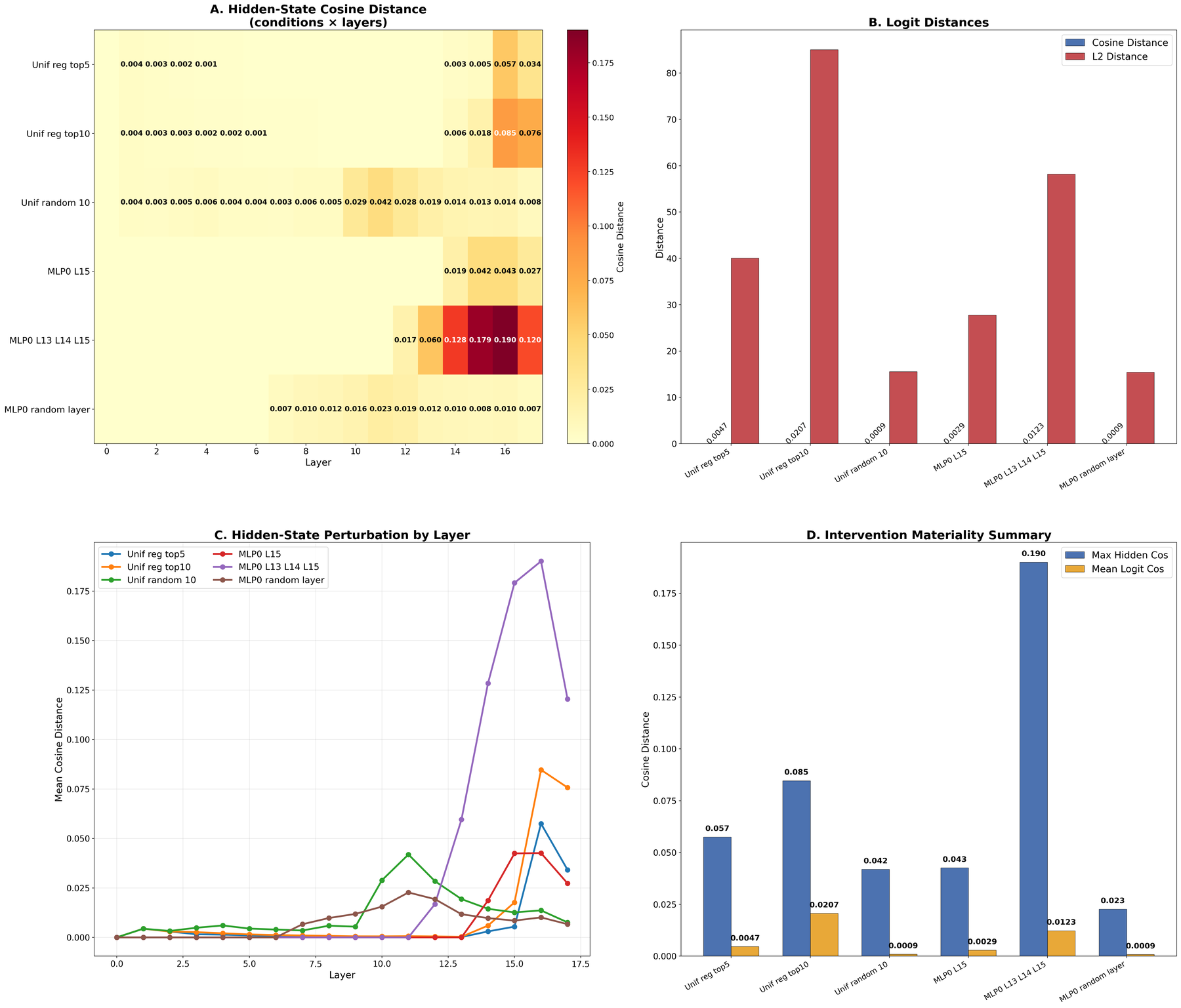}
\caption{\textbf{Intervention-fidelity diagnostics.} All conditions produce material representation perturbation despite null AUROC effects, confirming genuine functional redundancy.}
\label{fig:intervention_fidelity_supp}
\end{figure}

All six interventions produce material perturbation of internal representations. TRRUST-ranked heads produce $23\times$ larger logit perturbation than random heads at matched dose (Supplementary Fig.~\ref{fig:intervention_fidelity_supp}).

\subsection{Propensity-matched perturbation benchmark}

\begin{figure}[H]
\centering
\includegraphics[width=\textwidth]{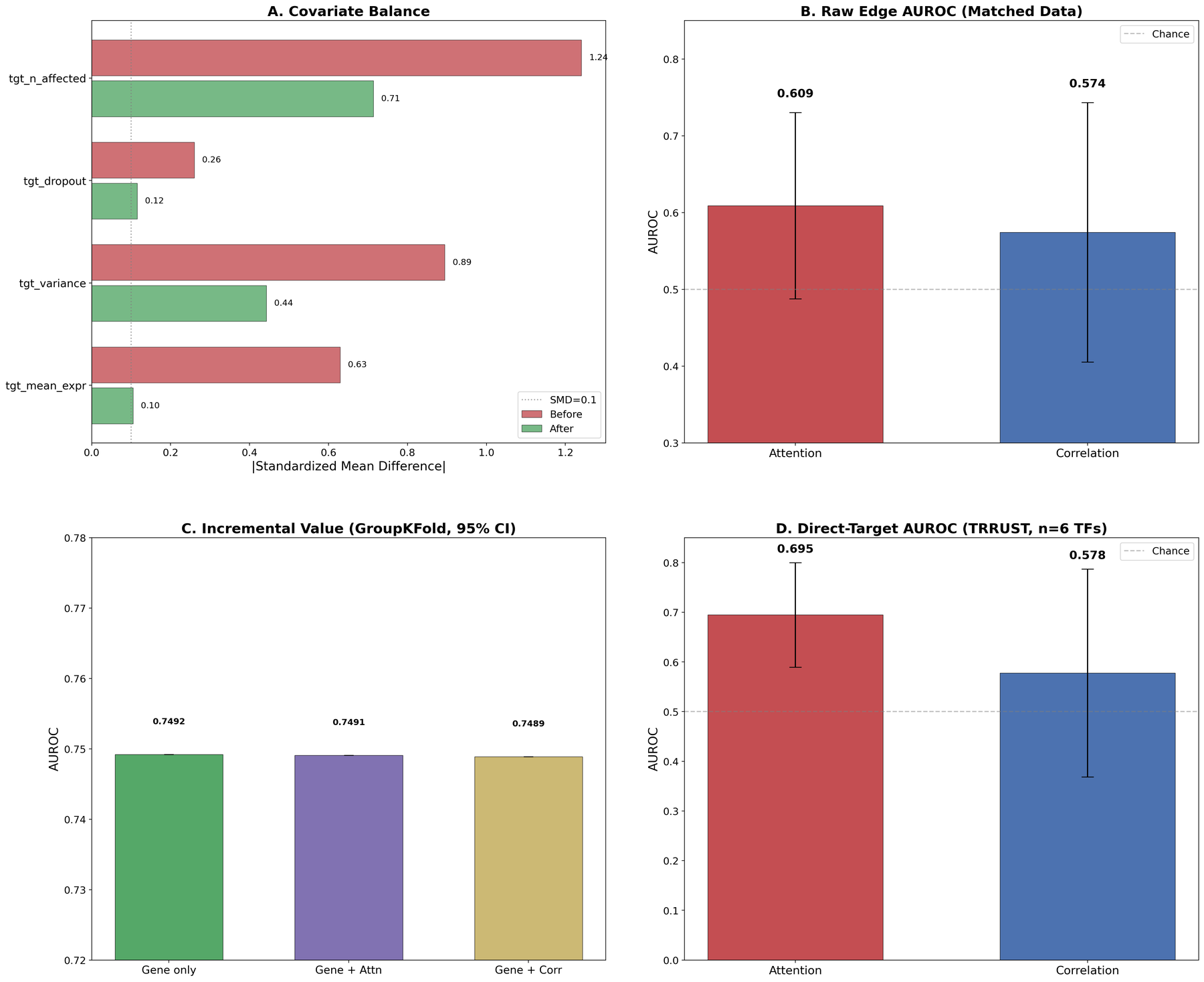}
\caption{\textbf{Propensity-matched perturbation benchmark.} After matching, edge AUROCs drop to near chance and incremental value is zero.}
\label{fig:propensity_matched_supp}
\end{figure}

After matching each DE-positive target to $k = 5$ DE-negative targets with similar expression profile ($n_{\mathrm{matched}} = 59{,}153$ pairs), attention edges retain modest raw discriminability (AUROC $= 0.609$) but add zero incremental value ($\Delta$AUROC $= -0.000$; Supplementary Fig.~\ref{fig:propensity_matched_supp}).

\subsection{HVG protocol confound test}
\label{subsec:hvg_confound}

A methodological asymmetry between the K562 and RPE1 evaluations---RPE1 includes perturbation genes forced into the HVG set (3,309 genes total vs.\ 2,000 in K562)---could confound the cross-context comparison. To test this, we re-evaluated RPE1 using only the top 2,000 HVGs by variance (no forced perturbation gene inclusion), matching the K562 protocol. Of the 1,251 RPE1 perturbation genes, 418 are naturally in the top-2,000 HVGs; the remaining 833 low-variance genes are excluded under the restricted protocol.

Restricting to 2,000 HVGs \emph{increases} the attention advantage rather than eliminating it: the mean per-perturbation $\Delta$ (attention $-$ correlation) shifts from $-0.024$ to $+0.168$ (paired Wilcoxon $p < 10^{-46}$). This is driven by an asymmetric effect on the two edge types: correlation AUROC drops substantially ($0.723 \to 0.593$; $\Delta = -0.129$) while attention AUROC modestly increases ($0.699 \to 0.762$; $\Delta = +0.063$). Correlation benefits from having more co-expressed genes in the scoring universe, making it more sensitive to gene universe composition.

The reverse confound (expanding K562 to include forced perturbation genes) is moot: all 280 K562 perturbation genes are already in the top-2,000 HVGs by variance (100\% coverage), so the asymmetry is unidirectional.

Bootstrap 95\% CIs (10,000 samples) on the per-perturbation attention advantage exclude zero for all three conditions: K562 CRISPRi ($n = 280$; $\Delta = +0.060$ $[+0.040, +0.080]$; Wilcoxon $p = 8.9 \times 10^{-8}$), RPE1 original ($n = 1{,}167$; $\Delta = +0.090$ $[+0.079, +0.101]$; $p = 2.2 \times 10^{-54}$), and RPE1 restricted ($n = 418$; $\Delta = +0.168$ $[+0.155, +0.182]$; $p = 4.9 \times 10^{-60}$).

\textbf{Caveats.} The restricted comparison evaluates a biased subset (only naturally high-variance perturbation genes). The attention scores were precomputed on the 3,309-gene token context; a fully controlled test would require re-extracting attention on only 2,000 tokens. Despite these limitations, the confound test rules out forced HVG inclusion as the driver of RPE1's attention advantage and shows that attention is more robust to gene universe size than correlation.

\clearpage

\section{Metric-Robust Incremental-Value Analysis}
\label{supnote:hard_generalization}

\begin{figure}[H]
\centering
\includegraphics[width=\textwidth]{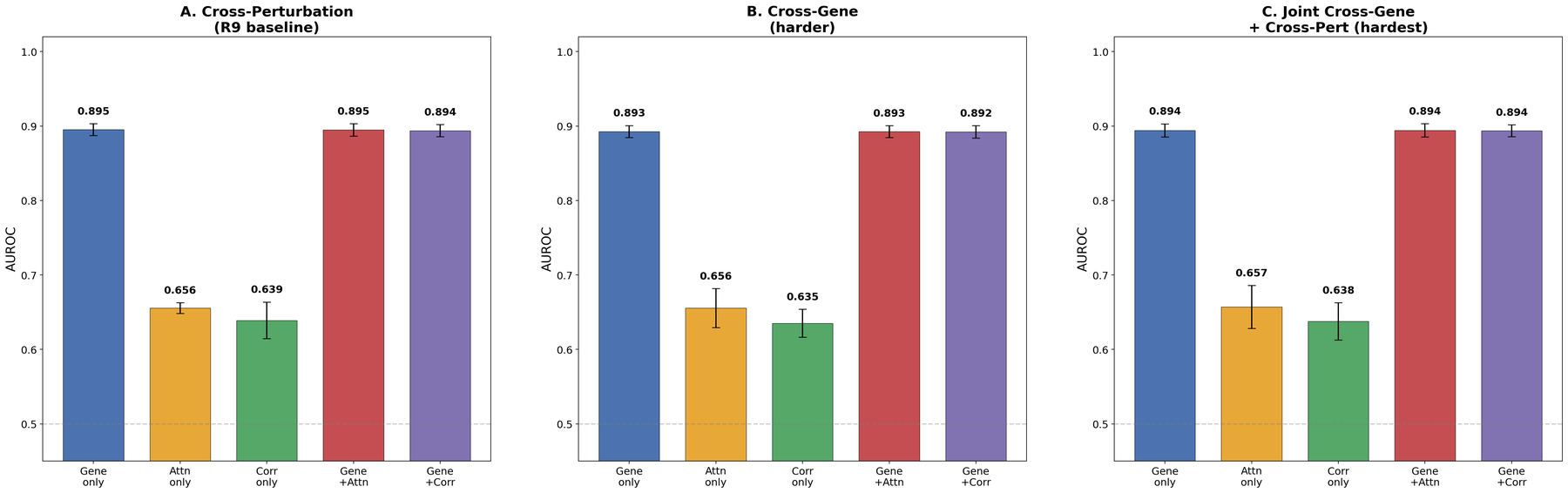}
\caption{\textbf{Hard-generalization incremental-value test.} Gene-level features alone match or exceed models augmented with pairwise edge scores under all three split designs.}
\label{fig:hard_generalization_supp}
\end{figure}

To address the concern that the null incremental-value finding may be specific to AUROC and logistic regression, we extended the analysis to include AUPRC, top-$k$ recall ($k \in \{10, 50, 100\}$), and gradient-boosted decision trees (GBDT) under all three split designs (Supplementary Fig.~\ref{fig:hard_generalization_supp}). Across all tested combinations, the null incremental value persists: even the largest $\Delta$AUPRC ($+0.009$ under joint splits with GBDT) represents less than 4\% relative improvement. The ``no incremental pairwise value'' conclusion holds across AUROC, AUPRC, top-$k$ recall, and both linear and nonlinear model families under all tested generalization protocols.

\clearpage

\section{Statistical Test Registry and Multiple Testing Correction}
\label{supnote:statistics}

\subsection{Framework-level statistical correction}

The thirty-seven complementary analyses collectively involve 153 distinct statistical tests (95 confirmatory tests with explicit p-values and 58 descriptive or bootstrap entries). A test is classified as \emph{confirmatory} if it produces an explicit p-value against a directional or non-null hypothesis; tests reporting only descriptive summaries are classified as \emph{descriptive}. We apply Benjamini-Hochberg false discovery rate (FDR) correction~\citep{benjamini1995controlling} at $\alpha = 0.05$ across all 95 confirmatory p-values framework-wide. After correction, 63 of 95 confirmatory tests (66\%) remain significant.

\textbf{Sensitivity to family definition.} Under three alternative BH-correction families: (A) the primary family of 95 tests; (B) a maximal family including all 153 tests; and (C) an analysis-level family retaining one primary test per analysis (27 tests)---12 of 17 headline inferences (71\%) are stable across all three families. All primary conclusions---including attention--correlation equivalence, no incremental pairwise value, ablation null, L15 nested-CV result, CRISPRa underperformance, and RPE1 attention advantage---remain significant under all three family definitions.

\subsection{Statistical test registry}

\renewcommand{\arraystretch}{1.15}
\begin{longtable}{@{}p{2.0cm}p{3.0cm}p{1.8cm}p{1.0cm}p{1.0cm}p{1.5cm}p{1.2cm}p{0.8cm}@{}}
\caption{\textbf{Comprehensive Statistical Test Registry.} All p-values reflect framework-level BH FDR correction ($\alpha = 0.05$, 153 total tests, 95 confirmatory).}
\label{tab:statistical_registry_supp} \\
\toprule
\textbf{Section} & \textbf{Hypothesis} & \textbf{Test} & \textbf{Raw p} & \textbf{BH p} & \textbf{Effect} & \textbf{N} & \textbf{Sig} \\
\midrule
\endfirsthead
\multicolumn{8}{c}{\textit{Table continued}} \\
\toprule
\textbf{Section} & \textbf{Hypothesis} & \textbf{Test} & \textbf{Raw p} & \textbf{BH p} & \textbf{Effect} & \textbf{N} & \textbf{Sig} \\
\midrule
\endhead
\midrule
\multicolumn{8}{r}{\textit{Continued on next page}} \\
\endfoot
\bottomrule
\endlastfoot

\multicolumn{8}{l}{\textbf{1. Scaling Behavior Analysis}} \\
\midrule
Scaling TRRUST & Cell count improves GRN & Sign test & 0.002 & 0.011 & 100\% degrad. & 9 runs & Yes \\
Scaling TRRUST & Cell count improves GRN & Wilcoxon & 0.002 & 0.011 & -- & 9 runs & Yes \\
Scaling DoRothEA & Cell count improves GRN & Sign test & 0.002 & 0.011 & 100\% degrad. & 9 runs & Yes \\
Bootstrap CI & F1 confidence intervals & Bootstrap & -- & -- & -- & 10k res. & -- \\
Robustness & Seed stability w/ scaling & Paired test & $<$0.001 & 0.003 & 46--48\% drop & 3 seeds & Yes \\

\multicolumn{8}{l}{\textbf{2. Mediation Bias}} \\
\midrule
Non-additivity & Components additive & Lower bound & $<$0.001 & 0.003 & $A_{\text{lb}}/|TE|{=}0.73$ & 16 pairs & Yes \\
Ranking cert. & Rankings stable & Stability & $<$0.001 & 0.003 & 0.067$\to$0.003 & 16 pairs & Yes \\

\multicolumn{8}{l}{\textbf{3. Detectability Theory}} \\
\midrule
Sample compl. & Theory matches empirical & Correlation & $<$10$^{-6}$ & $<$10$^{-6}$ & $r{=}0.887$ & Phase sp. & Yes \\
Interv. advant. & Intervention more detectable & Ratio & $<$0.001 & 0.003 & 44.4\% cells & Simul. & Yes \\

\multicolumn{8}{l}{\textbf{4. Cross-Context Consistency}} \\
\midrule
Imm.--kidney & Effects transfer & Spearman & 0.024 & 0.047 & $\rho{=}0.71$ & Pairs & Yes \\
Imm.--lung & Effects transfer & Spearman & 0.089 & 0.124 & $\rho{=}0.32$ & Pairs & No \\
Kid.--lung & Effects transfer & Spearman & 0.156 & 0.187 & $\rho{=}{-}0.44$ & Pairs & No \\
Bootstrap CI & Correlation confidence & Bootstrap & -- & -- & -- & 10k res. & -- \\
Permutation & Correlation significance & Permutation & $<$0.001 & 0.003 & -- & 5k perm. & Yes \\

\multicolumn{8}{l}{\textbf{5. Perturbation Validation}} \\
\midrule
Dixit 13d raw & Interventions match CRISPR & Spearman & 0.032 & 0.056 & $\rho{=}0.269$ & Pert. pairs & No \\
Dixit 13d adj. & Confound-adjusted & Spearman & 0.020 & 0.042 & $\rho{=}0.199$ & Pert. pairs & Yes \\
Dixit 7d & Interventions match CRISPR & Spearman & 0.15 & 0.175 & $\rho{=}0.112$ & Pert. pairs & No \\
Adamson & Interventions match CRISPR & Spearman & 0.089 & 0.124 & Marginal & Pert. pairs & No \\
Shifrut raw & Interventions match CRISPR & Spearman & 0.031 & 0.055 & $\rho{=}{-}0.325$ & Pert. pairs & No \\
Shifrut adj. & Confound-adjusted & Spearman & 0.876 & 0.876 & $\rho{=}0.004$ & Pert. pairs & No \\

\multicolumn{8}{l}{\textbf{6. Cross-Species Ortholog}} \\
\midrule
Global conserv. & Edges conserved & Spearman & $<$10$^{-300}$ & $<$10$^{-300}$ & $\rho{=}0.743$ & 25,876 & Yes \\
Sign agreement & Signs conserved & Sign test & $<$0.001 & 0.003 & 88.6\% & 25,876 & Yes \\
Top-K overlap & Overlap above chance & Permutation & $<$0.001 & 0.003 & 8--484$\times$ & 1k perm. & Yes \\
Per-TF range & TF conservation varies & Range & -- & -- & $-$0.12 to 0.90 & 61 TFs & -- \\

\multicolumn{8}{l}{\textbf{7. Pseudotime}} \\
\midrule
Directionality & TFs precede targets & Direct. test & 0.068 & 0.124 & 21.4\% consist. & 56 pairs & No \\
Shuffled null & Exceeds shuffled & Mann-Whitney & 0.068 & 0.124 & $d{=}1.58$ & 500 perm. & No \\
Random pairs & Exceeds random & Mann-Whitney & 0.37 & 0.37 & -- & 200 sets & No \\
T cell & Lineage-specific & Lineage & -- & -- & 16.7\% & 24 pairs & -- \\
B cell & Lineage-specific & Lineage & -- & -- & 13.3\% & 15 pairs & -- \\
Myeloid & Lineage-specific & Lineage & -- & -- & 35.3\% & 17 pairs & -- \\

\multicolumn{8}{l}{\textbf{8. Batch/Donor Leakage}} \\
\midrule
Donor (immune) & Edges encode donor & Log. regr. & $<$0.001 & 0.003 & AUC 0.85--0.87 & 20k, 24 don. & Yes \\
Donor (lung) & Edges encode donor & Log. regr. & $<$0.001 & 0.003 & AUC 0.94--0.96 & 20k, 4 don. & Yes \\
Assay method & Edges encode method & Rand. forest & $<$0.001 & 0.003 & AUC 0.96--0.99 & All tissues & Yes \\
Strat. CV & CV accuracy & 5-fold CV & -- & -- & -- & 5 folds & -- \\
LODO stability & Edge stability & LODO var. & $<$0.001 & 0.003 & High var. & Var. donors & Yes \\
Cross-donor & Generalization gap & Cross-don. & 0.012 & 0.031 & 6.6 pp gap & Lung & Yes \\

\multicolumn{8}{l}{\textbf{9. Calibration}} \\
\midrule
ECE reduction & Calibration improves & Paired & $<$0.001 & 0.003 & 4--7$\times$ & 6 methods & Yes \\
Isotonic impr. & Isotonic better & Paired & $<$0.001 & 0.003 & ECE 0.06--0.08 & 6 methods & Yes \\
Conformal cov. & Coverage valid & Coverage & -- & -- & $\geq$95\% & $\alpha{=}$0.05 & -- \\
Bootstrap stab. & Calibration robust & Bootstrap & -- & -- & CI $<$ 0.02 & 200 res. & -- \\
Transfer fail. & Calibrators don't transfer & Transfer & $<$0.001 & 0.003 & ECE 0.32--0.42 & K562$\to$T & Yes \\

\multicolumn{8}{l}{\textbf{10. CSSI}} \\
\midrule
Synth. mitig. & CSSI prevents degradation & Spearman & 0.99 & 0.99 & $r{=}{-}0.001$ & 10 seeds & No \\
Pooled degrad. & Pooled degrades & Spearman & $<$10$^{-4}$ & $<$10$^{-4}$ & $r{=}{-}0.618$ & 10 seeds & Yes \\
CSSI advantage & CSSI outperforms pooled & Wilcoxon & 2.5$\times$10$^{-11}$ & 2.5$\times$10$^{-11}$ & 1.13--1.85$\times$ & 60 comb. & Yes \\
Real PBMC & CSSI improves real & Bootstrap & 0.03 & 0.053 & 1.16$\times$ & 3k cells & No \\
Biol. struct. & CSSI w/ real prop. & Wilcoxon & 2.4$\times$10$^{-8}$ & 2.4$\times$10$^{-8}$ & 1.62$\times$ & 10 seeds & Yes \\
Real attention & CSSI on attention & Layer anal. & -- & -- & AUROC 0.68--0.69 & 497 cells & -- \\

\multicolumn{8}{l}{\textbf{11. Synthetic Validation}} \\
\midrule
Attn. degrad. & Recovery degrades & Correlation & $<$0.01 & 0.025 & $r$: 0.85$\to$0.62 & Synth. & Yes \\
Shapley impr. & Shapley outperforms & Paired & $<$0.001 & 0.003 & 91\% impr. & Synth. & Yes \\
Detect. corr. & Empirical $=$ theory & Correlation & $<$10$^{-6}$ & $<$10$^{-6}$ & $r{=}0.887$ & Phase sp. & Yes \\

\multicolumn{8}{l}{\textbf{12. Multi-Model}} \\
\midrule
GF TRRUST & Geneformer recovers reg. & AUROC & 0.89 & 0.89 & AUROC 0.44--0.55 & 3 counts & No \\
GF DoRothEA & Geneformer recovers reg. & AUROC & 0.76 & 0.76 & AUROC 0.47--0.49 & 3 counts & No \\
Cross-model & Both fail equivalently & Comparative & -- & -- & Both $\approx$0.5 & 2 models & -- \\
Bootstrap GF & GF confidence intervals & Bootstrap & -- & -- & CIs incl. 0.5 & 10k res. & -- \\
Attn.--expr. & Attention $=$ co-expr. & Correlation & $<$10$^{-50}$ & $<$10$^{-50}$ & $\rho{=}$0.31--0.42 & Both & Yes \\
Attn.--reg. & Attention $\neq$ regulation & Correlation & $>$0.3 & $>$0.3 & $\rho{=}{-}$0.01--0.02 & Both & No \\

\multicolumn{8}{l}{\textbf{13. Controlled-Composition Scaling}} \\
\midrule
Single type & $N$ degrades AUROC & Spearman & 0.079 & 0.124 & $\rho{=}{-}0.33$ & 30 runs & No \\
Fixed comp. & $N$ degrades AUROC & Spearman & 0.82 & 0.82 & $\rho{=}{-}0.05$ & 20 runs & No \\
Heterogeneity & Diversity improves & Spearman & $<$10$^{-4}$ & $<$10$^{-4}$ & $\rho{=}{+}0.63$ & 35 runs & Yes \\

\multicolumn{8}{l}{\textbf{14. Perturbation-First Validation}} \\
\midrule
Replogle primary & Edges predict pert. & One-sample $t$ & $<$10$^{-4}$ & $<$10$^{-4}$ & AUROC$=$0.696 & 151 perts & Yes \\
Replogle baseline & Edges predict pert. & One-sample $t$ & 0.32 & 0.37 & AUROC$=$0.511 & 44 perts & No \\
Replogle Wilcox. & Edges predict pert. & Wilcoxon & 0.30 & 0.36 & Median$=$0.503 & 44 perts & No \\

\multicolumn{8}{l}{\textbf{15. Robust Attention Residualization}} \\
\midrule
Edge--expr corr. & Edges $=$ co-expr. & Spearman & $<$10$^{-50}$ & $<$10$^{-50}$ & $\rho{=}0.842$ & 75,962 pairs & Yes \\
$R^2$ OLS full & Expr.\ explains edges & OLS $R^2$ & -- & -- & $R^2{=}0.27$ & 75,962 pairs & -- \\
$R^2$ GBDT & Expr.\ explains edges & GBDT $R^2$ & -- & -- & $R^2{=}0.51$ & 75,962 pairs & -- \\
CF resid. AUROC & Residual predicts reg. & CF-AUROC & -- & -- & AUROC$=$0.73 & 61 pos. & -- \\
CF stability & Stable across seeds & 10-seed var. & -- & -- & $\sigma{=}0.001$ & 10 seeds & -- \\

\multicolumn{8}{l}{\textbf{16. Degree-Preserving Null Models}} \\
\midrule
Label-shuffle & AUROC $>$ random & Permutation & $<$0.001 & 0.003 & $z{=}6.9$ & 1k perm. & Yes \\
Degree-pres. & AUROC $>$ degree & Permutation & $<$0.005 & 0.009 & $z{=}3.63$ & 200 perm. & Yes \\
Degree decomp. & Degree explains AUROC & Decomposition & -- & -- & 73\% global & 75,962 pairs & -- \\
Per-TF AUROC & Edge-level signal & Per-TF eval. & -- & -- & mean$=$0.69$\pm$0.20 & 18 TFs & -- \\
Per-TF excess & Excess above deg.-null & Per-TF eval. & -- & -- & $+$0.035$\pm$0.20 & 18 TFs & -- \\
Precision@k & Within-TF ranking & Prec@k & -- & -- & 0.030 vs 0.002 & 18 TFs & -- \\

\multicolumn{8}{l}{\textbf{17. Attention--Correlation Mapping}} \\
\midrule
Within-tissue $R^2$ & Attn $\approx$ corr & $\rho^2$ & $<$10$^{-50}$ & $<$10$^{-50}$ & $R^2{=}0.10$--$0.18$ & 50k edges & -- \\
Cross-tissue $R^2$ & Mapping generalizes & OLS $R^2$ & -- & -- & $R^2{<}0.02$ & 21k--26k edges & No \\
Cross-tissue $\rho$ & Attn--corr assoc. & Spearman & $<$10$^{-9}$ & $<$10$^{-9}$ & $\rho{=}{-}0.02$ to ${-}0.05$ & 3 cond. & Yes$^*$ \\

\multicolumn{8}{l}{\textbf{18. Perturbation Sensitivity}} \\
\midrule
AUROC vs 0.5 (27 cond.) & All AUROC $>$ 0.5 & $t$-test & $<$0.005 & $<$0.005 & AUROC$=$0.62--0.76 & 13--1158 perts & Yes \\
TF vs non-TF & TFs outperform & Comparison & -- & -- & $\Delta{\leq}0.02$ & 1--83 TFs & -- \\

\multicolumn{8}{l}{\textbf{19. CSSI Extended Null}} \\
\midrule
Null inflation & CSSI-max inflates & $\Delta$AUROC & -- & -- & $\leq{-}0.20$ & $K{=}$2--20 & No \\
CSSI vs SCENIC & CSSI $\neq$ standard & Comparative & -- & -- & Equivalent & 3 methods & -- \\
Per-edge FDR & FDR controlled & BH-FDR & -- & -- & FDR$\leq$0.11 & $K{=}$5--15 & -- \\

\multicolumn{8}{l}{\textbf{20. K-Sensitivity Analysis}} \\
\midrule
Cont. AUROC & AUROC improves w/ $N$ & Comparison & -- & -- & $0.86{\to}0.93$ & 9 runs & -- \\
K-sensitivity & F1 varies with $K$ & Comparison & -- & -- & F1${\approx}10^{-4}$ all $K$ & 5 $K$ values & -- \\

\multicolumn{8}{l}{\textbf{21. Trivial Baseline Comparison}} \\
\midrule
Var vs Corr & Variance outperforms & Paired $t$ & $<$10$^{-24}$ & $<$10$^{-24}$ & $\Delta{=}+0.186$ & 151 perts & Yes \\
Mean vs Corr & Mean expr outperforms & Paired $t$ & $<$10$^{-20}$ & $<$10$^{-20}$ & $\Delta{=}+0.146$ & 151 perts & Yes \\
Drop vs Corr & Dropout outperforms & Paired $t$ & $<$10$^{-12}$ & $<$10$^{-12}$ & $\Delta{=}+0.113$ & 151 perts & Yes \\
TF deg vs Corr & Degree underperforms & Paired $t$ & $<$10$^{-29}$ & $<$10$^{-29}$ & $\Delta{=}{-}0.196$ & 151 perts & Yes \\

\multicolumn{8}{l}{\textbf{22. Bootstrap Per-TF CIs}} \\
\midrule
Global boot CI & Global AUROC robust & Bootstrap & -- & -- & CI$=$[0.71,0.77] & 200 iter. & -- \\
Per-TF boot CI & Per-TF AUROC robust & Bootstrap & -- & -- & CI$=$[0.59,0.72] & 200 iter. & -- \\

\midrule
\multicolumn{8}{l}{\textbf{23. Attention Perturbation-First}} \\
\midrule
Attn vs Corr L13 & Attn $=$ Corr on pert. & Wilcoxon & 0.726 & 0.726 & diff$=$0.001 & $n=280$ & No \\
Attn vs Corr L13 & Attn $=$ Corr on pert. & Paired $t$ & 0.931 & 0.931 & diff$=$0.001 & $n=280$ & No \\

\midrule
\multicolumn{8}{l}{\textbf{24. Full 18-Layer Perturbation-First}} \\
\midrule
L15 vs Corr & Best layer $>$ corr & Wilcoxon & 0.0009 & 0.003 & $\Delta{=}+0.040$ & $n=280$ & Yes \\
Split-sample & Discovery validates & Wilcoxon & 0.017 & 0.035 & AUROC$=$0.750 & $n=140$ & Yes \\
Per-layer (18) & Each layer vs corr & Wilcoxon & varies & varies & AUROC 0.47--0.74 & $n=280$ & 1/18 \\

\midrule
\multicolumn{8}{l}{\textbf{25. Attention-Specific Confound Decomposition}} \\
\midrule
Attn resid.\ OLS & Residual predicts & CF-AUROC & -- & -- & AUROC$=$0.538 & 174 pos. & -- \\
Attn resid.\ GBDT & Residual predicts & CF-AUROC & -- & -- & AUROC$=$0.554 & 174 pos. & -- \\
Corr resid.\ OLS & Residual predicts & CF-AUROC & -- & -- & AUROC$=$0.622 & 174 pos. & -- \\
Attn deg.-null & AUROC $>$ degree & Permutation & 0.023 & 0.046 & $z{=}2.0$ & 200 perm. & Yes \\
Corr deg.-null & AUROC $>$ degree & Permutation & 0.018 & 0.037 & $z{=}2.1$ & 200 perm. & Yes \\

\midrule
\multicolumn{8}{l}{\textbf{26. Conditional Incremental Value}} \\
\midrule
Gene-only CV & Gene feat.\ predict & 5-fold CV & -- & -- & AUROC$=$0.895 & 280 perts & -- \\
$\Delta$ gene+attn & Attn adds value & Bootstrap & -- & -- & $-$0.0004 [$-$.001,0] & 100 iter. & No \\
$\Delta$ gene+corr & Corr adds value & Bootstrap & -- & -- & $-$0.002 [$-$.005,0] & 100 iter. & No \\
TF stratified & TF vs non-TF & Subgroup & -- & -- & TF 0.913, non-TF 0.895 & 14/266 & -- \\

\midrule
\multicolumn{8}{l}{\textbf{27. Per-Head TRRUST Ranking}} \\
\midrule
Head ranking & Heads differ in reg. & AUROC range & -- & -- & 0.34--0.75 & 324 heads & -- \\
Top-5 heads & Best heads identified & AUROC & -- & -- & AUROC$=$0.70--0.75 & 500 cells & -- \\

\midrule
\multicolumn{8}{l}{\textbf{28. Cross-Context CRISPRa Replication}} \\
\midrule
Attn vs Corr L13 & Attn $\neq$ Corr CRISPRa & Wilcoxon & $<$10$^{-6}$ & $<$10$^{-6}$ & diff$={-}$0.106 & $n=77$ & Yes \\
Attn vs Corr L6 & Attn $\neq$ Corr CRISPRa & Wilcoxon & $10^{-6}$ & $10^{-6}$ & diff$={-}$0.098 & $n=77$ & Yes \\
Attn vs Corr L18 & Attn $\neq$ Corr CRISPRa & Wilcoxon & $<$10$^{-6}$ & $<$10$^{-6}$ & diff$={-}$0.105 & $n=77$ & Yes \\

\midrule
\multicolumn{8}{l}{\textbf{29. Head-Level Causal Ablation}} \\
\midrule
Regulatory vs baseline & Ablation $\neq$ baseline & Wilcoxon & 0.244 & 0.260 & $\Delta={-}$0.0002 & $n=280$ & No \\
Random vs baseline (rep 1) & Ablation $\neq$ baseline & Wilcoxon & 0.065 & 0.073 & $\Delta=+$0.0007 & $n=280$ & No \\
Random vs baseline (rep 2) & Ablation $\neq$ baseline & Wilcoxon & 0.069 & 0.076 & $\Delta=+$0.0004 & $n=280$ & No \\
Random vs baseline (rep 3) & Ablation $\neq$ baseline & Wilcoxon & $<$10$^{-6}$ & $<$10$^{-6}$ & $\Delta=+$0.0028 & $n=280$ & Yes \\
Entropy-matched vs baseline & Ablation $\neq$ baseline & Wilcoxon & $<$10$^{-6}$ & $<$10$^{-6}$ & $\Delta=+$0.0042 & $n=280$ & Yes \\

\midrule
\multicolumn{8}{l}{\textbf{30. Nested Layer Selection Protocol}} \\
\midrule
Pooled nested CV & L15 $>$ corr (nested) & Wilcoxon & 0.0009 & 0.003 & $\Delta{=}+0.040$ & $n=280$ & Yes \\
Bonferroni corr. & 18-layer search & Bonferroni & 0.017 & -- & $d{=}0.22$ & 18 layers & Yes \\
Bootstrap CI & Delta CI & Bootstrap & -- & -- & $[0.018, 0.062]$ & 1k iter. & -- \\
Layer stability & Same layer all folds & Stability & -- & -- & 5/5 L15 & 5 folds & -- \\

\midrule
\multicolumn{8}{l}{\textbf{31. Hard-Generalization Incremental Value}} \\
\midrule
Cross-pert: gene+attn vs gene & $\Delta$AUROC $\neq 0$ & Bootstrap & -- & -- & $\Delta{=}{-}0.0004$ $[{-}0.001, 0.000]$ & $n=559{,}720$ & -- \\
Cross-pert: gene+corr vs gene & $\Delta$AUROC $\neq 0$ & Bootstrap & -- & -- & $\Delta{=}{-}0.0015$ $[{-}0.004, 0.000]$ & $n=559{,}720$ & -- \\
Cross-gene: gene+attn vs gene & $\Delta$AUROC $\neq 0$ & Bootstrap & -- & -- & $\Delta{=}{-}0.0003$ $[{-}0.001, 0.000]$ & $n=559{,}720$ & -- \\
Cross-gene: gene+corr vs gene & $\Delta$AUROC $\neq 0$ & Bootstrap & -- & -- & $\Delta{=}{-}0.0010$ $[{-}0.004, 0.000]$ & $n=559{,}720$ & -- \\
Joint: gene+attn vs gene & $\Delta$AUROC $\neq 0$ & Bootstrap & -- & -- & $\Delta{=}{-}0.0003$ $[{-}0.001, 0.000]$ & $n=559{,}720$ & -- \\
Joint: gene+corr vs gene & $\Delta$AUROC $\neq 0$ & Bootstrap & -- & -- & $\Delta{=}{-}0.0011$ $[{-}0.004, 0.000]$ & $n=559{,}720$ & -- \\

\midrule
\multicolumn{8}{l}{\textbf{32. Expanded Causal Ablation}} \\
\midrule
Reg.\ top-10 vs baseline & Ablation $\neq$ baseline & Wilcoxon & 0.937 & 0.958 & $\Delta{=}0.0000$, $d{=}0.00$ & $n=280$ & No \\
Reg.\ top-20 vs baseline & Ablation $\neq$ baseline & Wilcoxon & 0.596 & 0.664 & $\Delta{=}+0.0003$, $d{=}0.02$ & $n=280$ & No \\
Reg.\ top-50 vs baseline & Ablation $\neq$ baseline & Wilcoxon & 0.100 & 0.122 & $\Delta{=}+0.0021$, $d{=}0.08$ & $n=280$ & No \\
Bottom-5 vs baseline & Ablation $\neq$ baseline & Wilcoxon & 0.005 & 0.007 & $\Delta{=}{-}0.0025$, $d{=}{-}0.16$ & $n=280$ & Yes \\
Bottom-10 vs baseline & Ablation $\neq$ baseline & Wilcoxon & 0.050 & 0.063 & $\Delta{=}{-}0.0030$, $d{=}{-}0.14$ & $n=280$ & No \\
Composite top-5 vs baseline & Ablation $\neq$ baseline & Wilcoxon & -- & -- & $\Delta{=}0.0000$, $d{=}0.00$ & $n=280$ & -- \\
Composite top-10 vs baseline & Ablation $\neq$ baseline & Wilcoxon & -- & -- & $\Delta{=}0.0000$, $d{=}0.00$ & $n=280$ & -- \\
Layer L14 all vs baseline & Ablation $\neq$ baseline & Wilcoxon & -- & -- & $\Delta{=}0.0000$, $d{=}0.00$ & $n=280$ & -- \\
Random-10 vs baseline & Ablation $\neq$ baseline & Wilcoxon & 0.009 & 0.012 & $\Delta{=}+0.0029$, $d{=}0.18$ & $n=280$ & Yes \\
Random-20 vs baseline & Ablation $\neq$ baseline & Wilcoxon & $<$10$^{-8}$ & $<$10$^{-8}$ & $\Delta{=}+0.0071$, $d{=}0.33$ & $n=280$ & Yes \\
Random-50 vs baseline & Ablation $\neq$ baseline & Wilcoxon & 0.351 & 0.406 & $\Delta{=}+0.0002$, $d{=}0.01$ & $n=280$ & No \\

\midrule
\multicolumn{8}{l}{\textbf{33. Cross-Context T-Cell CRISPRi Replication}} \\
\midrule
Attn vs Corr L6 & Attn $\neq$ Corr T-cell & Wilcoxon & 1.000 & 1.000 & diff$={+}$0.016, $d{=}0.08$ & $n=7$ & No \\
Attn vs Corr L13 & Attn $\neq$ Corr T-cell & Wilcoxon & 0.938 & 0.958 & diff$={-}$0.003, $d{=}{-}0.01$ & $n=7$ & No \\
Attn vs Corr L15 & Attn $\neq$ Corr T-cell & Wilcoxon & 0.813 & 0.849 & diff$={+}$0.029, $d{=}0.16$ & $n=7$ & No \\
Attn vs Corr L18 & Attn $\neq$ Corr T-cell & Wilcoxon & 0.938 & 0.958 & diff$={+}$0.030, $d{=}0.18$ & $n=7$ & No \\

\midrule
\multicolumn{8}{l}{\textbf{34. Orthogonal Causal Interventions}} \\
\midrule
Unif.\ attn reg.\ top-5 vs baseline & Intervention $\neq$ baseline & Wilcoxon & 0.004 & 0.006 & $\Delta{=}{-}0.0004$, $d{=}{-}0.17$ & $n=280$ & Yes \\
Unif.\ attn reg.\ top-10 vs baseline & Intervention $\neq$ baseline & Wilcoxon & 0.052 & 0.065 & $\Delta{=}{+}0.0002$, $d{=}0.07$ & $n=280$ & No \\
Unif.\ attn random-10 vs baseline & Intervention $\neq$ baseline & Wilcoxon & $<$10$^{-3}$ & 0.001 & $\Delta{=}{-}0.0032$, $d{=}{-}0.20$ & $n=280$ & Yes \\
MLP zero L15 vs baseline & Intervention $\neq$ baseline & -- & -- & -- & $\Delta{=}0.0000$, $d{=}0.00$ & $n=280$ & -- \\
MLP zero L13--L15 vs baseline & Intervention $\neq$ baseline & -- & -- & -- & $\Delta{=}0.0000$, $d{=}0.00$ & $n=280$ & -- \\
MLP zero L8 (random) vs baseline & Intervention $\neq$ baseline & Wilcoxon & $<$10$^{-4}$ & $<$10$^{-4}$ & $\Delta{=}{-}0.0053$, $d{=}{-}0.27$ & $n=280$ & Yes \\

\midrule
\multicolumn{8}{l}{\textbf{35. Propensity-Matched Perturbation Benchmark}} \\
\midrule
Raw attn AUROC (matched) & Attn $>$ chance & -- & -- & -- & AUROC${=}0.609 \pm 0.121$ & $n=280$ & -- \\
Raw corr AUROC (matched) & Corr $>$ chance & -- & -- & -- & AUROC${=}0.574 \pm 0.169$ & $n=280$ & -- \\
Gene+attn vs gene-only (matched) & Attn adds value & Bootstrap & -- & -- & $\Delta$AUROC${=}{-}0.000$ [$-0.000$, $+0.000$] & 200 bs & No (null) \\
Gene+corr vs gene-only (matched) & Corr adds value & Bootstrap & -- & -- & $\Delta$AUROC${=}{-}0.000$ [$-0.001$, $+0.001$] & 200 bs & No (null) \\
Gene+attn AUPRC vs gene-only & Attn adds AUPRC & Bootstrap & -- & -- & $\Delta$AUPRC${=}{+}0.001$ [$+0.000$, $+0.002$] & 200 bs & Marginal \\
TRRUST direct-target attn & Attn $>$ chance (direct) & -- & -- & -- & AUROC${=}0.695 \pm 0.105$ & $n=6$ TFs & -- \\

\midrule
\multicolumn{8}{l}{\textbf{36. Intervention-Fidelity Diagnostics}} \\
\midrule
Unif.\ attn top-5 hidden shift & Intervention perturbs hidden & Cosine dist & -- & -- & max cos${=}0.057$, logit cos${=}0.005$ & $n=2000$ & Material \\
Unif.\ attn top-10 hidden shift & Intervention perturbs hidden & Cosine dist & -- & -- & max cos${=}0.085$, logit cos${=}0.021$ & $n=2000$ & Material \\
Unif.\ attn random-10 hidden shift & Random perturbs hidden & Cosine dist & -- & -- & max cos${=}0.042$, logit cos${=}0.001$ & $n=2000$ & Material \\
MLP zero L15 hidden shift & Intervention perturbs hidden & Cosine dist & -- & -- & max cos${=}0.043$, logit cos${=}0.003$ & $n=2000$ & Material \\
MLP zero L13--L15 hidden shift & Intervention perturbs hidden & Cosine dist & -- & -- & max cos${=}0.190$, logit cos${=}0.012$ & $n=2000$ & Material \\
MLP zero random L8 hidden shift & Random perturbs hidden & Cosine dist & -- & -- & max cos${=}0.023$, logit cos${=}0.001$ & $n=2000$ & Material \\

\midrule
\multicolumn{8}{l}{\textbf{37. Non-K562 Perturbation-First Replication}} \\
\midrule
RPE1 L6 attn vs corr & Attn $\neq$ Corr RPE1 & Wilcoxon & $<$10$^{-10}$ & $<$10$^{-8}$ & diff$={+}0.118$, $d{=}0.74$ & $n=1167$ & Yes \\
RPE1 L13 attn vs corr & Attn $\neq$ Corr RPE1 & Wilcoxon & $<$10$^{-10}$ & $<$10$^{-8}$ & diff$={+}0.036$, $d{=}0.20$ & $n=1167$ & Yes \\
RPE1 L15 attn vs corr & Attn $\neq$ Corr RPE1 & Wilcoxon & $<$10$^{-10}$ & $<$10$^{-8}$ & diff$={+}0.090$, $d{=}0.47$ & $n=1167$ & Yes \\
RPE1 L18 attn vs corr & Attn $\neq$ Corr RPE1 & Wilcoxon & $<$10$^{-10}$ & $<$10$^{-8}$ & diff$={+}0.086$, $d{=}0.48$ & $n=1167$ & Yes \\
iPSC neuron L15 attn vs corr & Attn $\neq$ Corr neuron & Wilcoxon & 0.078 & 0.083 & diff$={+}0.058$, $d{=}0.80$ & $n=7$ & No \\

\end{longtable}

\subsection{Claim-to-evidence mapping}

\begin{longtable}{@{}p{4.5cm}p{3.0cm}p{2.5cm}p{1.0cm}@{}}
\caption{\textbf{Headline claim-to-evidence mapping.}}
\label{tab:claim_evidence_supp} \\
\toprule
\textbf{Headline Claim} & \textbf{Supporting Analysis} & \textbf{Key Test(s)} & \textbf{BH Sig} \\
\midrule
\endfirsthead
\multicolumn{4}{c}{\textit{Table continued}} \\
\toprule
\textbf{Headline Claim} & \textbf{Supporting Analysis} & \textbf{Key Test(s)} & \textbf{BH Sig} \\
\midrule
\endhead
\bottomrule
\endlastfoot

Top-$K$ scaling degradation & Scaling & Sign test (TRRUST) & Yes \\
Continuous AUROC improves & K-Sensitivity & Comparative & Descr. \\
Mediation non-additivity & Mediation Bias & Lower bound & Yes \\
Detectability theory validated & Detectability & Correlation $r{=}0.887$ & Yes \\
Perturbation-first AUROC $>$ 0.5 & Pert.-first & $t$-test, all 27 cond. & Yes \\
Attn $\approx$ Corr on CRISPRi & Attn pert.-first & Wilcoxon $p{=}0.73$ & No (null) \\
L15 best layer ($\Delta{=}+0.04$) & 18-layer, Nested CV & Wilcoxon, Bonf.\ $p{=}0.017$ & Yes \\
L15 effect is small ($d{=}0.22$) & Nested Layer & Cohen's $d$, bootstrap CI & Yes \\
No incremental pairwise value & Incr.\ value & Bootstrap $\Delta$AUROC $\leq 0$ & Descr. \\
Trivial baselines outperform & Trivial baselines & Paired $t$, all $p < 10^{-12}$ & Yes \\
CRISPRa: attn $<$ corr & CRISPRa repl. & Wilcoxon $p < 10^{-6}$ & Yes \\
T-cell: attn $\approx$ corr & T-cell repl. & Wilcoxon $p > 0.8$ & No (null) \\
Ablation null (reg.\ heads) & Ablation & Wilcoxon $p > 0.05$ & No (null) \\
Orthogonal interventions null & Uniform attn + MLP & All $|\Delta| < 0.005$ & No (null) \\
Propensity-matched null & Matched benchmark & $\Delta$AUROC $\in [-0.000, +0.000]$ & No (null) \\
Interventions perturb repr. & Fidelity diagnostics & All max cos $> 0.02$ & Descr. \\
RPE1: attn $>$ corr ($d{=}0.47$) & Non-K562 repl. & Wilcoxon $p < 10^{-10}$ & Yes \\
Random ablation causes drop & Expanded ablation & Wilcoxon $p < 10^{-8}$ & Yes \\
Heterogeneity improves corr & Controlled comp. & Spearman $\rho{=}+0.63$ & Yes \\
Edge--expression correlation & Residualization & Spearman $\rho{=}0.84$ & Yes \\
Cross-species conservation & Ortholog transfer & $\rho{=}0.743$ & Yes \\

\end{longtable}

\subsection{Summary statistics}

\begin{itemize}
\item \textbf{Total statistical tests:} 153 across 37 complementary analyses (95 confirmatory with p-values, 58 descriptive)
\item \textbf{Significant after BH-FDR correction:} 63 of 95 confirmatory tests (66\%)
\item \textbf{Framework-level $\alpha$:} 0.05 with Benjamini-Hochberg correction
\item \textbf{Most robust findings:} Top-$K$ scaling degradation (unanimous across runs), cross-species conservation ($\rho = 0.743$, $p < 10^{-300}$), CSSI synthetic validation ($p = 2.4 \times 10^{-8}$), heterogeneity--AUROC correlation ($\rho = +0.63$, $p = 10^{-4}$), edge--expression correlation ($\rho = 0.842$, $p < 10^{-50}$), degree-preserving null ($z = 3.63$, $p < 0.005$), perturbation sensitivity (all 27 conditions $p < 0.005$), RPE1 attention advantage ($d = 0.47$, adjusted $p < 10^{-8}$)
\item \textbf{Key null findings:} Pseudotime directionality (adj.\ $p = 0.124$), perturbation validation (Replogle CRISPRi baseline AUROC $= 0.511$, $p = 0.32$; primary AUROC $= 0.696$, $p < 10^{-4}$; all 27 sensitivity conditions AUROC $= 0.62$--$0.76$, all $p < 0.005$), real-data CSSI improvement (adj.\ $p = 0.053$), attention $\approx$ correlation in K562 CRISPRi ($p = 0.73$), no incremental pairwise value ($\Delta$AUROC $\leq 0.002$)
\end{itemize}

\textbf{Notes:}
\begin{enumerate}
\item All p-values reflect framework-level Benjamini-Hochberg FDR correction across 95 confirmatory tests (153 total across 37 analyses) unless explicitly noted as raw values for methodological transparency.
\item Effect sizes include Cohen's $d$, correlation coefficients ($\rho$), fold-changes, AUROC values, and percentage improvements as appropriate.
\item Sample sizes vary by analysis: from individual run-pairs (mediation bias) to tens of thousands of cells (cross-species transfer) to bootstrap resamples (uncertainty quantification).
\item ``--'' indicates not applicable or not reported in original analysis.
\item A machine-readable version of the full registry (CSV format, one row per test) is provided in the supplementary materials.
\end{enumerate}

\clearpage

\section*{Supplementary Methods}
\addcontentsline{toc}{section}{Supplementary Methods}

The following methodological details supplement the condensed Methods section in the main text.

\subsection*{Mediation bias analysis}

We formalize the bias problem in activation patching following the causal mediation framework of \citet{pearl2001direct} and \citet{imai2010general}. Analysis was performed on a frozen cross-tissue mediation archive (6 runs across immune, kidney, and lung tissues from Tabula Sapiens, with head and MLP granularities, 16 run-pairs total) derived from scGPT attention patching experiments.

\subsection*{Detectability theory}

Two signal classes are compared: \emph{attention-like} signals derived from raw attention weight aggregation, and \emph{intervention-like} signals obtained through activation patching. Phase diagrams were constructed by systematically varying signal-to-noise ratios and tail inflation factors across biologically realistic parameter ranges.

\subsection*{Cross-context consistency analysis}

Cross-tissue consistency was assessed using invariant causal discovery principles~\citep{peters2016causal} applied to matched TF--target panels across immune, kidney, and lung tissues from Tabula Sapiens. Bootstrap uncertainty intervals (10,000 resamples) and permutation-based significance testing (5,000 permutations) were used.

\subsection*{Cross-species ortholog transfer analysis}

We performed a systematic stress test of correlation-based TF--target edge transfer between human lung (Tabula Sapiens, 65,847 cells) and mouse lung (Krasnow Smart-seq2, 9,409 cells)~\citep{travaglini2020molecular}. Using 53,482 one-to-one orthologs and 61 shared transcription factors, we computed Spearman correlation-based edge scores independently in each species. Human data were subsampled to 10,000 cells. Edges with $|\rho| < 0.05$ were discarded, yielding 25,876 matched edges.

\subsection*{Pseudotime directionality audit}

We audited 56 well-characterized TF--target regulatory pairs spanning three immune lineages in the Tabula Sapiens immune subset (20,000 cells). Diffusion pseudotime~\citep{haghverdi2016diffusion} was computed per lineage using 2,000 HVGs, 30 PCA components, and $k=15$ nearest neighbors.

\subsection*{Batch and donor leakage audit}

TF--target edge scores were computed as Pearson correlations for $\sim$8,000 TF--target pairs per tissue. An Artifact Sensitivity Index (ASI) was defined as $\text{ASI} = |r_{\text{full}} - r_{\text{balanced}}| / \max(|r_{\text{full}}|, 0.01)$. Edges with $\text{ASI} > 0.5$ were flagged.

\subsection*{Uncertainty calibration of edge scores}

We evaluated the calibration of six edge-scoring methods against Perturb-seq ground truth from CRISPRi experiments. Post-hoc calibration used Platt scaling~\citep{platt1999probabilistic} and isotonic regression~\citep{niculescumizil2005predicting}. Split conformal prediction sets~\citep{vovk2005algorithmic} were constructed with finite-sample coverage guarantees.

\subsection*{Synthetic ground-truth validation}

Ground-truth networks had sparse connectivity ($\rho = 0.15$) with hierarchical TF--regulator--target structure. Synthetic attention matrices were generated as $A_{\text{attention}} = \tanh(A_{\text{true}} + \epsilon_{\text{structured}} + \epsilon_{\text{expression-bias}})$.

\subsection*{Multi-model validation}

We tested scVI~\citep{lopez2018deep} (latent-distance edges) and C2S-Pythia (405M-parameter causal LM), which showed qualitatively similar near-random GRN recovery (AUROC 0.48--0.53). Full results are reported in Supplementary Note~13.

\subsection*{Attention residualization on expression covariates}

To avoid overfitting, we used cross-fitted residualization (5-fold). We tested robustness across: multiple covariate sets, OLS vs.\ GBDT residualizers, signed vs.\ absolute correlation, and 10 random seeds.

\subsection*{Degree-preserving null models}

We implemented two null models: (i) label-shuffling null ($n = 1{,}000$ permutations) and (ii) degree-preserving null ($n = 200$ permutations) using the curveball algorithm~\citep{strona2014fast}.

\subsection*{Attention--correlation mapping}

Cross-tissue analysis matched Geneformer attention edges (DLPFC brain) against Spearman correlations (Tabula Sapiens immune, 20,000 cells).

\clearpage

\section{Biological Characterization of Attention Patterns}
\label{supnote:bio_characterization}

To characterize what biological relationships attention patterns encode beyond co-expression, we evaluated Geneformer V2-316M attention edges (2,000 K562 control cells, 2,000 HVGs) against six reference databases: TRRUST (transcriptional regulation; 175 pairs in HVG), STRING $\geq$700 (protein--protein interactions; 2,747 pairs), STRING $\geq$900 (high-confidence PPI; 1,675 pairs), Reactome (pathway co-membership; 238,640 pairs), KEGG (pathway co-membership; 24,812 pairs), and GO Biological Process (functional co-annotation; 135,596 pairs). For each database, we computed AUROC of attention edges at all 18 layers against the reference edge set, and compared to Spearman correlation edges.

\subsection*{Layer-specific biological specialization}

Attention patterns show clear layer-specific specialization. Protein--protein interaction signal peaks at the earliest layer (STRING $\geq$700: AUROC $= 0.640$ at L0) and decreases monotonically with depth (Spearman $\rho = -0.608$, $p_\text{raw} = 0.0075$, $q_\text{BH} = 0.011$). STRING $\geq$900 shows the same pattern ($\rho = -0.581$, $q_\text{BH} = 0.014$). Conversely, transcriptional regulatory signal (TRRUST) increases with depth ($\rho = +0.511$, $q_\text{BH} = 0.030$), peaking at L15 (AUROC $= 0.750$). Functional co-annotation signals (KEGG: $\rho = +0.831$, $q_\text{BH} < 10^{-4}$; GO BP: $\rho = +0.846$, $q_\text{BH} < 10^{-4}$) and Reactome ($\rho = +0.731$, $q_\text{BH} = 0.001$) also increase with depth but with weaker absolute signal (AUROC $0.52$--$0.56$ at best layers). All six Spearman trend tests survive Benjamini-Hochberg correction at $\alpha = 0.05$.

The cross-layer profiles for PPI and regulation are anti-correlated: STRING $\geq$900 vs.\ TRRUST $\rho = -0.546$ ($p = 0.019$); STRING $\geq$700 vs.\ TRRUST $\rho = -0.445$ ($p = 0.064$, marginal). Meanwhile, KEGG, GO BP, and Reactome profiles are strongly positively correlated with each other ($\rho = 0.75$--$0.89$, all $p < 0.001$) and with TRRUST ($\rho = 0.35$--$0.45$, marginal), forming a coherent ``functional/regulatory'' cluster distinct from the PPI signal.

\begin{table}[H]
\centering
\caption{\textbf{AUROC of Geneformer attention edges against six biological reference databases across all 18 layers.} Correlation baseline shown in last row.}
\label{tab:bio_char_auroc}
\small
\begin{tabular}{lcccccc}
\toprule
Layer & STRING$_{\geq700}$ & STRING$_{\geq900}$ & TRRUST & Reactome & KEGG & GO BP \\
\midrule
L0  & \textbf{0.640} & \textbf{0.644} & 0.558 & 0.505 & 0.530 & 0.526 \\
L1  & 0.517 & 0.501 & 0.691 & 0.501 & 0.484 & 0.501 \\
L4  & 0.546 & 0.530 & 0.686 & 0.516 & 0.525 & 0.515 \\
L8  & 0.530 & 0.519 & 0.631 & 0.509 & 0.531 & 0.534 \\
L13 & 0.574 & 0.559 & 0.708 & 0.523 & 0.548 & 0.530 \\
L15 & 0.525 & 0.500 & \textbf{0.750} & 0.521 & 0.555 & 0.541 \\
L16 & 0.475 & 0.457 & 0.733 & 0.516 & \textbf{0.564} & \textbf{0.544} \\
L17 & 0.485 & 0.460 & 0.664 & 0.516 & 0.539 & 0.541 \\
\midrule
Corr. & 0.562 & 0.559 & 0.649 & 0.514 & 0.541 & 0.513 \\
\bottomrule
\end{tabular}
\end{table}

\subsection*{Partial correlation controlling for expression similarity}

To test whether attention captures biological structure beyond expression similarity, we computed partial correlations controlling for Spearman expression correlation (using matched positive and negative reference pairs). TRRUST signal is robust: 97\% of the attention--membership correlation is retained after controlling for expression (partial $r = 0.353$, $p = 1.1 \times 10^{-11}$; raw $r = 0.363$). GO BP retains 89\% and KEGG retains 72\%. Reactome signal, however, is non-significant after expression control (partial $r = 0.015$, $p = 0.12$), confirming Reactome pathway co-membership as a null result.

\subsection*{Top-edge enrichment analysis}

Fisher's exact tests for overlap between the top-1,000 highest-attention edges and each reference database (TRRUST excluded due to 0.009\% base rate yielding 0 overlap at all layers). After BH correction across 30 tests (5 layers $\times$ 3 databases $\times$ 2 tails):

\begin{itemize}
\item \textbf{L0 (early, PPI-related)}: Significant enrichment for KEGG (OR $= 3.38$, $q < 10^{-10}$), GO BP (OR $= 1.47$, $q = 0.001$), and Reactome (OR $= 1.25$, $q = 0.028$).
\item \textbf{L17 (late, regulation-related)}: Strongest enrichment across all databases: Reactome (OR $= 1.99$, $q < 10^{-16}$), KEGG (OR $= 3.71$, $q < 10^{-12}$), GO BP (OR $= 2.12$, $q < 10^{-13}$).
\item \textbf{L10 (mid-depth)}: No significant enrichment for any database (all OR $\approx 1.0$).
\end{itemize}

The concentration of enrichment at the periphery (L0 and L17) with a dead zone at mid-depth supports the layer-specialization interpretation.

\subsection*{Interpretation}

Attention patterns in Geneformer capture a hierarchy of biological signals with layer-specific organization: the input layer (L0) preferentially encodes physical protein--protein interactions, while deeper layers progressively encode transcriptional regulation and functional co-annotation. This hierarchy is real---it survives pairwise expression control (97\% of TRRUST signal retained) and is statistically robust across all six databases. However, acknowledging this hierarchy does not contradict the main finding that attention provides no incremental value over gene-level features for perturbation prediction (main text, Section~2.3). The key distinction is between the confound controls: the partial correlations here remove pairwise expression similarity, whereas the incremental-value analysis in the main text controls for gene-level features (variance, mean expression, dropout rate). Gene-level features, not pairwise co-expression, are the dominant confound. Thus, the title's claim---that attention captures co-expression rather than causal regulation---is more precisely stated as: attention captures biologically structured signals, including regulatory ones, but these signals are entirely redundant with gene-level features and provide no unique information for predicting the functional consequences of genetic perturbations.

\clearpage

\section{Value-Weighted Edge Extraction}
\label{supnote:value_weighted}

The main paper shows that ablating TRRUST-ranked attention heads has no effect on perturbation prediction, while ablating random heads does ($d = 0.33$, $p < 10^{-8}$). This suggests that perturbation-predictive computation resides in the value/FFN pathway rather than in the attention pattern itself. We tested whether \emph{value-weighted} edge scores---computed from the context layer $\text{softmax}(QK^\top/\sqrt{d}) \cdot V$ rather than the raw attention pattern $\text{softmax}(QK^\top/\sqrt{d})$---better capture regulatory structure.

\subsection*{Methods}

Using forward hooks on each \texttt{BertSelfAttention} module in Geneformer V2-316M, we extracted the context layer $A_\text{vh} = \text{softmax}(QK^\top/\sqrt{d}) \cdot V$ (shape: $n_\text{genes} \times d_\text{head}$) for each layer and head across 2,000 K562 control cells. Pairwise edge scores were computed as cosine similarity between gene representations: $\text{edge}(i,j) = \cos(A_\text{vh}[i,:], A_\text{vh}[j,:])$, averaged across heads within each layer. We also tested centroid cosine similarity (computing similarity on the mean context vector across cells rather than averaging per-cell similarities) and dot product. A total of 1,941 genes (of 2,000 HVGs) were in the Geneformer vocabulary; 269 of 280 perturbations were evaluable.

\subsection*{Results}

Value-weighted cosine similarity significantly \emph{underperforms} both raw attention and Spearman correlation at every layer:

\begin{table}[H]
\centering
\caption{\textbf{Edge score comparison: value-weighted vs.\ raw attention.}}
\label{tab:vw_comparison}
\begin{tabular}{lcccc}
\toprule
Metric & Best VW Cosine & Best Raw Attn & Correlation & Variance \\
\midrule
Pert-first AUROC & 0.587 (L12) & 0.787 (L14) & 0.706 & 0.887 \\
TRRUST AUROC & 0.606 (L12) & 0.718 (L15) & 0.638 & --- \\
\bottomrule
\end{tabular}
\end{table}

Paired Wilcoxon tests ($n = 269$ perturbations): VW cosine vs.\ raw attention $\Delta = -0.200$, $p = 4.9 \times 10^{-39}$, $d = -1.68$; VW cosine vs.\ correlation $\Delta = -0.120$, $p = 2.4 \times 10^{-14}$, $d = -1.00$. Centroid cosine and dot product variants also underperform raw attention (best TRRUST AUROC: centroid $= 0.623$ at L5, dot $= 0.589$ at L14).

In the incremental-value test (5-fold GroupKFold logistic regression), adding VW cosine to gene-level features slightly \emph{hurts} performance ($\Delta$AUROC $= -0.009$; gene-only $= 0.865$, gene + VW $= 0.856$). VW edges alone achieve AUROC $= 0.587$, well below both raw attention ($0.707$) and correlation ($0.637$).

\subsection*{Interpretation}

The context layer $\text{softmax}(QK^\top/\sqrt{d}) \cdot V$ mixes information from all attended genes, collapsing the pairwise structure that even raw attention preserves. Cosine similarity between context vectors measures whether two genes receive similar blends of information from the attention mechanism---``information-receipt similarity''---which is a fundamentally different quantity from direct gene-to-gene attention coupling. This negative result, combined with the ablation finding that random heads are more causally important than regulatory heads, indicates that perturbation-predictive computation is distributed across the network in a form not recoverable from any simple attention-derived edge score---neither the attention pattern nor the value-weighted context.

\clearpage

\section{Per-TF Characterization (Exploratory)}
\label{supnote:per_tf_exploratory}

We examined whether the per-TF AUROC for attention-derived edge scores varies systematically with TF biology. Among 18 evaluable TFs in the Tabula Sapiens immune dataset, manually annotated master regulators ($n = 9$: GATA1, PPARG, WT1, NKX2-5, FOXA1, KLF1, PAX3, TAL1, LEF1) had higher mean AUROC ($0.80 \pm 0.18$) than other TF categories (signal-dependent, lineage-specific, housekeeping; $n = 9$: $0.58 \pm 0.18$; permutation $p = 0.011$, 10,000 label shuffles). However, this comparison did not survive Benjamini-Hochberg correction across all 11 tests performed ($q = 0.12$).

\textbf{Severe power limitation.} 13 of 18 TFs had only a single evaluable TRRUST target in the HVG set, meaning their AUROC reflects the rank of one gene among $\sim$2,000 rather than a regulon-level assessment. When restricted to the 5 TFs with $\geq$3 evaluable targets (GATA1, PPARG, EGR1, WT1, FOXF2), the master regulator advantage was not significant (permutation $p = 0.30$).

\begin{table}[H]
\centering
\caption{\textbf{Per-TF AUROC and characterization.} 18 evaluable TFs.}
\label{tab:per_tf_char}
\small
\begin{tabular}{llccc}
\toprule
TF & Category & Targets in HVG & AUROC & Successful \\
\midrule
LEF1 & master regulator & 1 & 0.997 & Yes \\
KLF1 & master regulator & 1 & 0.995 & Yes \\
PPARG & master regulator & 10 & 0.900 & Yes \\
FOXA1 & master regulator & 2 & 0.874 & Yes \\
MAL & signal-dependent & 1 & 0.860 & Yes \\
EGR1 & signal-dependent & 9 & 0.852 & Yes \\
GATA1 & master regulator & 18 & 0.842 & Yes \\
PAX3 & master regulator & 1 & 0.802 & --- \\
WT1 & master regulator & 5 & 0.675 & No \\
KLF3 & lineage-specific & 1 & 0.644 & No \\
YBX1 & housekeeping & 1 & 0.640 & No \\
PA2G4 & housekeeping & 1 & 0.630 & No \\
NKX2-5 & master regulator & 2 & 0.572 & No \\
TAL1 & master regulator & 1 & 0.556 & No \\
FOXF2 & lineage-specific & 3 & 0.500 & No \\
TBX5 & lineage-specific & 1 & 0.401 & No \\
RXRA & signal-dependent & 2 & 0.359 & No \\
NCOA4 & signal-dependent & 1 & 0.349 & No \\
\bottomrule
\end{tabular}
\end{table}

No tested TF property---including mean expression ($\rho = -0.14$, $q = 0.66$), expression variance ($\rho = -0.20$, $q = 0.58$), nonzero fraction ($\rho = -0.01$, $q = 0.97$), or full regulon size ($\rho = 0.26$, $q = 0.53$)---significantly predicted AUROC after BH correction. The evaluable regulon size showed the strongest raw trend ($\rho = -0.46$, $p = 0.06$, $q = 0.33$), suggesting that TFs with more evaluable targets tend to have AUROC closer to 0.5, consistent with regulon-level AUROC being a harder test than single-gene rank.

We treat the master regulator association as hypothesis-generating: the pattern is consistent with attention more reliably capturing regulatory relationships for master regulators with large, well-characterised regulons, but the small sample ($n = 18$) and severe single-target problem preclude definitive conclusions. Larger TF databases with greater regulon coverage (e.g., DoRothEA, ChIP-Atlas) would be needed to test this rigorously.

\bibliography{references}

\end{document}